\documentclass[sn-mathphys,Numbered]{sn-jnl}

\usepackage{graphicx}%
\usepackage{fullpage}
\usepackage{multirow}%
\usepackage{amsmath,amssymb,amsfonts}%
\usepackage{amsthm}%
\usepackage{mathrsfs}%
\usepackage[title]{appendix}%
\usepackage{xcolor}%
\usepackage{textcomp}%
\usepackage{manyfoot}%
\usepackage{booktabs}%
\usepackage{bm}
\usepackage{algorithm}%
\usepackage{algorithmicx}%
\usepackage{algpseudocode}%
\usepackage{listings}%
\usepackage{physics}
\usepackage{siunitx}
\usepackage{upgreek}
\usepackage{orcidlink}

\begin{document}

\title[Laser spectroscopy and CP-violation sensitivity of actinium monofluoride]{Laser spectroscopy and CP-violation sensitivity of actinium monofluoride}


\author*[1,2,3]{\fnm{M.} \sur{Athanasakis-Kaklamanakis\orcidlink{0000-0003-0336-5980}}}\email{m.athkak@cern.ch}
\author*[4,5]{\fnm{M.} \sur{Au}\orcidlink{0000-0002-8358-7235}}\email{mia.au@cern.ch}
\author[6]{\fnm{A.} \sur{Kyuberis}\orcidlink{0000-0001-7544-3576}}
\author[7]{\fnm{C.} \sur{Z\"ulch}\orcidlink{0009-0007-2563-1342}}
\author[7,8,9,10]{\fnm{K.} \sur{Gaul}\orcidlink{0000-0002-6990-6949}}
\author[11]{\fnm{H.} \sur{Wibowo}\orcidlink{0000-0003-4093-0600}}
\author[12]{\fnm{L.} \sur{Skripnikov}\orcidlink{0000-0002-2062-684X}}

\author[2,1]{\fnm{L.} \sur{Lalanne}\orcidlink{0000-0003-1207-9038}}
\author[13,4]{\fnm{J.~R.} \sur{Reilly}}
\author[1,2]{\fnm{\'A.} \sur{Koszor\'us}\orcidlink{0000-0001-7959-8786}}
\author[2]{\fnm{S.} \sur{Bara}\orcidlink{0000-0001-7129-3486}}
\author[9,5]{\fnm{J.} \sur{Ballof}\orcidlink{0000-0003-0874-500X}}
\author[7,1]{\fnm{R.} \sur{Berger}}
\author[4]{\fnm{C.} \sur{Bernerd}\orcidlink{0000-0002-2183-9695}}
\author[6]{\fnm{A.} \sur{Borschevsky}\orcidlink{0000-0002-6558-1921}}
\author[14,15]{\fnm{A.~A.} \sur{Breier}\orcidlink{0000-0003-1086-9095}}
\author[4]{\fnm{K.} \sur{Chrysalidis}\orcidlink{0000-0003-2908-8424}}
\author[2]{\fnm{T.~E.} \sur{Cocolios}\orcidlink{0000-0002-0456-7878}}
\author[2]{\fnm{R.~P.} \sur{de Groote}\orcidlink{0000-0003-4942-1220}}
\author[2]{\fnm{A.} \sur{Dorne}}
\author[11]{\fnm{J.} \sur{Dobaczewski}\orcidlink{0000-0002-4158-3770}}
\author[2]{\fnm{C.~M.} \sur{Fajardo Zambrano}\orcidlink{0000-0002-6088-6726
}}
\author[13]{\fnm{K.~T.} \sur{Flanagan}\orcidlink{0000-0003-0847-2662}}
\author[16,17]{\fnm{S.} \sur{Franchoo}}
\author[2]{\fnm{J.~D.} \sur{Johnson}\orcidlink{0000-0003-4397-5732}}
\author[18,19]{\fnm{R.~F.} \sur{Garcia Ruiz}\orcidlink{0000-0002-2926-5569}}
\author[20]{\fnm{D.} \sur{Hanstorp}}
\author[21]{\fnm{S.} \sur{Kujanp\"a\"a}\orcidlink{0000-0002-5709-3442}}
\author[22]{\fnm{Y.~C.} \sur{Liu}}
\author[13]{\fnm{K.~M.} \sur{Lynch}\orcidlink{0000-0001-8591-2700}}
\author[13]{\fnm{A.} \sur{McGlone}\orcidlink{0000-0003-4424-865X}}
\author[12]{\fnm{N.~S.} \sur{Mosyagin}\orcidlink{0000-0002-9158-494X}}
\author*[2]{\fnm{G.} \sur{Neyens}\orcidlink{0000-0001-8613-1455}}\email{gerda.neyens@kuleuven.be}
\author[20]{\fnm{M.} \sur{Nichols}\orcidlink{0000-0003-3693-7295}}
\author[1]{\fnm{L.} \sur{Nies}\orcidlink{0000-0003-2448-3775}}
\author[18]{\fnm{F.} \sur{Pastrana}\orcidlink{0009-0008-7469-7513}}
\author[4]{\fnm{S.} \sur{Rothe}}
\author[23,24]{\fnm{W.} \sur{Ryssens}\orcidlink{0000-0002-4775-4403}}
\author[2]{\fnm{B.} \sur{van den Borne}\orcidlink{0000-0003-3348-7276}}
\author[4,13]{\fnm{J.} \sur{Wessolek}}
\author[18,19]{\fnm{S.~G.} \sur{Wilkins}\orcidlink{0000-0001-8897-7227}}
\author[22]{\fnm{X.~F.} \sur{Yang}\orcidlink{0000-0002-1633-4000}}

\affil[1]{\orgdiv{Experimental Physics Department}, \orgname{CERN}, \orgaddress{{Geneva}, \postcode{CH-1211}, \country{Switzerland}}}

\affil[2]{\orgdiv{Instituut voor Kern- en Stralingsfysica}, \orgname{KU Leuven}, \orgaddress{{Leuven}, \postcode{B-3001}, \country{Belgium}}}

\affil[3]{\orgdiv{Centre for Cold Matter}, \orgname{Imperial College London}, \orgaddress{{London}, \postcode{SW7 2AZ}, \country{United Kingdom}}}

\affil[4]{\orgdiv{Systems Department}, \orgname{CERN}, \orgaddress{{Geneva}, \postcode{CH-1211}, \country{Switzerland}}}

\affil[5]{\orgdiv{Department of Chemistry}, \orgname{Johannes Gutenberg-Universit\"{a}t Mainz}, \orgaddress{{Mainz}, \postcode{55099}, \country{Germany}}}

\affil[6]{\orgdiv{Van Swinderen Institute of Particle Physics and Gravity}, \orgname{University of Groningen}, \orgaddress{{Groningen}, \postcode{9712 CP}, \country{Netherlands}}}

\affil[7]{\orgdiv{Fachbereich Chemie}, \orgname{Philipps-Universit\"{a}t Marburg}, \postcode{35032}, \orgaddress{{Marburg},  \country{Germany}}}

\affil[8]{\orgname{Helmholtz Institute Mainz}, \postcode{55099}, \orgaddress{Mainz}, \country{Germany}}

\affil[9]{\orgdiv{GSI Helmholtzzentrum für Schwerionenforschung GmbH}, \orgname{GSI}, \postcode{64291}, \orgaddress{{Darmstadt},  \country{Germany}}}

\affil[10]{\orgdiv{Institut f\"ur Physik}, \orgname{Johannes Gutenberg-Universit\"{a}t Mainz}, \orgaddress{{Mainz}, \postcode{55099}, \country{Germany}}}

\affil[11]{\orgdiv{School of Physics, Engineering and Technology}, \orgname{University of York}, \orgaddress{{York}, \postcode{YO10 5DD}, \country{United Kingdom}}}

\affil[12]{\orgdiv{Affiliation covered by a cooperation agreement with CERN at the time of the experiment}}

\affil[13]{\orgdiv{Department of Physics and Astronomy}, \orgname{The University of Manchester}, \orgaddress{{Manchester}, \postcode{M13 9PL}, \country{United Kingdom}}}

\affil[14]{\orgdiv{Institut f\"ur Physik und Astronomie},  \orgname{Technische Universit\"at Berlin}, \orgaddress{\postcode{10623}, \country{Germany}}}

\affil[15]{\orgdiv{Laboratory for Astrophysics, Institute of Physics}, \orgname{University of Kassel}, \orgaddress{{Kassel}, \postcode{34132}, \country{Germany}}}

\affil[16]{\orgname{Laboratoire Ir\`{e}ne Joliot-Curie}, \orgaddress{{Orsay}, \postcode{F-91405}, \country{France}}}

\affil[17]{\orgname{University Paris-Saclay}, \orgaddress{{Orsay}, \postcode{F-91405}, \country{France}}}

\affil[18]{\orgdiv{Department of Physics}, \orgname{Massachusetts Institute of Technology}, \orgaddress{{Cambridge}, \postcode{MA 02139}, \country{USA}}}

\affil[19]{\orgdiv{Laboratory for Nuclear Science}, \orgname{Massachusetts Institute of Technology}, \orgaddress{{Cambridge}, \postcode{MA 02139}, \country{USA}}}

\affil[20]{\orgdiv{Department of Physics}, \orgname{University of Gothenburg}, \orgaddress{{Gothenburg}, \postcode{SE-41296}, \country{Sweden}}}

\affil[21]{\orgdiv{Department of Physics}, \orgname{University of Jyväskylä}, \orgaddress{{Jyväskylä}, \postcode{40351}, \country{Finland}}}

\affil[22]{\orgdiv{School of Physics and State Key Laboratory of Nuclear Physics and Technology}, \orgname{Peking University}, \orgaddress{{Beijing}, \postcode{100971}, \country{China}}}

\affil[23]{\orgdiv{Institut d’Astronomie et d’Astrophysique}, \orgname{Universit\'{e} libre de Bruxelles}, \orgaddress{Brussels} \postcode{1050}, \country{Belgium}}

\affil[24]{\orgdiv{Brussels Laboratory of the Universe - BLU-ULB}, \orgname{Universit\'{e} libre de Bruxelles}, \orgaddress{Brussels} \postcode{1050}, \country{Belgium}}


\abstract{
The apparent invariance of the strong nuclear force under combined charge conjugation and parity (\textit{CP}) remains an open question in modern physics. Precision experiments with heavy atoms and molecules can provide stringent constraints on \textit{CP} violation via searches for effects due to permanent electric dipole moments and other \textit{CP}-odd properties in leptons, hadrons, and nuclei. Radioactive molecules have been proposed as highly sensitive probes for such searches, but experiments with most such molecules have so far been beyond technical reach. Here we report the first production and spectroscopic study of a gas-phase actinium molecule, $^{227}$AcF. We observe the predicted strongest electronic transition from the ground state, which is necessary for efficient readout in searches of symmetry-violating interactions. 
Furthermore, we perform electronic- and nuclear-structure calculations for $^{227}$AcF to determine its sensitivity to various \textit{CP}-violating parameters, and find that a realistic, near-term experiment with a precision of 1\,mHz would improve current constraints on the \textit{CP}-violating parameter hyperspace by three orders of magnitude. Our results thus highlight the potential of $^{227}$AcF for exceptionally sensitive searches of \textit{CP} violation.
}

\maketitle

\section*{Introduction}\label{sec:intro}
Providing a satisfactory explanation for the cosmological baryon asymmetry requires that the combined symmetry of charge conjugation and parity (\textit{CP}) is broken, following Sakharov's conditions~\cite{Sakharov1991}. Within the Standard Model (SM) of particle physics, \textit{CP} violation emerges in the weak nuclear force via the complex phase of the Cabibbo–Kobayashi–Maskawa matrix, which is responsible for the observed \textit{CP}-violating decay of $K$ mesons, and in the strong nuclear force via a term quantified by the quantum chromodynamics (QCD) $\Bar{\theta}$ phase. The upper bounds to \textit{CP}-violating terms within the SM are considered not to be able to account for the observed baryon asymmetry in the universe~\cite{Bodeker2021Baryogenesis}, and thus new sources of \textit{CP} violation are theorized to exist. Signatures of leptonic, hadronic, and nuclear \textit{CP}-odd properties manifest as symmetry-violating electromagnetic moments, which in turn lead to transition frequency shifts in atoms and molecules that can in principle be observed via high-precision measurements. These systems thus offer a sensitive approach to search for \textit{CP} violation and physics beyond the SM~\cite{Safronova2018}.

Due to the electrostatic screening by the bound electrons, neutral atoms and molecules are not suitable probes for measurements of the \textit{CP}-odd electric dipole moment (EDM) of the nucleus~\cite{Schiff1963}. Instead, experiments search for the closely related nuclear Schiff moment~\cite{Flambaum2002}, which is linked to the QCD Lagrangian via a chiral effective field theory of the pion-nucleon interaction~\cite{Haxton1983b,Maekawa2011,Dobaczewski2018}. Despite global experimental efforts to measure the Schiff moment in $^{199}$Hg~\cite{Graner2016}, $^{205}$Tl~\cite{Grasdijk2021}, $^{225}$Ra~\cite{Bishof2016}, and other nuclei, only upper bounds have been determined so far. Future efforts to improve experimental precision are thus accompanied by theoretical investigations to identify nuclei with an enhanced Schiff moment, thus maximizing the chances of observing a non-zero signal in the laboratory.

\begin{figure}
    \centering
    \includegraphics[width=0.99\textwidth]{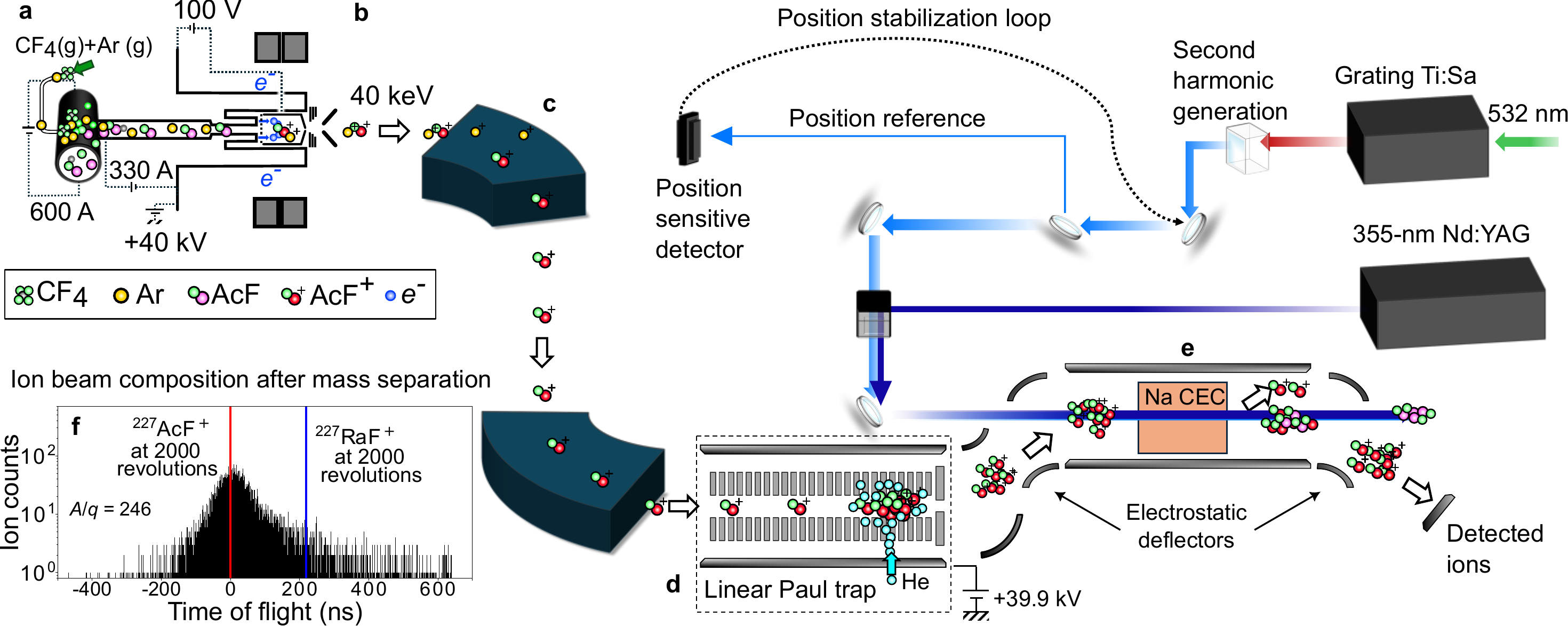}
    \caption{\color{black}{Schematic of the experiment. From top left: \textbf{(a)} Target and ion-source unit releasing Ac to form AcF$_x$ molecules from injected CF$_4$. \textbf{(b)} All species are ionized by accelerated electrons within the anode volume and extracted across a voltage difference (40\,kV) to ground potential. The ion beam is then \textbf{(c)} mass-separated, \textbf{(d)} cooled and bunched before \textbf{(e)} neutralization via charge exchange with sodium vapor and laser resonance ionization. Inset \textbf{(f)} shows the time-of-flight (TOF) spectrum of the ion beam after 2000 revolutions in a multi-reflection TOF mass spectrometer~\cite{Wolf2013,Au2024AcFx}. Expected TOF for $^{227}$Ac$^{19}$F$^+$ ($t_{1/2}=21.7$\,y) and $^{227}$Ra$^{19}$F$^+$ ($t_{1/2}=42.5$\,min) shown in red and blue lines, respectively. The x-axis is offset by the expected TOF for $^{227}$Ac$^{19}$F$^+$. No molecules other than $^{227}$Ac$^{19}$F$^+$ were identified in the beam.}}
    \label{fig:schematic}
\end{figure}

Heavy radioactive nuclei with static octupole deformation possess enhanced Schiff moments in their intrinsic (body-fixed) frame of reference, owing to the close-lying opposite-parity states of same angular momentum that arise from their reflection-asymmetric shape~\cite{Flambaum2020a,Dobaczewski2018}. Of the nuclei that have been theoretically investigated so far, the longest-lived actinium isotopes $^{225,227}$Ac are expected to possess the largest nuclear Schiff moments~\cite{Flambaum2020a}\footnote{The prediction for the exceptionally large Schiff moment of $^{229}$Pa relies on an unconfirmed value for the spacing of the opposite-parity doublet and a tentative spin assignment for its ground state.}, and M\"ossbauer spectroscopy of $^{227}$Ac has recently been proposed as a possible pathway to study its \textit{CP}-violating odd-electric and even-magnetic nuclear moments~\cite{Scheck2023Mossbauer227Ac}. Experiments with heavy polar molecules are currently the most sensitive approach to search for hadronic, leptonic, and nuclear \textit{CP}-violating properties~\cite{Safronova2018}, thanks to their enormous internal electric field and high polarizability in the presence of small applied fields. Therefore, precision experiments using polar actinium molecules, such as actinium monofluoride (AcF), offer the most compelling direction for searches of \textit{CP} violation with heavy and deformed actinium nuclei.

The longest-lived actinium isotope $^{227}$Ac has a half-life of 21.8~years and a long decay chain of radioactive products. Thus, manufacturing macroscopic actinium samples and transporting them for use with existing experimental setups for EDM searches would be prohibitively radioactive. Radioactive ion beam (RIB) facilities such as CERN-ISOLDE~\cite{Catherall2017} are able to deliver actinium molecules for study, offering the ability to perform experiments with such compounds in environments with appropriate radiation-protection measures in place. So far, molecular formation of radioactive species produced via the isotope separation on-line (ISOL) method~\cite{Au2023InsourceIntrap,Au2023Thesis} used at the CERN-ISOLDE facility has proven to reliably deliver isotopically pure ion beams of radioactive molecules. Controlled formation of molecules within the ISOL source environment is additionally used for purification of radioactive ion beams from contaminants, and for improving the extraction efficiency of the radioactive isotopes~\cite{Koester2007,Au2023InsourceIntrap,Ballof2021Thesis}. Past investigations using the ISOL method demonstrate that in addition to exhibiting limited extraction efficiency due to its refractory chemical properties~\cite{jajcisinovaProductionStudyFr2024}, actinium also forms highly refractory ceramics in an oxygen-rich environment~\cite{Johnson2023}. This limits extraction rates of actinium itself as well as its oxides and nitrides, motivating developments to form and extract the more volatile fluoride ions to reach higher extraction rates~\cite{Au2024AcFx}. As the shot-noise-limited precision scales with the square root of the number of probed molecules~\cite{Lasner2018}, the availability of AcF contributes significantly to its potential as a compelling system for precision spectroscopy.

\section*{Production and spectroscopy of AcF}
No experimental spectroscopic study of AcF has been reported so far. Detailed knowledge of the strongest electronic transition from the $X$~$^1 \Sigma^+$ ground state is critical, as the $X$~$^1 \Sigma^+$ state is envisioned for precision measurements and is not limited by a finite radiative lifetime like the metastable $^3\Delta_1$ electronic states in ThO and HfF$^+$ that have been used in electron EDM studies. Knowledge of the strongest transition from $X$~$^1 \Sigma^+$ is necessary to assess the state readout efficiency via optical detection and the potential for laser cooling. State-readout efficiency is in turn crucial for a sensitive high-precision experiment~\cite{Lasner2018,Ho2020a,Grasdijk2021}. Therefore, spectroscopic experiments are required to characterize the relevant properties of AcF, while also providing a versatile benchmark of predictive relativistic ab initio quantum chemistry for radioactive molecules~\cite{Opportunities2024Progressb,AthKak2025ElectronCorrelation}, as recently applied on AcF~\cite{Skripnikov2023AcF}.

{
In this work, intense, chemically and isotopically pure beams of $^{227}$Ac$^{19}$F$^+$ were produced at the CERN-ISOLDE facility, as shown in Fig.~\ref{fig:schematic}. Actinium nuclides were produced in nuclear reactions induced by 1.4-GeV protons from the CERN Proton Synchrotron Booster bombarding a thick, room-temperature target of UC$_\textrm{x}$. Two weeks after proton irradiation, the radiogenic actinium nuclides were extracted by resistively heating the target to $>$1300\,$^{\circ}$C, facilitating nuclide diffusion and effusion through the UC$_\textrm{x}$ matrix~\cite{Ramos2020}. Carbon tetrafluoride (CF$_4$) was supplied to the heated target at a rate of \SI{0.065 }{\nano\mol\per\second} for the formation of fluoride molecules (Fig.~\ref{fig:schematic}a). Ionization techniques using hot cavities and resonance laser ionization are most frequently used at ISOLDE, but without further information on its chemical properties, these techniques could not be used to produce AcF. A forced-electron-beam-induced arc discharge (FEBIAD) ion source~\cite{Penescu2010} was connected to the target to create ions of the extracted species through electron impact, plasma ionization, and molecular dissociation (Fig.~\ref{fig:schematic}b).

The ion beam was accelerated to 40\,keV and the $^{227}$Ac$^{19}$F$^+$ ions were separated from all other radiogenic products using two magnetic dipole separators in series (Fig.~\ref{fig:schematic}c). The temperature of the target was controlled to regulate the mass-separated $^{227}$Ac$^{19}$F$^+$ ion beam at an intensity between 6$\times$10$^6$ and 2$\times$10$^7$ ions per second for the duration of the experiment. While an intensity of more than 6$\times$10$^7$ ions per second could be achieved at higher temperature, the higher rates correspondingly deplete the in-target actinium inventory faster. The continuous, isotopically purified beam of $^{227}$Ac$^{19}$F$^+$ was then accumulated in a linear Paul trap that reduced the internal temperature of the molecules via collisions with room-temperature helium and released the cooled ion beam in 5-$\upmu$s bunches every 10\,ms (Fig.~\ref{fig:schematic}d).

\begin{figure}
    \centering
    \includegraphics[width=0.99\textwidth]{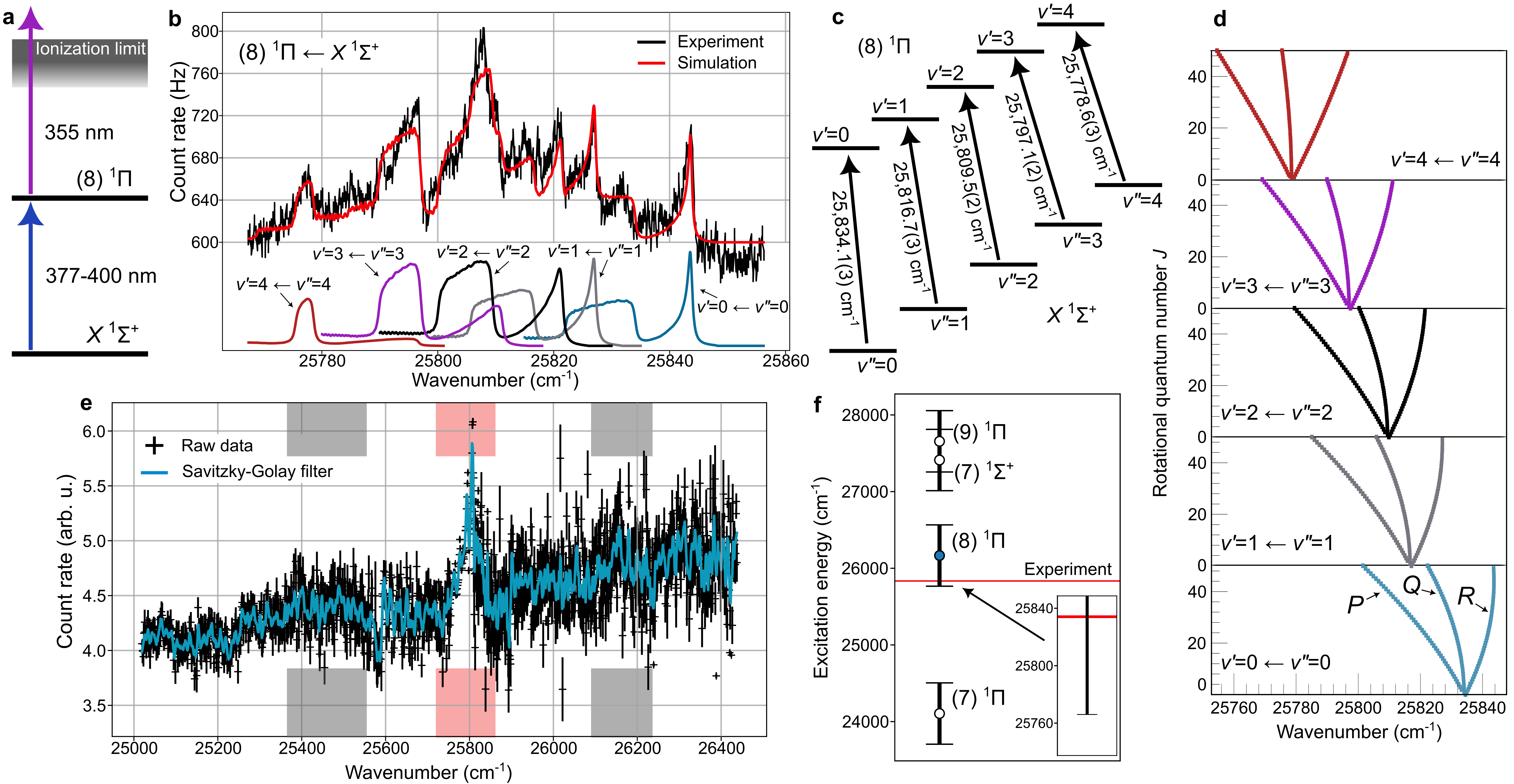}
    \caption{Spectroscopy of AcF and molecular sensitivity to nuclear \textit{CP}-odd properties. \textbf{(a)} Two-step laser resonance ionization scheme used to search for electronic transitions in AcF. \textbf{(b)} Spectrum of the measured vibronic transitions from the ground state in AcF, and simulation based on the contour-fitted molecular constants. \textbf{(c)} Assigned vibronic transitions observed in the spectrum shown in (b). \textbf{(d)} Fortrat diagrams for the measured vibronic transitions, showing the evolution of the $P$, $Q$, and $R$ branches along the diagonal vibrational progression. \textbf{(e)} Ion count rate as a function of the wavenumber of the first excitation laser across the full scanning range. A cubic Savitzky-Golay de-noising filter with a 10-point window length is applied to the raw data to assist in the visualization of data trends. The complete spectrum is constructed from multiple shorter scans. A vertical offset is applied to each scan to compensate for their different background rates due to long-term variations in the beam intensity. The $y$-axis is thus in arbitrary units. The red band marks the region of the spectrum in (b), and the gray bands mark regions investigated and excluded as resonant features. \textbf{(f)} Comparison of excitation wavenumbers to different electronic states calculated with FS-RCC~\cite{Skripnikov2023AcF} and the upper state in the spectrum of (b). The error bars correspond to the theoretical uncertainty in the excitation wavenumber calculations, as reported in Ref.~\cite{Skripnikov2023AcF}.}
    \label{fig:figAcF}
\end{figure}

The bunched ion beam was delivered to the collinear resonance ionization spectroscopy (CRIS) experiment~\cite{Cocolios2013CRIS} (Fig.~\ref{fig:schematic}e). Following in-flight neutralization in a charge-exchange cell of sodium vapor ($\sim$210\,$^\circ$C), the ions that were not neutralized were discarded by electrostatically deflecting them away from the neutral beam. Spectroscopy of $^{227}$AcF was performed at CRIS to search for electronic transitions from the ground state using a two-step pulsed laser scheme (Fig.~\ref{fig:figAcF}a). Ultraviolet light was produced by single-pass second-harmonic generation (SHG) of a pulsed titanium:sapphire (Ti:Sa) laser and was scanned between 377-400~nm to search for a resonant transition from the $X$~$^1 \Sigma^+$ ground state of AcF. A high-power third-harmonic pulsed Nd:YAG laser at 355~nm followed as the second step to non-resonantly ionize the molecules found in an excited electronic state when the first laser step was at the resonant frequency of an electronic transition from $X$~$^1 \Sigma^+$. To continuously scan a broad wavenumber range ($\sim$1,400\,cm$^{-1}$) with the second-harmonic Ti:Sa light, a position-stabilization loop was implemented along with stabilization of the SHG crystal angle to optimize the SHG output power. Further information on the production and spectroscopy can be found in the \textit{Methods}.

Extensive relativistic Fock-space coupled cluster calculations (FS-RCC)~\cite{Skripnikov2023AcF} predict only one transition from $X$~$^1 \Sigma^+$ to an excited electronic state in the scanned wavenumber range, and its calculated transition dipole moment is the largest among all transitions from $X$~$^1 \Sigma^+$. The only resonance across the full scan range (see Fig.~\ref{fig:figAcF}e) was observed from 25,770 to 25,860\,cm$^{-1}$, shown in Fig.~\ref{fig:figAcF}b. Based on the FS-RCC calculations, this spectrum corresponds to the diagonal vibrational progression of $(8)^1 \Pi \leftarrow X$~$^1\Sigma^+$, calculated to be the strongest transition from the ground state, and thus the optimal choice for a high-efficiency optical state-readout scheme in a future precision experiment. With the assistance of calculated molecular constants and Franck-Condon factors (FCFs) for AcF$^+$ and AcF, the peaks in the spectrum were assigned as shown in Figs.~\ref{fig:figAcF}c,d (see \textit{Methods} for details). The calculated FCFs for allowed electronic transitions to other electronic states in the vicinity of $(8)^1 \Pi$ are highly diagonal~\cite{Skripnikov2023AcF}. Therefore, it appears unlikely that the observed spectrum belongs to an overtone and non-diagonal vibrational progression from $X$~$^1 \Sigma^+$ to a state other than $(8)^1 \Pi$, while the diagonal vibrational progression to $(8)^1 \Pi$ is not observed. Other wavenumber regions were further investigated, shown as gray bands in Fig.~\ref{fig:figAcF}e around 25,400 and 26,150\,cm$^{-1}$. In these regions, the number of resonantly-ionized molecules was not affected by the presence or absence of the scanning laser in the interaction region. These regions were thus excluded as transitions corresponding to a resonant excitation from the ground state. The measured excitation energy for the upper state in the spectrum in Fig.~\ref{fig:figAcF}b is in agreement with the ab initio calculations for the $(8)^1 \Pi$ state~\cite{Skripnikov2023AcF}, as shown in Fig.~\ref{fig:figAcF}f.

The contour of the different $v' \leftarrow v''$, $\Delta v =0$ transitions of the measured vibrational progression visibly changes shape as $v$ increases (Fig.~\ref{fig:figAcF}b). This is the result of a gradual change in the slope of the \textit{P}, \textit{Q}, and \textit{R} rotational branches for increasing $v$, as shown in Fig.~\ref{fig:figAcF}d. Additionally, the observed spacing between vibrational transitions becomes uneven for $v>1$ (Fig.~\ref{fig:figAcF}b,c). At the present time, the origin of this is unclear. Rotationally resolved spectroscopy in the future is necessary to provide additional information regarding the anharmonicity of the vibrational potential in the electronic states involved.

The radiative lifetime of the $(8)^1 \Pi$ state was calculated to be $\tau_{\rm{calc}}=6.65$\,ns~\cite{Skripnikov2023AcF}. The lifetime of the observed upper electronic state was investigated experimentally, but only an upper limit (equal to the pulse width of the excitation laser) could be determined at $\tau_{\rm{exp}} \leq 38(4)$\,ns, via the experimental approach previously applied to RaF~\cite{AthKak2024Lifetime}. Consequently, a lower limit on the radiative decay rate is set at $\Gamma \geq 2.6(3) \times 10^7$\,s$^{-1}$. The strength of the measured transition corroborates the assignment of the upper state as $(8)^1 \Pi$. Future experimental study is of interest to provide a direct lifetime measurement.

\begin{figure}
    \centering
    \includegraphics[width=0.99\textwidth]{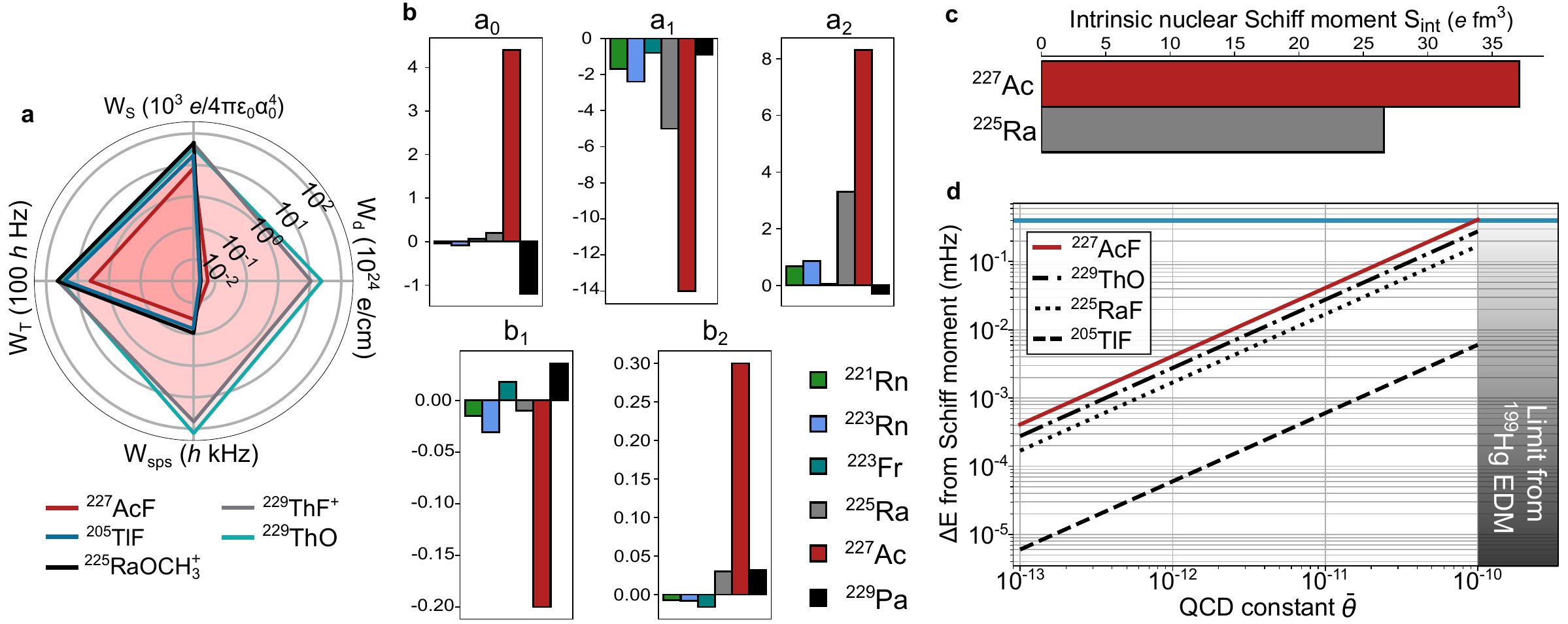}
    \caption{Sensitivity of the $^{227}$Ac nucleus to \textit{CP} violation. \textbf{(a)} Comparison of molecular sensitivity to the nuclear Schiff moment ($W_\mathcal{S}$), tensor-pseudotensor electron-nucleon interaction ($W_\mathrm{T}$), electric dipole moment of the electron ($W_\mathrm{d}$), and scalar-pseudoscalar electron-nucleon interaction ($W_\mathrm{sps}$) for the $^3\Delta_1$ electronic states in $^{229}$ThO and $^{229}$ThF$^+$, and for the $^1\Sigma^+$ electronic ground states in $^{227}$AcF, $^{205}$TlF, and $^{225}$RaOCH$^+_3$. \textbf{(b)} Comparison of the sensitivity coefficients (in units of $e$\,fm$^3$) of the laboratory nuclear Schiff moment to the \textit{CP}-odd pion-nucleon (isoscalar $a_0$, isovector $a_1$, isotensor $a_2$) and heavy-meson-exchange ($b_1$, $b_2$) interactions for different nuclei. Values for $^{227}$Ac are from this work, and for all other nuclei from Ref.~\cite{Dobaczewski2018}. \textbf{(c)} Comparison of the nuclear Schiff moment in the intrinsic body-fixed frame ($S_{\rm{int}}$) for $^{227}$Ac and $^{225}$Ra, calculated with nuclear density functional theory in this work and in Ref.~\cite{Dobaczewski2018}. \textbf{(d)} Expected magnitude of the \textit{CP}-odd molecular energy shift in the laboratory due to the Schiff moment as a function of the QCD constant $\bar{\theta}$ for different molecules, taking into account the calculated $W_\mathcal{S}$ constant for each species. The teal line marks the precision needed to place a new bound on $\bar{\theta}$ using $^{227}$AcF.}
    \label{fig:figCoefficients}
\end{figure}

To assess the potential of a precision experiment with $^{227}$AcF to constrain \textit{CP} violation, electronic- and nuclear-structure calculations of the molecular sensitivity to \textit{CP}-odd properties are necessary. The sensitivity $W_\mathcal{S}$ of the $X$~$^1\Sigma^+$ state to the nuclear Schiff moment was calculated in Refs.~\cite{Skripnikov2020,Chen2024SchiffMoments}, where the prominent role of electron correlation in the magnitude of $W_\mathcal{S}$ was highlighted. Due to cancellation effects of the individual molecular orbital contributions, the impact of electron correlation on $W_\mathcal{S}$ was found to be more pronounced ($>$20$\%$) in AcF than in other actinium molecules. Relativistic coupled cluster calculations were thus performed in this work for dedicated investigations of $W_\mathcal{S}$ in AcF, using current state-of-the-art quantum chemistry techniques at a high degree of accuracy. The obtained final value of $W_\mathcal{S}=-8017(574)$}\,$e/4\pi\epsilon_0 a_0^4$ for the $X$~$^1\Sigma^+$ state incorporates higher-order effects than the results reported in the previous works~\cite{Skripnikov2020,Chen2024SchiffMoments}, and provides a transparent and robust uncertainty analysis. A detailed uncertainty breakdown is given in \textit{Methods} and comparison with past works in the Supplementary Information.

\section*{Impact on the \textit{CP} violation hyperspace}
The nuclear Schiff moment is not the only source of the molecular \textit{CP}-odd signal that experiments would seek to measure in the laboratory. Other leptonic, hadronic, and nuclear \textit{CP}-odd properties, such as the electron EDM, would similarly give rise to a molecular EDM, and must be accounted for by calculating the sensitivity of the molecule to the different sources of \textit{CP} violation. Utilizing the toolbox approach described in Ref.~\cite{gaul:2020}, the sensitivity of the $X$~$^1\Sigma^+$ ground state of AcF to the \textit{CP}-odd nuclear Schiff moment ($W_\mathcal{S}$), tensor-pseudotensor electron-nucleon interaction ($W_{\rm{T}}$), electron EDM ($W_{\rm{d}}$), and scalar-pseudoscalar electron-nucleon interaction ($W_{\rm{sps}}$) were calculated in this work within the two-component zeroth-order regular approximation complex generalized Hartree–Fock (2c-ZORA-cGHF) framework. The factors are compared to those of other molecules in Fig.~\ref{fig:figCoefficients}a. In closed-shell molecules like AcF and RaOCH$_3^+$ ($^1\Sigma^+$), $W_{\rm{d}}$ and $W_{\rm{sps}}$ are strongly suppressed by the electron-to-proton mass ratio compared to open-shell molecules like HfF$^+$ and ThO ($^3\Delta_1$). Sensitivity to these interactions can only emerge due to electron magnetization by the nuclear magnetic dipole moment, giving rise to the second-order properties $W_{\rm{d}}^\mathrm{m}$ and $W_{\rm{sps}}^\mathrm{m}$. For $^{227}$AcF, they are calculated to be $W_{\rm{d}}^\mathrm{m}=-5\times10^{-4}\,{\mathrm{GV}}/{\mathrm{cm}\mu_\mathrm{N}}$ and $W_{\rm{sps}}^\mathrm{m}=-2\times10^{-3}\,{\mathrm{peV}}/{\mu_\mathrm{N}}$. Closed- and open-shell molecules thus provide complementary information on sources of \textit{CP} violation~\cite{chupp:2015,Gaul2024GlobalCP_new,degenkolb:2024}, and experiments using different molecules are necessary.

To estimate the impact of an experiment using $^{227}$AcF within the global landscape of \textit{CP}-odd searches with atoms and molecules, the value of $W_\mathcal{S}$ has to be combined with nuclear-structure calculations of the Schiff moment of the $^{227}$Ac nucleus. Previous theoretical investigations of the nuclear Schiff moment in the intrinsic, body-fixed frame ($S_{\rm{int}}$) of $^{227}$Ac relied on nuclear density functional theory (NDFT) calculations of $S_{\rm{int}}$ in the neighboring $^{225}$Ra~\cite{Dobaczewski2005}, and semi-empirical scaling factors based on the intrinsic quadrupole and octupole deformation and parity spacing of $^{225}$Ra and $^{227}$Ac~\cite{Flambaum2020a}. This estimation indicated that $^{227}$Ac has the largest laboratory-frame Schiff moment $S_{\rm{lab}}$ across all investigated nuclei. To confirm this result within a rigorous theoretical framework, a direct NDFT calculation of both $S_{\rm{int}}$ and $S_{\rm{lab}}$ in $^{227}$Ac was performed in this work. The calculated value $S_{\rm{int}}(^{227}\rm{Ac})=37.1(16)$\,$e$\,fm$^3$ is 40$\%$ higher than $S_{\rm{int}}(^{225}\rm{Ra})=26.6(19)$\,$e$\,fm$^3$~\cite{Dobaczewski2018} (see Fig.~\ref{fig:figCoefficients}c).

For a nucleus with a low-lying state $\Bar{\Psi}_0$ with the same angular momentum but opposite parity as the ground state ${\Psi}_0$, the laboratory-frame nuclear Schiff moment $S_{\rm{lab}}$ can be approximated as~\cite{Dobaczewski2018}
\begin{equation}\label{eq:full_Schiff}
    S_{\rm{lab}} \approx -2\Re{ \frac{\langle {\Psi}_0 |\hat{S}|{\Bar{\Psi}_0 \rangle \langle \Bar{\Psi}_0 |\hat{V}_{\rm{P,T}} |{\Psi}_0\rangle}}{E_{\Bar{\Psi}_0} - E_{{\Psi}_0}}},
\end{equation}
where $\hat{S}$ is the Schiff operator, $E_{{\Bar{\Psi}}_0}$ is the energy of state ${\Bar{\Psi}}_0$, and $\hat{V}_{\rm{P,T}}$ is the \textit{CP}-violating nuclear potential that precision experiments aim to elucidate. $S_{\rm{lab}}$ can also be expressed parametrically in terms of the \textit{CP}-odd coupling constants within $\hat{V}_{\rm{P,T}}$ as
\begin{equation}\label{eq:Slab_gg}
    S_{\rm{lab}} = a_0 g \Bar{g}_0 + a_1 g \Bar{g}_1 + a_2 g \Bar{g}_2 + b_1 \Bar{c}_1 + b_2 \Bar{c}_2,
\end{equation}
where $g$ is the QCD pion-nucleon coupling constant, $\Bar{g}_i$ are unknown isoscalar, isovector, and isotensor \textit{CP}-odd pion-nucleon coupling constants, $\Bar{c}_i$ are unknown \textit{CP}-odd zero-range heavy-meson-exchange coupling constants, and $a_i$, $b_i$ are sensitivity coefficients. The constants $\Bar{g}_i$ and $\Bar{c}_i$ are sought to be extracted from precision experiments, while the sensitivity coefficients $a_i$, $b_i$ need to be determined from nuclear-structure calculations.

\begin{table}[]
\begin{tabular}{cccccc}
\hline\hline
      & $a_0$     & $a_1$       & $a_2$     & $b_1$        & $b_2$        \\\hline
&     &      &     &    &  \\
$^{227}$Ac & 4.4(10) & $-14.0$(18) & 8.3(31) & $-0.2$(2) & 0.3(3)\\
&     &      &     &    &  \\
\hline\hline
\end{tabular}
\caption{Sensitivity coefficients (in units of $e$\,fm$^3$) of the laboratory Schiff moment of $^{227}$Ac to the isoscalar ($a_0$), isovector ($a_1$), and isotensor ($a_2$) \textit{CP}-odd pion-nucleon coupling constants and \textit{CP}-odd heavy-meson-exchange coupling constants ($b_1$, $b_2$), calculated with nuclear density functional theory in this work.}\label{tab:Schiff_coefficients_227Ac}
\end{table}

In $^{227}$Ac, the energy difference ${E_{\Bar{\Psi}_0} - E_{{\Psi}_0}}$ between the $I^\pi = \tfrac{3}{2}^{-}$ ground state ${{\Psi}_0}$ and the lowest-lying $I^\pi = \tfrac{3}{2}^{+}$ partner state ${\Bar{\Psi}_0}$ is only $27.369$\,keV~\cite{Maples1977}. Table~\ref{tab:Schiff_coefficients_227Ac} shows the $S_{\rm{lab}}$ sensitivity coefficients in $^{227}$Ac as determined from the NDFT calculation, and Fig.~\ref{fig:figCoefficients}b compares the sensitivity coefficients in $^{227}$Ac to those in $^{221,223}$Rn, $^{223}$Fr, $^{225}$Ra, and $^{229}$Pa~\cite{Dobaczewski2018}. Evidently, the Schiff moment in $^{227}$Ac is significantly more sensitive to all \textit{CP}-odd interactions than the other nuclei, including $^{225}$Ra, which is currently under experimental investigation using ultracold atoms~\cite{Bishof2016}. Inversely, for a given magnitude of $\Bar{g}_i$ and $\Bar{c}_i$, the laboratory Schiff moment of $^{227}$Ac will be larger than for the other nuclides, which makes actinium molecules that are sensitive to $S_{\rm{lab}}(^{227}\rm{Ac)}$ highly promising probes for the first measurement of a nuclear \textit{CP}-odd property.

The $S_{\rm{lab}}$ moment can also be expressed in terms of the QCD $\Bar{\theta}$ phase, based on the relationships $g\Bar{g}_0=0.21\Bar{\theta}$ and $g\Bar{g}_1=-0.046\Bar{\theta}$~\cite{deVries2015Theta,Bsaisou2015Theta,Yamanaka2017Theta,Flambaum2020a}. Combined with the nuclear sensitivity coefficients $a_0, a_1$ and the molecular sensitivity to the Schiff moment $W_{\mathcal{S}}$, the expected molecular energy shift in the laboratory due to the nuclear Schiff moment can be estimated as a function of $\Bar{\theta}$. Figure~\ref{fig:figCoefficients}d compares the expected energy shift in $^{227}$AcF, $^{229}$ThO, $^{225}$RaF, and $^{205}$TlF. Evidently, while $W_{\mathcal{S}}$ is of moderate magnitude in AcF as compared to other molecules (see Fig.~\ref{fig:figCoefficients}a), a precision experiment using $^{227}$AcF would have the highest sensitivity to $\bar{\theta}$ across the considered species. Specifically, an experiment using $^{227}$AcF with an uncertainty of 0.1\,mHz, as achieved in 1989 using $^{205}$TlF~\cite{Cho1989TlF}, would place a new and more stringent limit on $\bar{\theta}$, which is currently set at $\bar{\theta}<1.5\times10^{-10}$ via the $^{199}$Hg EDM~\cite{Graner2016}. It must be noted that since no calculations have been reported on the relationship between $\bar{\theta}$ and $g\bar{g}_2$, $\bar{c}_1$, and $\bar{c}_2$, their contribution to the results shown in Fig.~\ref{fig:figCoefficients}d are set to zero.

\begin{figure}[h!]
    \centering
    \includegraphics[width=.8\textwidth]{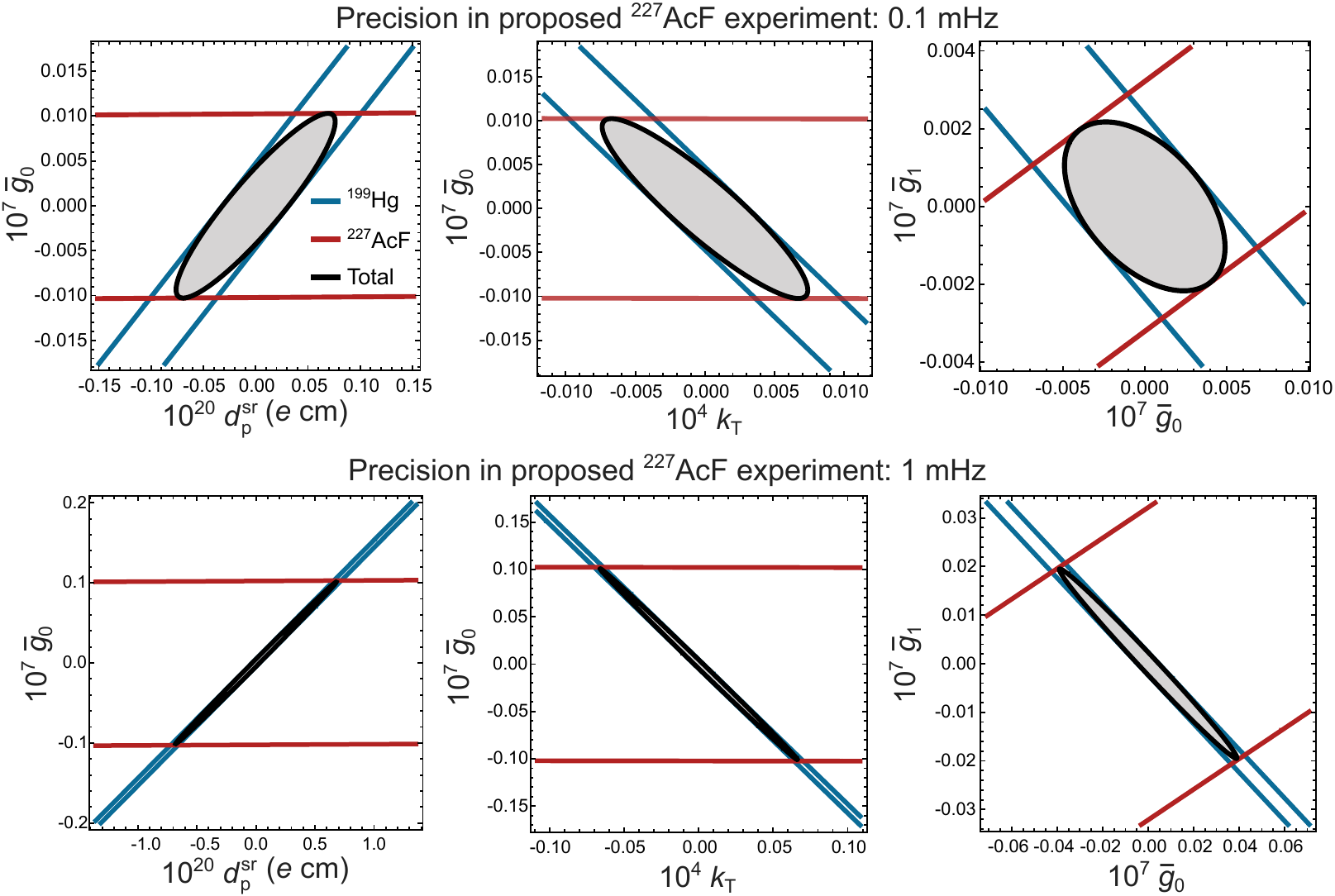}
    \caption{
    Cuts of selected two-dimensional subspaces of the full
 seven-dimensional \textit{CP}-odd parameter space for different values of experimental precision of a proposed $^{227}$AcF experiment -- top: 0.1~mHz, bottom: 1~mHz. The electronic sensitivity factors for $^{227}$AcF used in the analysis, computed at the level of ZORA-cGKS-BHandH, are shown in Table~\ref{tab:ptodd_cghf}, the calculated nuclear structure parameters are given in Table~\ref{tab:Schiff_coefficients_227Ac}, and for the volume interaction with the short-range contribution to the proton EDM we use $R_\mathrm{vol} \approx 3.22\,\mathrm{fm}^2$, as discussed in the text. A theoretical uncertainty of 20\%\ is considered,  and the molecular sensitivity factors to all \textit{CP}-odd properties in the hyperspace are conservatively scaled by a factor of $0.8$. The impact of a proposed $^{227}$AcF experiment is shown in a global analysis including existing experiments using $^{129}$Xe~\cite{allmendinger:2019,sachdeva:2019}, $^{171}$Yb~\cite{zheng:2022}, $^{133}$Cs~\cite{murthy:1989}, $^{199}$Hg~\cite{Graner2016}, $^{205}$Tl~\cite{regan:2002}, $^{255}$Ra~\cite{parker:2015,Bishof2016}, $^{174}$YbF~\cite{hudson:2002,hudson:2011}, $^{180}$HfF$^{+}$~\cite{roussy:2023}, $^{205}$TlF~\cite{cho:1991}, $^{207}$PbO~\cite{eckel:2013}, and $^{232}$ThO~\cite{andreev:2018}, employing the experimental data given in the respective references within the global analysis. All electronic structure parameters for these experiments are taken from Ref.~\cite{Gaul2024GlobalCP_new} and were combined, where available, with nuclear structure data from Ref.~\cite{Chupp2019}. In other cases, rough estimates of nuclear structure parameters from Ref.~\cite{Gaul2024GlobalCP_new} were employed. The seven-dimensional ellipsoid is computed at the 95$\%$ confidence level.
 }
    \label{fig:figGlobalAnalysis}
\end{figure}

The computed $S_{\rm{lab}}$ sensitivity coefficients $a_0$, $a_1$ for $^{227}$Ac can be combined with a rough order-of-magnitude estimate for the volume interaction $R_\mathrm{vol}$ with the short-range contribution to the proton EDM as $R_\mathrm{vol} \approx \frac{3(1.2\,\mathrm{fm})^2 A^{2/3}}{50}$ following Ref.~\cite{Ginges2004}, and the electronic structure calculations of $W_\mathcal{S}$, $W_\mathrm{m}$, $W_\mathrm{T}$, $W_\mathrm{p}$ $W_\mathrm{d}$, and $W_\mathrm{sps}$ for AcF. The expected impact that an experiment with $^{227}$AcF would have on the constraints of different \textit{CP}-violating parameters can then be extracted from a global analysis~\cite{chupp:2015,Gaul2024GlobalCP_new} that incorporates previously reported EDM experiments using $^{129}$Xe, $^{171}$Yb, $^{133}$Cs, $^{199}$Hg, $^{205}$Tl, $^{225}$Ra, $^{174}$YbF, $^{180}$HfF$^{+}$, $^{205}$TlF, $^{207}$PbO, and $^{232}$ThO. A global analysis of \textit{CP} violation was performed in this work as per Ref.~\cite{Gaul2024GlobalCP_new}, assuming two separate experiments with $^{227}$AcF with respective precision of 1~mHz and 0.1~mHz. 

The unconstrained level of \textit{CP} violation is expressed as a region enclosed by experimental limits within a multi-dimensional hyperspace, whose dimensions correspond to the experimentally measurable \textit{CP}-odd properties. Here, a seven-dimensional hyperspace is considered, formed by the tensor-pseudoscalar $k_{\rm{T}}$, scalar-pseudoscalar $k_{\rm{sps}}$, and pseudoscalar-scalar $k_{\rm{p}}$ nucleon-electron current interaction coupling constants, the electron EDM $d_{\rm{e}}$, the short-range nucleon EDM $d^{\rm{sr}}_{\rm{p}}$, and the isoscalar $\bar{g}_{0}$ and isovector $\bar{g}_{1}$ pion-nucleon interaction coupling constants. This results in the following effective Hamiltonian for each atom and molecule in the global analysis, including $^{227}$AcF
\begin{equation}
\begin{aligned}[t]
{H^{P,T}} &=                                     
\Omega \left[W_\mathrm{d}d_\mathrm{e}+W_\mathrm{sps}k_\mathrm{sps}\right] + \Theta W_\mathcal{M} \left[\mathcal{M}_\mathrm{EDM} d^\mathrm{sr}_\mathrm{p} + g a_{\mathcal{M},0} \Bar{g}_0+g a_{\mathcal{M},1}\Bar{g}_1\right]\\
&+\mathcal{I}
\left[
W_{\mathrm{T}} k_\mathrm{T}
+W_{\mathrm{p}} k_\mathrm{p}
+W^\mathrm{m}_{\mathrm{sps}}\gamma k_\mathrm{sps}
+W_{\mathcal{S}}\left(ga_0 \Bar{g}_0+ga_1\Bar{g}_1\right)
+W^\mathrm{m}_{\mathrm{d}}\gamma d_\mathrm{e}
+W_{\mathrm{m}}\eta_{\mathrm{p}} d^\mathrm{sr}_\mathrm{p}
+W_{\mathcal{S}}R_\mathrm{vol}d^\mathrm{sr}_\mathrm{p}
\right],
\label{eq:ptodd_spinrot}
\end{aligned}
\end{equation}
where $\Omega$ and $\mathcal{I}$ are the projections of the total electronic $\vec{J}_\mathrm{e}$ and nuclear angular momentum $\vec{I}$ on the molecular axis, $\Theta$ is the projection of the product of $\vec{J}_\mathrm{e}$ with the second order tensor of $\mathcal{I}$ on the molecular axis, $W_\mathcal{M}$ is the electronic structure constant for the interaction with a nuclear magnetic quadrupole moment (NMQM), $\mathcal{M}_\mathrm{EDM}$ and $a_{\mathcal{M},0}$ $a_{\mathcal{M},1}$ are nuclear structure constants for different contributions to the NMQM, $\eta_{\mathrm{p}}=\frac{\mu_\mathrm{N}}{A}+\frac{\mu}{Z}$, and $\gamma$ is the nuclear gyromagnetic ratio $\gamma=\mu/I$. For $^{227}$AcF, $\Omega=\Theta=0$, $\eta_{\mathrm{p}}=0.01811\,\mu_\mathrm{N}$, and $\gamma=0.8133\,\mu_\mathrm{N}$. The \textit{CP} violation parameters, electronic-structure coefficients $W_i$ and nuclear-structure coefficients are defined as in Ref.~\cite{Gaul2024GlobalCP_new}.

In Fig.~\ref{fig:figGlobalAnalysis}, selected two-dimensional subspaces of the seven-dimensional hyperspace are presented, in which $^{227}$AcF would play a determining role, for the two assumed values of $^{227}$AcF precision $\delta f$ at 1\,mHz and 0.1\,mHz. Plots of all two-dimensional subspaces are provided in the Supplementary Information. It is found that a precision experiment with $^{227}$AcF would reduce the volume of the seven-dimensional coverage region by a factor of $6\times10^{3}$ for $\delta f = 1$\,mHz, and by $6\times10^{4}$ for $\delta f = 0.1$\,mHz.  This can be understood as a result of the complementarity between an experiment with $^{227}$AcF and existing results using other closed-shell systems, such as $^{199}$Hg, which can be seen in Fig.~\ref{fig:figGlobalAnalysis}, and the high sensitivity of the $^{227}$Ac nucleus to \textit{CP} violation.

The precision assumed for the $^{227}$AcF experiment in the global analysis can be realistically achieved with the demonstrated production rates of $^{227}$AcF at CERN-ISOLDE in this work and conservative estimates for a precision setup. The statistical uncertainty of a spin-precession frequency measurement using Ramsey interferometry at the shot-noise limit is~\cite{Grasdijk2021}
\begin{equation} \label{eq:snl_uncertainty}
    \delta f = \frac{1}{2\pi \tau C \sqrt{N}}b,
\end{equation}
where $\tau$ is the time the $^{227}$AcF beam would take to travel across the interaction region, $C$ is the contrast of the interference fringes, $N$ is the total number of molecules detected across the experimental campaign, and $b=\sqrt{2}$ is a conservative factor that accounts for excess noise in partially closed optical transitions~\cite{Lasner2018}, such as the 387-nm transition reported here~\cite{Skripnikov2023AcF}. For an experiment with a dedicated setup using $^{227}$AcF beams at a kinetic energy of 0.1\,keV, an interaction region of 1\,m, a fringe contrast of 0.55, the demonstrated $^{227}$AcF production rate of 6$\times$10$^7$ ions per second, and a combined efficiency of 2.5$\%$ in cooling to the rovibronic ground state, charge exchange, quantum state control, and detection, an uncertainty of $\delta f = 1$\,mHz would be achieved in 100 days worth of measurements. An uncertainty of $\delta f = 0.1$\,mHz could be achieved in the same number of measurement days via a hundred-fold increase of the production rate, hundred-fold decrease of the kinetic energy, or a combination of such improvements accompanied by improvements in efficiency and fringe contrast. As a molecule homoelectronic to TlF, a similar response to external electric and magnetic fields~\cite{Grasdijk2021} can be considered for AcF. Therefore, the necessary degree of polarization and the control of stray magnetic fields to limit systematic errors below 1\,mHz are achievable with existing technology.

To conclude, the results of this work allow assessing the potential of $^{227}$AcF for searches of \textit{CP} violation. The first spectroscopic study of AcF is demonstrated, discovering an electronic transition at 387\,nm, identified to be the strongest transition from the ground state. Rotationally resolved laser spectroscopy of this transition in the future will be a stepping stone for designing dedicated experiments to search for the nuclear Schiff moment of $^{227}$Ac and as input for the calculations of the electronic sensitivity factors to \textit{CP}-odd properties in AcF, which exhibit significant dependence on the equilibrium bond length (Suppl. Fig.~4 and Suppl. Table~7).

Due to the radioactivity of $^{227}$Ac, $^{227}$AcF beams produced at RIB facilities may be envisioned for both further spectroscopy and future precision experiments. Developments are needed for new high-precision experimental techniques using beams directly produced at RIB facilities. Gas-jet techniques that have been developed for the study of radioactive atoms, such as in-gas jet laser ionization spectroscopy (IGLIS)~\cite{Ferrer2017,Ferrer2021IGLIS,Lantis2024Jetris}, may be adapted for a precision experiment using $^{227}$AcF, as the gas-jet approach was used in the past to set a limit to the electron EDM using $^{174}$YbF~\cite{hudson:2011} and the Schiff moment of $^{205}$Tl with $^{205}$TlF~\cite{Cho1989TlF}. The low rotational temperature of molecules produced with this approach would likely be key for improving the signal-to-noise ratio in the spectra compared to this work, and therefore it would also be suited for spectroscopy of excited electronic states. Importantly, further spectroscopic studies on AcF represent essential developments needed to expand current techniques towards also studying other radioactive molecules with complex electronic structure, with impact on fundamental, nuclear, and chemical physics~\cite{Opportunities2024Progressb}.

\section*{Methods}\label{sec:methods}
\subsection*{Experiment}
\subsubsection*{Production}
The ionization potential of AcF was predicted to lie above the threshold for efficient ionization by contact with a hot surface. Additionally, without knowledge of its electronic structure, resonance laser ionization could not be used to produce ion beams of AcF$^+$. The FEBIAD-type ion source was chosen and operating parameters were identified for the production of AcF$^+$. The UC$_\textrm{x}$ target in the FEBIAD-type ion source unit was irradiated for 114.4\,h (4.77\,days) prior to the start of the experiment, receiving $1.7\times 10^{18}$ protons, or a total of 73.774\,$\upmu$Ah. During irradiation, the target unit was kept under vacuum ($\sim$$1\times 10^{-6}$\,mbar) and the target container was resistively heated to slightly above room temperature to prevent condensation during irradiation. At the start of the experiment, the tantalum cathode of the ion source was resistively heated to \SI{1950}{\degreeCelsius} to facilitate electron emission. An anode voltage of \SI{100}{\volt} was applied to the anode grid to accelerate the electrons and induce ionization within the anode volume, which was also maintained at \SI{100}{\volt} with a magnetic-confinement field induced by applying a current of \SI{2.8}{\ampere} to the ion source electromagnet. A bias voltage of \SI{40}{\kilo\volt} was applied to the target and ion source unit, such that the ion beam was extracted to the ground potential of the beamline with an energy of \SI{40.1}{\kilo\electronvolt}.

The target temperature was increased from $\sim$\SI{1300}{\degreeCelsius} at the start of the experiment up to \SI{2000}{\degreeCelsius} by heating in steps on the order of \SI{10}{\ampere} to maintain a continuous supply of AcF$^+$. A mix of \SI{10}{\percent}~CF$_4$ and \SI{90}{\percent}~Ar gas was added to the target via a leak of $1.5\times 10^{-4}$~\SI{}{\milli\bar\liter\per\second} calibrated for He, injecting \SI{0.065 }{\nano\mol\per\second} of CF$_4$ for the formation of fluoride molecules.

An extensive beam purity investigation was performed using $\alpha$-decay spectroscopy of implanted ions and multi-reflection time-of-flight mass spectrometry using the ISOLTRAP apparatus~\cite{Wolf2013,Au2024AcFx}. The main expected isobaric contaminant, $^{227}$Ra$^{19}$F$^+$, was eliminated thanks to the asynchronous proton irradiation and nuclide extraction, taking advantage of the drastically longer half-life of $^{227}$Ac (21.8~years) compared to $^{227}$Ra (42~minutes). The TOF spectra in Fig.~\ref{fig:schematic}g and Suppl.~Fig.~3 show that the ion beam delivered for study was purely composed of $^{227}$Ac$^{19}$F$^+$, with no identifiable contaminants above background.

\subsubsection*{Collinear Resonance Ionization Spectroscopy}
At CRIS, the molecular beam was temporally and spatially overlapped in a collinear geometry with pulsed lasers that step-wise excited the molecular electron to ionization. At the end of the laser-molecule interaction region, the ionized molecules were deflected from the path of the residual neutral beam onto a single-ion detector. The excitation spectra were produced by monitoring the ion count rate on the detector as a function of the laser excitation wavenumbers.

Prior ab initio calculations of the excitation energies in AcF~\cite{Skripnikov2023AcF} predicted the (8)$^1\Pi$ state to lie at 26,166(450)\,cm$^{-1}$ above $X$~$^1\Sigma^+$. The 1$\sigma$ error of $\pm$450\,cm$^{-1}$ required a scanning range of 1,800\,cm$^{-1}$ to have 95$\%$ probability of discovering the predicted transition. Such a wide range is challenging for continuous scanning of light produced from a single-pass $\beta$-barium borate SHG crystal, as the SHG crystal angle requires active  stabilization to ensure optimal frequency doubling for all fundamental wavenumbers, while small deviations from the optimal crystal angle also lead to the doubled light exiting the crystal at an angle. The latter issue is exacerbated by the distance between the laser table and the beamline, exceeding 15\,m, which means that small exit angles from the SHG crystal lead to the laser light not entering the CRIS beamline.

To compensate for both issues, an active crystal-angle stabilization system was constructed using a ThorLabs PIAK10 piezoelectric inertia actuator, controlled with a proportional–integral–derivative loop reading a fraction of the SHG power output with the help of a beam sampler. To ensure that the second-harmonic light always followed the optimal trajectory for interaction with the molecules in the CRIS interaction region, a commercial active laser-beam stabilization system from MRC Systems was also installed, as shown in Fig.~\ref{fig:schematic}. This extended the continuous scanning range from 5\,cm$^{-1}$ without stabilization to $\sim$1,450\,cm$^{-1}$.

The observed excitation wavenumber for $(8)^1\Pi \leftarrow X$~$^1\Sigma^+$ was determined by simultaneously monitoring the laser wavenumber and the ion acceleration voltage, defining the kinetic energy of the beam. The wavenumber of the fundamental Ti:Sa laser was monitored with a four-channel HighFinesse WSU-2 wavemeter and the acceleration voltage of the $^{227}$AcF$^+$ ions delivered by CERN-ISOLDE was monitored with a 7.5-digit digital multimeter (Keithley DMM7510) with a precision of 100\,mV. To trace long-term drifts of the wavemeter, a grating-stabilized diode laser (TOPTICA dlpro) locked to a hyperfine transition in a Rb vapor cell (TEM CoSy) was also continuously monitored by the wavemeter. The small difference in wavelength between the Rb line ($\sim$780\,nm) and the fundamental wavelength of the Ti:Sa in this work ($\sim$774\,nm) provided confidence that the drift correction is valid in the fundamental wavelength region where the $(8)^1\Pi \leftarrow X$~$^1\Sigma^+$ transition was discovered.

The arc-discharge ion source was chosen to ionize AcF because the molecule's ionization potential (calculated at 6.06(2)\,eV~\cite{Skripnikov2023AcF}) exceeds the work function of rhenium, tantalum, and tungsten that are used for high-temperature surface ionization. Ionization by electron bombardment was expected to lead to rotationally hot molecules that may lead to prohibitively low spectroscopic efficiency. To measure the rotational temperature of the beam, the fundamental harmonic of the Ti:Sa laser at 752\,nm (in the molecular rest frame) was used in conjunction with the 355-nm non-resonant ionization step to perform spectroscopy of the $A$~$^2\Pi_{1/2} \leftarrow X$~$^2\Sigma_{1/2}$ transition in $^{226}$RaF. Using the molecular constants extracted from the rotationally resolved spectroscopy of the molecule~\cite{Udrescu2023}, the rotational temperature was fitted to extract $T_{\rm{rot}}=1,200(80)$\,K. In absence of any prior information of the AcF spectra, this temperature was used in the spectroscopic analysis.

\subsection*{Relativistic coupled cluster calculations}
The relativistic coupled cluster calculations of the sensitivity $W_\mathcal{S}$ of AcF to the nuclear Schiff moment of Ac were carried out using the development version of the DIRAC program~\cite{DIRAC23}, which allows the use of relativistic methods within the 4-component Dirac--Coulomb (DC) Hamiltonian. The single-reference coupled cluster approach with single, double, and perturbative triple excitations [CCSD(T)] was used within the finite field approach, following the implementation presented in Ref. \cite{gaul:2024b}. 

Throughout this study, we used the relativistic Dyall basis sets~\cite{dyall3,dyall2} of different quality for the actinium and fluorine atoms; for the $W_\mathcal{S}$ calculations, the basis sets for actinium were also manually optimized following the recommendation of Ref.~\cite{gaul:2024b} and described in the Supplementary Information. 

The calculations were carried out at the equilibrium bond distance ($R_e=3.974 \, a_0$) obtained via structure optimization carried out at the DC-CCSD(T) level, correlating 50 electrons with active space from $-20 \, E_\mathrm{h}$ to $50 \, E_\mathrm{h}$ (corresponding to $4f$, $5s$, $5p$, $5d$, $6s$, $6p$, $7s$ for Ac and $2s$, $2p$ for F). The $R_e$ optimization was performed using the s-aug-dyall.cv$n$z ($n$=2,3,4) basis sets and the results were extrapolated to the complete basis set limit (CBSL). For basis set extrapolation, we used the scheme of Helgaker~\textit{et al.}~\cite{cbs} (H-CBSL). 

An extensive computational study was carried out to evaluate the effect of different parameters on the obtained value of the $W_\mathcal{S}$ factor and to evaluate the final uncertainty, following procedures used in the past for various atomic and molecular properties~\cite{prop1,prop2,prop3,LeiKarGuo20,kyuPas23}. The three main sources of uncertainty in these calculations are the incompleteness of the employed basis set, the approximations in the treatment of electron correlation, and the missing relativistic effects. As we are considering high-order effects, these error sources are assumed to be largely independent, and hence they are treated separately. The Supplementary Information provides details of the computational study and the uncertainty evaluation. 
Our conservative uncertainty estimate is 7$\%$, dominated by the basis set incompleteness and the uncertainty in the calculated equilibrium geometry. 
The final recommended value of $W_\mathcal{S}$ is $-8017 \pm 574 \,e/4\pi\epsilon_0 a_0^4$.

       

\subsection*{2c-ZORA-cGHF calculations}
On the level of two-component (2c) complex generalized Hartree--Fock (cGHF) and 2c complex generalized Kohn--Sham (cGKS) employing the hybrid LDA functional, which includes 50\,\%\  Fock exchange, proposed by Becke (BHandH)~\cite{becke:1993}, quasirelativistic effects are incorporated within the zeroth-order-regular-approximation (ZORA) framework~\cite{chang:1986,lenthe:1993}. As suggested by Visscher and Dyall~\cite{visscher:1997}, a normalized spherical Gaussian nuclear charge density distribution is employed for the isotopes $^{227}$Ac and $^{19}$F. The gauge dependence of the ZORA framework is alleviated by a model potential as suggested by van W\"ullen~\cite{wullen:1998} with additional damping of the atomic Coulomb contribution to the model potential~\cite{liu:2002}.

The wavefunction was obtained self-consistently with convergence achieved 
when the change in energy between two self-consistent field (SCF) cycles was less than $10^{-9}\,E_\mathrm{h}$ and the relative change in the spin-orbit coupling energy contribution was lower than $10^{-13}\,\%$. Energy optimizations of the bond lengths were performed up to a change in the Cartesian gradient of $10^{-5}\,E_\mathrm{h}/a_0$. Excited electronic states were obtained self-consistently, employing the maximum overlap method (MOM)~\cite{gilbert:2008}, where the occupation numbers were chosen according to the determinant's overlap with the previous SCF cycle or the initial guess (IMOM,~\cite{barca:2018}). At the actinium center, Dyall's core-valence triple-$\zeta$ basis set with mono-augmentation was employed with added \textit{s} and \textit{p} functions in an even-tempered manner up to an exponent of $6\times 10^9$ and a multiplicator of 3 (see Supplementary Information). At the fluorine center, Dyall's core-valence triple-$\zeta$ basis set was employed without additional modifications. All basis sets were employed without contrations.

A characterization of the ground state of AcF and three exemplary excited states, which have not been experimentally observed so far, are shown in Table~\ref{tab:cghf_states}.

\begin{table}[!htb]
\centering
\caption{Qualitative characterization at the level of 2c-ZORA-cGHF for the electronic ground state and some excited electronic states in AcF, for the expectation values for the projection of the total electron angular momentum $\left< \Omega \right>$
and electron orbital angular momentum $\left< \Lambda \right>$ along the
molecular axis, the equilibrium bond lengths $R_\mathrm{e}$, the harmonic vibrational wavenumber
$\tilde{\omega}_\mathrm{e}$, the transition wavenumber $\Delta\tilde{\nu}_\mathrm{T}$ from the electronic ground state, and the spinor contributions $M_i$ on the Ac center from an atomic picture for the two highest occupied spinors ($i=$ 1, 2) obtained via a Mulliken analysis.}
\label{tab:cghf_states}
\begin{tabular}{
c
c
S[table-format=-1.2,round-mode=places,round-precision=2]
S[table-format=-1.2,round-mode=places,round-precision=2]
S[table-format=1.2,round-mode=figures,round-precision=3]
S[table-format=3.0,round-mode=figures,round-precision=3]
S[table-format=5.0,round-mode=figures,round-precision=3]
l
l
}
\toprule
 State 
& Term symbol
& {$\left< \Omega \right>$}
& {$\left< \Lambda \right>$}
& {$R_\mathrm{e}\,/\,a_0$}
& {$\tilde{\omega}_\mathrm{e}\,/\,\si{cm^{-1}}$}
& {$\Delta\tilde{\nu}_\mathrm{T}\,/\,\si{cm^{-1}}$}
& {$M_1 / \%$}
& {$M_2 / \%$}
\\
\midrule
$X$  & $^1\Sigma^+$ & 0.00  &  0.00001  &  4.05  & 527.73 & 0     &  89s,5p,6d & 89s,5p,6d \\
$X$ (BHandH)  & $^1\Sigma^+$ & 0.00  &  0.00  &  3.98  & {\textemdash} & {\textemdash}     &  89s,4p,7d & 89s,4p,7d \\
Excited state 1  & $^3\Delta_1$ & 1.00  &  2.00163  &  4.11  & 500.55 & 4758  &  87s,7p,6d & 100d      \\
Excited state 2  & $^3\Phi_2$   & 2.00  &  2.92735  &  4.13  & 483.08 & 17083 &  99d       & 55p,42d   \\
Excited state 3  & $^1\Sigma^+$ & 0.00  &  0.00001  &  4.14  & 480.33 & 23588 &  98d       & 98d       \\
\bottomrule
\end{tabular}
\end{table}

Utilizing the toolbox approach described in Ref.~\cite{gaul:2020}, the \textit{P,T}-odd properties of the electronic states were 
obtained as expectation values and from a linear-response ansatz \cite{bruck:2023,colombojofre:2022}, respectively, the latter for $W_\mathrm{d}$ and $W_\mathrm{sps}$ (labeled $W_\mathrm{d}^\mathrm{m}$ and $W_\mathrm{sps}^\mathrm{m}$). A complete list of the equations for each property is given in Table 6 of Ref.~\cite{Gaul2024GlobalCP_new}. 

In Table~\ref{tab:ptodd_cghf} \textit{P,T}-odd properties computed with this method are shown for the 
states presented in Table~\ref{tab:cghf_states} and compared to the $^3\Delta_1$ states in HfF$^+$, ThO, and ThF$^+$, and the $^1\Sigma^+$ states in RaOCH$_3^+$ and TlF.

\begin{table}[!htb]
\centering
\caption{\textit{P,T}-odd properties obtained on the level of
2c-cGHF and 2c-cGKS/BHandH computed on the energy-optimized structure of the respective methods and electronic states. For closed-shell systems with term symbol $^1\Sigma^+$, the magnetic contributions to $W_\mathrm{d}$ and $W_\mathrm{sps}$ are given. The computation for RaOCH$^+_3$ from Ref.~\cite{gaul:2024b} was done at the level of 4c Dirac--Hartree--Fock (4c-DHF).}
\label{tab:ptodd_cghf}
\begin{tabular}{
c
c
S[table-format=1.2,round-mode=places,round-precision=2]
S[table-format=-5.0]
S[table-format=-2.4,round-mode=figures,round-precision=3]
S[table-format=-3.4,round-mode=figures,round-precision=3]
S[table-format=-2.2,round-mode=figures,round-precision=3]
S[table-format=-2.2,round-mode=figures,round-precision=3]
S[table-format=-3.1,round-mode=figures,round-precision=3] 
}
\toprule
& State
& {$\left< \Omega \right>$}
& {$W_\mathcal{S} / \frac{e}{4\pi \epsilon_0 {a_0}^4}$}
& {$W_\mathrm{d}/ \frac{\mathrm{GV}}{\mathrm{cm}}$}
& {$W_\mathrm{sps}/\mathrm{peV}$}
& {$W_\mathrm{T} / \mathrm{peV}$}
& {$W_\mathrm{p} / \mathrm{feV}$}
& {$W_\mathrm{m} / \frac{\mathrm{kV}\eta_\mathrm{p}}{\mathrm{cm}\mu_\mathrm{N}}$}
\\
\midrule
HfF$^+$\cite{simpson:2025}          & $^3\Delta_1$  & 1.00  &-15000   & 26.47   & 95.1      & -4.55     & -16.96& 309       \\
ThO    \cite{simpson:2025}          & $^3\Delta_1$  & 1.00  &-35000   & 99.85   & 583.9     & -14.57    & -57.5 & 166.486   \\
ThF$^+$~\cite{Gaul2024GlobalCP_new} & $^3\Delta_1$  & 1.00  &-45600   & 43.5    & 248       & -17.4     & -69.1 & 603       \\
RaOCH$_3^+$ (4c-DHF)~\cite{gaul:2024b}       & $^1\Sigma^+$  & 0.00  &-56300   & -0.0112 & -0.298    & -20.72    & -81.47& 959       \\
TlF~\cite{Gaul2024GlobalCP_new}     & $^1\Sigma^+$& 0.00  & 45850   & 0.008   & 0.247     &  19.40    & 72.2  & -777      \\
AcF: $X$                            & $^1\Sigma^+$  & 0.00  & -8400   & -0.00045& -0.0215   & -1.634    & -6.73 & 38.802    \\
AcF: $X$ (BHandH)                   & $^1\Sigma^+$  & 0.00  & -8700   & -0.0012 & -0.0293   & -1.937    & -7.90 & 32.984    \\
AcF: Excited state 1                            & $^3\Delta_1$  & 1.00  &-28000   & -64.191 & -365.45   & -9.757    & -38.7 & 394.9742  \\
AcF: Excited state 2                            & $^3\Phi_2$    & 2.00  &-49000   & -19.71  & -131.87   & -20.41    & -80.3 & 842.0844  \\
AcF: Excited state 3                            & $^1\Sigma^+$  & 0.00  &-53000   & {\textemdash}  & {\textemdash} & -19.33 & -76.4 & 874.2670   \\
\bottomrule 
\end{tabular}
\end{table}

Vibrational corrections were estimated by solving the vibrational Schr\"odinger equation within a discrete variable representation (DVR) on a one-dimensional grid \cite{meyer:1970} ranging from $3.18 \, a_0$ to $4.77 \, a_0$, divided into 1000 equidistant points. The properties were interpolated as a 6$^{\rm{th}}$-order polynomial function of the bond length, and effects were estimated as described in the Supplementary Material of Ref.~\cite{Udrescu2021}. The change in the properties as a function of the bond length was computed and is shown in Suppl. Fig.~4a in the Supplementary Information.

The enhancement factor $W_\mathcal{S}$ for the nuclear Schiff moment with 2c-ZORA-cGHF evaluates to $W_\mathcal{S}=-8400\,e/4\pi\epsilon_0 a_0^4$, while further treatment of electron correlation at the level 2c-ZORA-cGKS with the BHandH functional leads to $W_\mathcal{S}=-8700\,e/4\pi\epsilon_0 a_0^4$, which is close to the results from CCSD(T). The 2c-ZORA values were obtained after minimization of the energy with respect to the bond length ($R_\mathrm{e}=4.05\,a_0$ compared to CCSD(T) computations $R_e=3.97\,a_0$ from this work), suggesting that electron correlation is of importance for the bond length and hence the rotational constant, and, thus, indirectly, also for the sensitivity to \textit{CP}-odd properties in the ground state of AcF. 

When comparing the molecular sensitivity $W_\mathcal{S}$ to the nuclear Schiff moment in AcF with that in other closed-shell systems, such as RaOCH$^+_3$ and TlF (Supp. Fig.~4b in the Supplementary Information), a $Z^2$-scaling is expected \cite{Sushkov1984}, but a reduced enhancement is computed for AcF. Contributions from the individual spinors to $W_\mathcal{S}$ show that the bonding spinors -- that is, the highest occupied molecular orbital (HOMO) -- contribute with sign opposite to the other spinors but with a similar magnitude. In the case of the ground state computed on the level of 2c-cGHF, the HOMO contributes with $29067 \, e/4\pi\epsilon_0 a_0^4$ and the remaining orbitals sum up to $-37481 \, e/4\pi\epsilon_0 a_0^4$.  For excited electronic states, the HOMO contributes to $W_\mathcal{S}$ with the same sign but lower magnitude, resulting in a larger value of $W_\mathcal{S}$ (see discussion in the main text and Fig.~2b of the Supplementary Material in Ref.~\cite{Zulch2022}, also discussion in Ref.~\cite{Chen2024SchiffMoments}). The suitability of these states for precision experiments depends on their radiative lifetime, among other factors, and thus a discussion is pending experimental observation.

\subsection*{Nuclear DFT calculations}
The nuclear DFT calculations were performed using the HFODD program~\cite{Dobaczewski2021HFODD}. The laboratory Schiff moment, $S_{\text{lab}}$, can be calculated using second-order perturbation theory as
\begin{equation}\label{eq:2nd order perturb S}
    S_{\rm{lab}}\approx\sum_{k\neq 0}\frac{\langle\Psi_{0}|\hat{S}_{0}|\Psi_{k}\rangle\langle\Psi_{k}|\hat{V}_{\text{P,T}}|\Psi_{0}\rangle}{E_{0}-E_{k}}+\text{c.c.},
\end{equation}
where $\hat{S}_{0}$ is the Schiff operator and $\hat{V}_{\text{P,T}}$ stands for the $P,T$-violating potential. The index $k$ refers to the excited states with the same angular momentum quantum numbers as the ground state $|\Psi_{0}\rangle$ but opposite parity. To leading order, the Schiff operator $\hat{S}_{0}$ is defined as 
\begin{equation}\label{eq: Schiff operator}
    \hat{S}_{0}=\frac{e}{10}\sqrt{\frac{4\pi}{3}}\sum_{p}\left(r^{3}_{p}-\frac{5}{3}\overline{r^{2}_{\text{ch}}}r_{p}\right)Y_{10}(\Omega_{p}),
\end{equation}
where the sum ranges over all protons (index $p$), $Y_{10}(\Omega)$ is the spherical harmonics $Y_{\ell m}(\Omega)$ with $\ell=1,m=0$, and $\overline{r^{2}_{\text{ch}}}$ is the mean-squared charge radius. In the coordinate representation, $\hat{V}_{\text{P,T}}$ takes the form~\cite{Maekawa2011, Haxton1983b, Herczeg1988}
\begin{align}
    \hat{V}_{\text{P,T}}(\textbf{r}_{1}-\textbf{r}_{2})&=-\frac{gm^{2}_{\pi}}{8\pi m_{N}}\bigg\{(\boldsymbol{\sigma}_{1}-\boldsymbol{\sigma}_{2})\cdot(\textbf{r}_{1}-\textbf{r}_{2})\bigg[\bar{g}_{0}\Vec{\tau}_{1}\cdot\Vec{\tau}_{2}-\frac{\bar{g}_{1}}{2}(\tau_{1z}+\tau_{2z})+\bar{g}_{2}(3\tau_{1z}\tau_{2z}-\Vec{\tau}_{1}\cdot\Vec{\tau}_{2})\bigg]\nonumber\\
    &-\frac{\bar{g}_{1}}{2}(\boldsymbol{\sigma}_{1}+\boldsymbol{\sigma}_{2})\cdot(\textbf{r}_{1}-\textbf{r}_{2})(\tau_{1z}-\tau_{2z})\bigg\}\frac{\exp(-m_{\pi}|\textbf{r}_{1}-\textbf{r}_{2}|)}{m_{\pi}|\textbf{r}_{1}-\textbf{r}_{2}|^{2}}\bigg[1+\frac{1}{m_{\pi}|\textbf{r}_{1}-\textbf{r}_{2}|}\bigg]\nonumber\\
    &+\frac{1}{2m^{3}_{N}}[\bar{c}_{1}+\bar{c}_{2}\Vec{\tau}_{1}\cdot\Vec{\tau}_{2}](\boldsymbol{\sigma}_{1}-\boldsymbol{\sigma}_{2})\cdot\boldsymbol{\nabla}\delta^{3}(\textbf{r}_{1}-\textbf{r}_{2}),
\end{align}
where isovector operators are denoted by arrows, $\tau_{z}$ is $+1(-1)$ for neutrons(protons), the pion mass is $m_{\pi}=0.7045$\,fm$^{-1}$, the nucleon mass is $m_{N}=4.7565$\,fm$^{-1}$, and $\bar{g}_{0}$, $\bar{g}_{1}$, and $\bar{g}_{2}$ are the unknown isoscalar, isovector, and isotensor \textit{CP}-odd pion-nucleon coupling constants, respectively. The strong $\pi NN$ coupling constant is denoted by $g$, and $\bar{c}_{1}$ and $\bar{c}_{2}$ are the coupling constants of a \textit{CP}-odd short-range interaction.

The $I^\pi=\tfrac{3}{2}^{-}$ ground state $|\Psi_{0}\rangle$ of $^{227}$Ac has a $I^\pi=\tfrac{3}{2}^{+}$ partner state $|\bar{\Psi}_{0}\rangle$ at $\Delta E\equiv E_{\bar{\Psi}_{0}}-E_{\Psi_{0}}=27.369$ keV \cite{Maples1977}. Due to the small energy difference $\Delta E$ and in the absence of other reported low-lying $\tfrac{3}{2}^{+}$ states, Eq. \eqref{eq:2nd order perturb S} reduces to
\begin{equation}\label{eq:Approximation of S}
    S_{\rm{lab}} \approx-2\Re{\frac{\langle\Psi_{0}|\hat{S}_{0}|\bar{\Psi}_{0}\rangle\langle\bar{\Psi}_{0}|\hat{V}_{\text{P,T}}|\Psi_{0}\rangle}{\Delta E}}.
\end{equation}
In octupole deformed nuclei, the NDFT calculation breaks parity and rotational symmetries, and the obtained nuclear wave function is the deformed quasiparticle vacuum, $|\Phi_{0}\rangle$, defined in the body-fixed frame of the nucleus. The intrinsic state, $|\Phi_{0}\rangle$, needs to be projected onto the laboratory states with well-defined angular momentum and parity. In the rigid deformation approximation, the matrix elements in the numerator of Eq. \eqref{eq:Approximation of S} read:
\begin{align}\label{eq:S0 and VPT}
    \langle\Psi_{0}|\hat{S}_{0}|\bar{\Psi}_{0}\rangle_{\text{rigid}}&=\frac{J}{J+1}S_{\rm{int}},\\
    \langle\bar{\Psi}_{0}|\hat{V}_{\text{P,T}}|\Psi_{0}\rangle_{\text{rigid}}&=\langle\hat{V}_{\text{P,T}}\rangle
\end{align}
where $J=\tfrac{3}{2}$ for $^{227}\text{Ac}$, $S_{\rm{int}}$ is the intrinsic Schiff moment, and $\langle\hat{V}_{\text{P,T}}\rangle$ is the intrinsic expectation value of the operator $\hat{V}_{\text{P,T}}$. In terms of the coupling constants $g$, $\bar{g}$'s, and $\bar{c}$'s, the expectation value $\langle\hat{V}_{\text{P,T}}\rangle$ can be rewritten as 
\begin{equation}\label{eq:decomposition of VPT}
    \langle\hat{V}_{\text{P,T}}\rangle=v_{0}g\bar{g}_{0}+v_{1}g\bar{g}_{1}+v_{2}g\bar{g}_{2}+w_{1}\bar{c}_{1}+w_{2}\bar{c}_{2}.
\end{equation}
The insertion of Eqs. \eqref{eq:S0 and VPT}-\eqref{eq:decomposition of VPT} into Eq. \eqref{eq:Approximation of S} leads to the expression of $S_{\rm{lab}}$ as
\begin{equation}
    S_{\rm{lab}}=a_{0}g\bar{g}_{0}+a_{1}g\bar{g}_{1}+a_{2}g\bar{g}_{2}+b_{1}\bar{c}_{1}+b_{2}\bar{c}_{2},
\end{equation}
where 
\begin{equation}\label{eq:Coefficients of S}
    a_{i}=-\frac{2J}{J+1}\frac{S_{\rm{int}}v_{i}}{\Delta E}\;\;\;\text{and}\;\;\;b_{i}=-\frac{2J}{J+1}\frac{S_{\rm{int}}w_{i}}{\Delta E}
\end{equation}
have units of $e$ fm$^{3}$.

As shown in Ref. \cite{Dobaczewski2018}, there is a strong correlation between the intrinsic Schiff moment $S_{\rm{int}}$ in one odd nucleus and the intrinsic matrix element of the octupole charge operator~\cite{Ring1980}
\begin{equation}
    \hat{Q}_{30}=e\sum_{p}r^{3}_{p}Y_{30}(\Omega_{p})
\end{equation}
in another even-even nucleus with close proton and neutron numbers. The correlation is visualized in Suppl. Fig.~5 in the Supplementary Information using several Skyrme functionals within the Hartree-Fock-Bogoliubov (HFB) method used in this work. For each functional, the neutron and proton pairing strengths were adjusted to reproduce the experimental pairing gaps of $^{229}$Th and $^{227}$Ac, respectively. The strong correlation between these two observables were used in a scheme to determine the intrinsic Schiff moment of a nucleus without relying on the results obtained with a specific functional. Using a variety of functionals, the intrinsic Schiff moment of $^{227}$Ac and the intrinsic electric octupole moment of $^{226}$Ra were calculated, and linear regression analysis was performed to determine the relationship between the two correlated observables. The resulting regression line was then used along with the measured value of the electric octupole moment of $^{226}$Ra, yielding the value of the intrinsic nuclear Schiff moment of $^{227}$Ac reported in this work. Details on the procedure and uncertainty estimation are given in the Supplementary Information.

\subsection*{Global analysis}
The measurement model of Ref.~\cite{Gaul2024GlobalCP_new} was expanded to explicitly contain two different pion-nucleon coupling constants $\Bar{g}_0$, $\Bar{g}_1$ and the assumption $d^\mathrm{sr}_\mathrm{p}=-d^\mathrm{sr}_\mathrm{n}$, where the suggestions of Ref.~\cite{degenkolb:2024} are followed. 

In the Hamiltonian in Eq.~\ref{eq:ptodd_spinrot}, seven fundamental $P,T$-odd parameters contribute to the total atomic or molecular EDM. Considering experiments using $^{129}$Xe~\cite{allmendinger:2019,sachdeva:2019}, $^{171}$Yb~\cite{zheng:2022}, $^{133}$Cs~\cite{murthy:1989}, $^{199}$Hg~\cite{Graner2016}, $^{205}$Tl~\cite{regan:2002}, $^{255}$Ra~\cite{parker:2015,Bishof2016}, $^{174}$YbF~\cite{hudson:2002,hudson:2011}, $^{180}$HfF$^{+}$~\cite{roussy:2023}, $^{205}$TlF~\cite{cho:1991}, $^{207}$PbO~\cite{eckel:2013},$^{232}$ThO~\cite{andreev:2018}, and $^{227}$AcF, this results in a $12\times7$ matrix $\bm{W}$, where the element in row $i$ and column $j$ corresponds to the Hamiltonian term in Eq.~\ref{eq:ptodd_spinrot} for atom/molecule $i$ and $P,T$-odd constant $j$. Electronic-structure coefficients for all atoms and molecules except AcF were taken from Ref.~\cite{Gaul2024GlobalCP_new} and nuclear-structure coefficients were taken from Ref.~\cite{Chupp2019}, where available. Otherwise nuclear-structure estimates reported in Ref.~\cite{Gaul2024GlobalCP_new} were employed.  Combined with the vector of the $P,T$-odd constants $\Vec{x}_{\rm{PT}}^{\rm{T}}=[k_{\rm{T}}, k_{\rm{p}}, k_{\rm{sps}},\Bar{g}_0,\Bar{g}_1, d_{\rm{e}}, d_{\rm{p}}^{\rm{sr}}]$ and the vector of the experimental frequency shifts $\Vec{f}^{\rm{T}}=[\delta f_1,...\delta f_{7}]$, the model is a system of linear equations of the form
\begin{equation}
    h \Vec{f} = \bm{W} \Vec{x}_{\rm{PT}}.
\end{equation}
Following Ref.~\cite{Gaul2024GlobalCP_new}, the restrictive power of a set of experiments can be determined by the volume enclosed by an ellipsoid in the seven-dimensional parameter hyperspace for a given confidence level. The global constraints on individual parameters are given by the extrema of the ellipsoid and are provided in Table~\ref{tab: limits}. It should be noted that a precision better than 1\,mHz in the proposed $^{227}$AcF experiment would not further improve global constraints on individual parameters, as these are limited by the experiments with the largest uncertainties (see also Ref.~\cite{degenkolb:2024} for a discussion).

\begin{table}[h!]
\caption{Comparison of global bounds on sources of \textit{P,T}-violation with and without the proposed $^{227}$AcF experiment with an experimental uncertainty of 1\,mHz. Global bounds are obtained
from seven-dimensional ellipsoidal coverage regions at 95\%\ confidence level, including existing experiments with $^{129}$Xe~\cite{allmendinger:2019,sachdeva:2019}, $^{171}$Yb~\cite{zheng:2022}, $^{133}$Cs~\cite{murthy:1989}, $^{199}$Hg~\cite{Graner2016}, $^{205}$Tl~\cite{regan:2002}, $^{255}$Ra~\cite{parker:2015,Bishof2016}, $^{174}$YbF~\cite{hudson:2002,hudson:2011}, $^{180}$HfF$^{+}$~\cite{roussy:2023}, $^{205}$TlF~\cite{cho:1991}, $^{207}$PbO~\cite{eckel:2013}, and $^{232}$ThO~\cite{andreev:2018}. All electronic-structure parameters for these experiments were taken from Ref.~\cite{Gaul2024GlobalCP_new} and were combined, where available, with nuclear-structure data from Ref.~\cite{Chupp2019}. In other cases, estimates of nuclear-structure parameters from Ref.~\cite{Gaul2024GlobalCP_new} were employed. A conservative theoretical uncertainty of 20\%\ is assumed and the electronic-structure sensitivity factors to all sources are scaled by a factor of $0.8$ to account for a worst-case scenario.} 
\label{tab: limits}
\begin{tabular}{l
S[round-precision=0,round-mode=places,table-format=1e-2,scientific-notation=true]
S[round-precision=0,round-mode=places,table-format=1e-2,scientific-notation=true]
}
\toprule
Source
&\multicolumn{1}{c}{Limit without $^{227}$AcF}
&\multicolumn{1}{c}{Limit with $^{227}$AcF}
\\
\midrule
$d_\mathrm{e}/(e\,\mathrm{cm})$                  & 4e-29     & 4e-29   \\
$d^\mathrm{sr}_\mathrm{p}/(e\,\mathrm{cm})$      & 4e-22     & 2e-23  \\
$k_\mathrm{sps}$                  & 1e-8      & 1e-8   \\
$k_\mathrm{T}$                  & 8e-6      & 4e-6  \\
$k_\mathrm{p}$                  & 3e-3      & 9e-4   \\
$\bar{g}_\mathrm{0}$            & 6e-9      & 6e-10  \\
$\bar{g}_\mathrm{1}$            & 4e-9      & 2e-10  \\
\bottomrule     
\end{tabular}       
\end{table}

\section*{Data availability}
The raw data are available upon request to the authors. Further information about data analysis is provided in the Supplementary Information.

\section*{Code availability}
The analysis code used to pre-process raw data prior to contour fitting with {\sc pgopher} can be provided upon request to the authors.


\begin{thebibliography}{89}
\ifx \bisbn   \undefined \def \bisbn  #1{ISBN #1}\fi
\ifx \binits  \undefined \def \binits#1{#1}\fi
\ifx \bauthor  \undefined \def \bauthor#1{#1}\fi
\ifx \batitle  \undefined \def \batitle#1{#1}\fi
\ifx \bjtitle  \undefined \def \bjtitle#1{#1}\fi
\ifx \bvolume  \undefined \def \bvolume#1{\textbf{#1}}\fi
\ifx \byear  \undefined \def \byear#1{#1}\fi
\ifx \bissue  \undefined \def \bissue#1{#1}\fi
\ifx \bfpage  \undefined \def \bfpage#1{#1}\fi
\ifx \blpage  \undefined \def \blpage #1{#1}\fi
\ifx \burl  \undefined \def \burl#1{\textsf{#1}}\fi
\ifx \doiurl  \undefined \def \doiurl#1{\url{https://doi.org/#1}}\fi
\ifx \betal  \undefined \def \betal{\textit{et al.}}\fi
\ifx \binstitute  \undefined \def \binstitute#1{#1}\fi
\ifx \binstitutionaled  \undefined \def \binstitutionaled#1{#1}\fi
\ifx \bctitle  \undefined \def \bctitle#1{#1}\fi
\ifx \beditor  \undefined \def \beditor#1{#1}\fi
\ifx \bpublisher  \undefined \def \bpublisher#1{#1}\fi
\ifx \bbtitle  \undefined \def \bbtitle#1{#1}\fi
\ifx \bedition  \undefined \def \bedition#1{#1}\fi
\ifx \bseriesno  \undefined \def \bseriesno#1{#1}\fi
\ifx \blocation  \undefined \def \blocation#1{#1}\fi
\ifx \bsertitle  \undefined \def \bsertitle#1{#1}\fi
\ifx \bsnm \undefined \def \bsnm#1{#1}\fi
\ifx \bsuffix \undefined \def \bsuffix#1{#1}\fi
\ifx \bparticle \undefined \def \bparticle#1{#1}\fi
\ifx \barticle \undefined \def \barticle#1{#1}\fi
\bibcommenthead
\ifx \bconfdate \undefined \def \bconfdate #1{#1}\fi
\ifx \botherref \undefined \def \botherref #1{#1}\fi
\ifx \url \undefined \def \url#1{\textsf{#1}}\fi
\ifx \bchapter \undefined \def \bchapter#1{#1}\fi
\ifx \bbook \undefined \def \bbook#1{#1}\fi
\ifx \bcomment \undefined \def \bcomment#1{#1}\fi
\ifx \oauthor \undefined \def \oauthor#1{#1}\fi
\ifx \citeauthoryear \undefined \def \citeauthoryear#1{#1}\fi
\ifx \endbibitem  \undefined \def \endbibitem {}\fi
\ifx \bconflocation  \undefined \def \bconflocation#1{#1}\fi
\ifx \arxivurl  \undefined \def \arxivurl#1{\textsf{#1}}\fi
\csname PreBibitemsHook\endcsname

\bibitem[\protect\citeauthoryear{Sakharov}{1991}]{Sakharov1991}
\begin{barticle}
\bauthor{\bsnm{Sakharov}, \binits{A.}}:
\batitle{{Violation of CP invariance, C asymmetry, and baryon asymmetry of the
  universe}}.
\bjtitle{Soviet Physics Uspekhi}
\bvolume{34}(\bissue{5}),
\bfpage{392}
(\byear{1991})
\doiurl{10.1070/PU1991v034n05ABEH002497}
\end{barticle}
\endbibitem

\bibitem[\protect\citeauthoryear{B{\"{o}}deker and
  Buchm{\"{u}}ller}{2021}]{Bodeker2021Baryogenesis}
\begin{barticle}
\bauthor{\bsnm{B{\"{o}}deker}, \binits{D.}},
\bauthor{\bsnm{Buchm{\"{u}}ller}, \binits{W.}}:
\batitle{{Baryogenesis from the weak scale to the grand unification scale}}.
\bjtitle{Reviews of Modern Physics}
\bvolume{93}(\bissue{3}),
\bfpage{35004}
(\byear{2021})
\doiurl{10.1103/RevModPhys.93.035004}
\end{barticle}
\endbibitem

\bibitem[\protect\citeauthoryear{Safronova et~al.}{2018}]{Safronova2018}
\begin{botherref}
\oauthor{\bsnm{Safronova}, \binits{M.S.}},
\oauthor{\bsnm{Budker}, \binits{D.}},
\oauthor{\bsnm{DeMille}, \binits{D.}},
\oauthor{\bsnm{Kimball}, \binits{D.F.J.}},
\oauthor{\bsnm{Derevianko}, \binits{A.}},
\oauthor{\bsnm{Clark}, \binits{C.W.}}:
{Search for new physics with atoms and molecules}.
Reviews of Modern Physics
\textbf{90}(2)
(2018)
\doiurl{10.1103/RevModPhys.90.025008}
\end{botherref}
\endbibitem

\bibitem[\protect\citeauthoryear{Schiff}{1963}]{Schiff1963}
\begin{barticle}
\bauthor{\bsnm{Schiff}, \binits{L.I.}}:
\batitle{{Measurability of nuclear electric dipole moments}}.
\bjtitle{Physical Review}
\bvolume{132}(\bissue{5}),
\bfpage{2194}--\blpage{2200}
(\byear{1963})
\doiurl{10.1103/PhysRev.132.2194}
\end{barticle}
\endbibitem

\bibitem[\protect\citeauthoryear{Flambaum and Ginges}{2002}]{Flambaum2002}
\begin{barticle}
\bauthor{\bsnm{Flambaum}, \binits{V.V.}},
\bauthor{\bsnm{Ginges}, \binits{J.S.M.}}:
\batitle{{Nuclear Schiff moment and time-invariance violation in atoms}}.
\bjtitle{Physical Review A}
\bvolume{65}(\bissue{3}),
\bfpage{9}
(\byear{2002})
\doiurl{10.1103/PhysRevA.65.032113}
\end{barticle}
\endbibitem

\bibitem[\protect\citeauthoryear{Haxton and Henley}{1983}]{Haxton1983b}
\begin{barticle}
\bauthor{\bsnm{Haxton}, \binits{W.C.}},
\bauthor{\bsnm{Henley}, \binits{E.M.}}:
\batitle{{Enhanced T-Nonconserving Nuclear Moments}}.
\bjtitle{Physical Review Letters}
\bvolume{51}(\bissue{21}),
\bfpage{1937}--\blpage{1940}
(\byear{1983})
\doiurl{10.1103/PhysRevLett.51.1937}
\end{barticle}
\endbibitem

\bibitem[\protect\citeauthoryear{Maekawa et~al.}{2011}]{Maekawa2011}
\begin{barticle}
\bauthor{\bsnm{Maekawa}, \binits{C.M.}},
\bauthor{\bsnm{Mereghetti}, \binits{E.}},
\bauthor{\bsnm{Vries}, \binits{J.}},
\bauthor{\bsnm{Kolck}, \binits{U.}}:
\batitle{{The time-reversal- and parity-violating nuclear potential in chiral
  effective theory}}.
\bjtitle{Nucl. Phys. A}
\bvolume{872}(\bissue{1}),
\bfpage{117}--\blpage{160}
(\byear{2011})
\doiurl{10.1016/j.nuclphysa.2011.09.020}
\end{barticle}
\endbibitem

\bibitem[\protect\citeauthoryear{Dobaczewski et~al.}{2018}]{Dobaczewski2018}
\begin{barticle}
\bauthor{\bsnm{Dobaczewski}, \binits{J.}},
\bauthor{\bsnm{Engel}, \binits{J.}},
\bauthor{\bsnm{Kortelainen}, \binits{M.}},
\bauthor{\bsnm{Becker}, \binits{P.}}:
\batitle{{Correlating Schiff Moments in the Light Actinides with Octupole
  Moments}}.
\bjtitle{Physical Review Letters}
\bvolume{121}(\bissue{23}),
\bfpage{232501}
(\byear{2018})
\doiurl{10.1103/PhysRevLett.121.232501}
\end{barticle}
\endbibitem

\bibitem[\protect\citeauthoryear{Graner et~al.}{2016}]{Graner2016}
\begin{barticle}
\bauthor{\bsnm{Graner}, \binits{B.}},
\bauthor{\bsnm{Chen}, \binits{Y.}},
\bauthor{\bsnm{Lindahl}, \binits{E.G.}},
\bauthor{\bsnm{Heckel}, \binits{B.R.}}:
\batitle{{Reduced Limit on the Permanent Electric Dipole Moment of {{$^{199}$Hg}}}}.
\bjtitle{Physical Review Letters}
\bvolume{116}(\bissue{16}),
\bfpage{1}--\blpage{5}
(\byear{2016})
\doiurl{10.1103/PhysRevLett.116.161601}
\end{barticle}
\endbibitem

\bibitem[\protect\citeauthoryear{Grasdijk et~al.}{2021}]{Grasdijk2021}
\begin{barticle}
\bauthor{\bsnm{Grasdijk}, \binits{O.}},
\bauthor{\bsnm{Timgren}, \binits{O.}},
\bauthor{\bsnm{Kastelic}, \binits{J.}},
\bauthor{\bsnm{Wright}, \binits{T.}},
\bauthor{\bsnm{Lamoreaux}, \binits{S.}},
\bauthor{\bsnm{DeMille}, \binits{D.}},
\bauthor{\bsnm{Wenz}, \binits{K.}},
\bauthor{\bsnm{Aitken}, \binits{M.}},
\bauthor{\bsnm{Zelevinsky}, \binits{T.}},
\bauthor{\bsnm{Winick}, \binits{T.}},
\bauthor{\bsnm{Kawall}, \binits{D.}}:
\batitle{{CeNTREX: a new search for time-reversal symmetry violation in the
  {{$^{205}$Tl}} nucleus}}.
\bjtitle{Quantum Science and Technology}
\bvolume{6}(\bissue{4}),
\bfpage{044007}
(\byear{2021})
\doiurl{10.1088/2058-9565/abdca3}
\end{barticle}
\endbibitem

\bibitem[\protect\citeauthoryear{Bishof et~al.}{2016}]{Bishof2016}
\begin{barticle}
\bauthor{\bsnm{Bishof}, \binits{M.}},
\bauthor{\bsnm{Parker}, \binits{R.H.}},
\bauthor{\bsnm{Bailey}, \binits{K.G.}},
\bauthor{\bsnm{Greene}, \binits{J.P.}},
\bauthor{\bsnm{Holt}, \binits{R.J.}},
\bauthor{\bsnm{Kalita}, \binits{M.R.}},
\bauthor{\bsnm{Korsch}, \binits{W.}},
\bauthor{\bsnm{Lemke}, \binits{N.D.}},
\bauthor{\bsnm{Lu}, \binits{Z.T.}},
\bauthor{\bsnm{Mueller}, \binits{P.}},
\bauthor{\bsnm{O'Connor}, \binits{T.P.}},
\bauthor{\bsnm{Singh}, \binits{J.T.}},
\bauthor{\bsnm{Dietrich}, \binits{M.R.}}:
\batitle{{Improved limit on the {{$^{225}$Ra}} electric dipole moment}}.
\bjtitle{Physical Review C}
\bvolume{94}(\bissue{2}),
\bfpage{1}--\blpage{17}
(\byear{2016})
\doiurl{10.1103/PhysRevC.94.025501}
\end{barticle}
\endbibitem

\bibitem[\protect\citeauthoryear{Wolf et~al.}{2013}]{Wolf2013}
\begin{barticle}
\bauthor{\bsnm{Wolf}, \binits{R.N.}},
\bauthor{\bsnm{Wienholtz}, \binits{F.}},
\bauthor{\bsnm{Atanasov}, \binits{D.}},
\bauthor{\bsnm{Beck}, \binits{D.}},
\bauthor{\bsnm{Blaum}, \binits{K.}},
\bauthor{\bsnm{Borgmann}, \binits{C.}},
\bauthor{\bsnm{Herfurth}, \binits{F.}},
\bauthor{\bsnm{Kowalska}, \binits{M.}},
\bauthor{\bsnm{Kreim}, \binits{S.}},
\bauthor{\bsnm{Litvinov}, \binits{Y.A.}},
\bauthor{\bsnm{Lunney}, \binits{D.}},
\bauthor{\bsnm{Manea}, \binits{V.}},
\bauthor{\bsnm{Neidherr}, \binits{D.}},
\bauthor{\bsnm{Rosenbusch}, \binits{M.}},
\bauthor{\bsnm{Schweikhard}, \binits{L.}},
\bauthor{\bsnm{Stanja}, \binits{J.}},
\bauthor{\bsnm{Zuber}, \binits{K.}}:
\batitle{{ISOLTRAP's multi-reflection time-of-flight mass
  separator/spectrometer}}.
\bjtitle{International Journal of Mass Spectrometry}
\bvolume{349-350},
\bfpage{123}--\blpage{133}
(\byear{2013})
\doiurl{10.1016/j.ijms.2013.03.020}
\end{barticle}
\endbibitem

\bibitem[\protect\citeauthoryear{Au et~al.}{2024}]{Au2024AcFx}
\begin{barticle}
\bauthor{\bsnm{Au}, \binits{M.}},
\bauthor{\bsnm{Nies}, \binits{L.}},
\bauthor{\bsnm{Stegemann}, \binits{S.}},
\bauthor{\bsnm{Athanasakis-Kaklamanakis}, \binits{M.}},
\bauthor{\bsnm{Cocolios}, \binits{T.E.}},
\bauthor{\bsnm{Fischer}, \binits{P.}},
\bauthor{\bsnm{Giesel}, \binits{P.F.}},
\bauthor{\bsnm{Johnson}, \binits{J.D.}},
\bauthor{\bsnm{K{\"{o}}ster}, \binits{U.}},
\bauthor{\bsnm{Lange}, \binits{D.}},
\bauthor{\bsnm{Mougeot}, \binits{M.}},
\bauthor{\bsnm{Reilly}, \binits{J.}},
\bauthor{\bsnm{Schlaich}, \binits{M.}},
\bauthor{\bsnm{Schweiger}, \binits{C.}},
\bauthor{\bsnm{Schweikhard}, \binits{L.}},
\bauthor{\bsnm{Wienholtz}, \binits{F.}},
\bauthor{\bsnm{Wojtaczka}, \binits{W.}},
\bauthor{\bsnm{D{\"{u}}llmann}, \binits{C.E.}},
\bauthor{\bsnm{Rothe}, \binits{S.}}:
\batitle{{Production and purification of molecular {{$^{225}$Ac}} at CERN-ISOLDE}}.
\bjtitle{Journal of Radioanalytical and Nuclear Chemistry}
\bvolume{334},
\bfpage{367}--\blpage{379}
(\byear{2024})
\doiurl{10.1007/s10967-024-09811-0}
\end{barticle}
\endbibitem

\bibitem[\protect\citeauthoryear{Flambaum and Dzuba}{2020}]{Flambaum2020a}
\begin{barticle}
\bauthor{\bsnm{Flambaum}, \binits{V.V.}},
\bauthor{\bsnm{Dzuba}, \binits{V.A.}}:
\batitle{{Electric dipole moments of atoms and molecules produced by enhanced
  nuclear Schiff moments}}.
\bjtitle{Physical Review A}
\bvolume{101},
\bfpage{042504}
(\byear{2020})
\doiurl{10.1103/PhysRevA.101.042504}
\end{barticle}
\endbibitem

\bibitem[\protect\citeauthoryear{Scheck
  et~al.}{2023}]{Scheck2023Mossbauer227Ac}
\begin{barticle}
\bauthor{\bsnm{Scheck}, \binits{M.}},
\bauthor{\bsnm{Chapman}, \binits{R.}},
\bauthor{\bsnm{Dobaczewski}, \binits{J.}},
\bauthor{\bsnm{Ederer}, \binits{C.}},
\bauthor{\bsnm{Ivanov}, \binits{P.}},
\bauthor{\bsnm{Lorusso}, \binits{G.}},
\bauthor{\bsnm{O’Donnell}, \binits{D.}},
\bauthor{\bsnm{Schr{\"{o}}der}, \binits{C.}}:
\batitle{{A new avenue in the search for CP violation: M{\"{o}}ssbauer
  spectroscopy of {{$^{227}$Ac}}}}.
\bjtitle{The European Physical Journal A}
\bvolume{59}(\bissue{5}),
\bfpage{116}
(\byear{2023})
\doiurl{10.1140/epja/s10050-023-01000-z}
\end{barticle}
\endbibitem

\bibitem[\protect\citeauthoryear{Catherall et~al.}{2017}]{Catherall2017}
\begin{barticle}
\bauthor{\bsnm{Catherall}, \binits{R.}},
\bauthor{\bsnm{Andreazza}, \binits{W.}},
\bauthor{\bsnm{Breitenfeldt}, \binits{M.}},
\bauthor{\bsnm{Dorsival}, \binits{A.}},
\bauthor{\bsnm{Focker}, \binits{G.J.}},
\bauthor{\bsnm{Gharsa}, \binits{T.P.}},
\bauthor{\bsnm{Giles}, \binits{T.J.}},
\bauthor{\bsnm{Grenard}, \binits{J.-L.}},
\bauthor{\bsnm{Locci}, \binits{F.}},
\bauthor{\bsnm{Martins}, \binits{P.}},
\bauthor{\bsnm{Marzari}, \binits{S.}},
\bauthor{\bsnm{Schipper}, \binits{J.}},
\bauthor{\bsnm{Shornikov}, \binits{A.}},
\bauthor{\bsnm{Stora}, \binits{T.}}:
\batitle{{The ISOLDE facility}}.
\bjtitle{Journal of Physics G: Nuclear and Particle Physics}
\bvolume{44}(\bissue{9}),
\bfpage{094002}
(\byear{2017})
\doiurl{10.1088/1361-6471/aa7eba}
\end{barticle}
\endbibitem

\bibitem[\protect\citeauthoryear{Au et~al.}{2023}]{Au2023InsourceIntrap}
\begin{barticle}
\bauthor{\bsnm{Au}, \binits{M.}},
\bauthor{\bsnm{Athanasakis-Kaklamanakis}, \binits{M.}},
\bauthor{\bsnm{Nies}, \binits{L.}},
\bauthor{\bsnm{Ballof}, \binits{J.}},
\bauthor{\bsnm{Berger}, \binits{R.}},
\bauthor{\bsnm{Chrysalidis}, \binits{K.}},
\bauthor{\bsnm{Fischer}, \binits{P.}},
\bauthor{\bsnm{Heinke}, \binits{R.}},
\bauthor{\bsnm{Johnson}, \binits{J.}},
\bauthor{\bsnm{K{\"{o}}ster}, \binits{U.}},
\bauthor{\bsnm{Leimbach}, \binits{D.}},
\bauthor{\bsnm{Marsh}, \binits{B.}},
\bauthor{\bsnm{Mougeot}, \binits{M.}},
\bauthor{\bsnm{Reich}, \binits{B.}},
\bauthor{\bsnm{Reilly}, \binits{J.}},
\bauthor{\bsnm{Reis}, \binits{E.}},
\bauthor{\bsnm{Schlaich}, \binits{M.}},
\bauthor{\bsnm{Schweiger}, \binits{C.}},
\bauthor{\bsnm{Schweikhard}, \binits{L.}},
\bauthor{\bsnm{Stegemann}, \binits{S.}},
\bauthor{\bsnm{Wessolek}, \binits{J.}},
\bauthor{\bsnm{Wienholtz}, \binits{F.}},
\bauthor{\bsnm{Wilkins}, \binits{S.G.}},
\bauthor{\bsnm{Wojtaczka}, \binits{W.}},
\bauthor{\bsnm{D{\"{u}}llmann}, \binits{C.E.}},
\bauthor{\bsnm{Rothe}, \binits{S.}}:
\batitle{{In-source and in-trap formation of molecular ions in the actinide
  mass range at CERN-ISOLDE}}.
\bjtitle{Nuclear Instruments and Methods in Physics Research Section B: Beam
  Interactions with Materials and Atoms}
\bvolume{541},
\bfpage{375}--\blpage{379}
(\byear{2023})
\doiurl{10.1016/j.nimb.2023.05.015}
\end{barticle}
\endbibitem

\bibitem[\protect\citeauthoryear{Au}{2023}]{Au2023Thesis}
\begin{botherref}
\oauthor{\bsnm{Au}, \binits{M.}}:
{Production of actinide atomic and molecular ion beams at CERN-ISOLDE}.
PhD thesis,
Johannes Gutenberg University Mainz,
Mainz
(2023).
\url{https://cds.cern.ch/record/2878875?ln=en}
\end{botherref}
\endbibitem

\bibitem[\protect\citeauthoryear{K{\"{o}}ster et~al.}{2007}]{Koester2007}
\begin{barticle}
\bauthor{\bsnm{K{\"{o}}ster}, \binits{U.}},
\bauthor{\bsnm{Carbonez}, \binits{P.}},
\bauthor{\bsnm{Dorsival}, \binits{A.}},
\bauthor{\bsnm{Dvorak}, \binits{J.}},
\bauthor{\bsnm{Eichler}, \binits{R.}},
\bauthor{\bsnm{Fernandes}, \binits{S.}},
\bauthor{\bsnm{Fr{\aa}nberg}, \binits{H.}},
\bauthor{\bsnm{Neuhausen}, \binits{J.}},
\bauthor{\bsnm{Novackova}, \binits{Z.}},
\bauthor{\bsnm{Wilfinger}, \binits{R.}},
\bauthor{\bsnm{Yakushev}, \binits{A.}}:
\batitle{{(Im-) possible ISOL beams}}.
\bjtitle{European Physical Journal: Special Topics}
\bvolume{150}(\bissue{1}),
\bfpage{285}--\blpage{291}
(\byear{2007})
\doiurl{10.1140/epjst/e2007-00326-1}
\end{barticle}
\endbibitem

\bibitem[\protect\citeauthoryear{Ballof}{2021}]{Ballof2021Thesis}
\begin{botherref}
\oauthor{\bsnm{Ballof}, \binits{J.}}:
{Radioactive Molecular Beams at CERN-ISOLDE}.
PhD thesis,
Johannes Gutenberg University Mainz,
Mainz
(2021)
\end{botherref}
\endbibitem

\bibitem[\protect\citeauthoryear{Jaj{\v c}i{\v s}inov{\'a}
  et~al.}{2024}]{jajcisinovaProductionStudyFr2024}
\begin{barticle}
\bauthor{\bsnm{Jaj{\v c}i{\v s}inov{\'a}}, \binits{E.}},
\bauthor{\bsnm{Dockx}, \binits{K.}},
\bauthor{\bsnm{Au}, \binits{M.}},
\bauthor{\bsnm{Bara}, \binits{S.}},
\bauthor{\bsnm{Cocolios}, \binits{T.E.}},
\bauthor{\bsnm{Chrysalidis}, \binits{K.}},
\bauthor{\bsnm{{Farooq-Smith}}, \binits{G.J.}},
\bauthor{\bsnm{Fedorov}, \binits{D.V.}},
\bauthor{\bsnm{Fedosseev}, \binits{V.N.}},
\bauthor{\bsnm{Flanagan}, \binits{K.T.}},
\bauthor{\bsnm{Heines}, \binits{M.}},
\bauthor{\bsnm{Houngbo}, \binits{D.}},
\bauthor{\bsnm{Johnson}, \binits{J.D.}},
\bauthor{\bsnm{Kellerbauer}, \binits{A.}},
\bauthor{\bsnm{Kraemer}, \binits{S.}},
\bauthor{\bsnm{Marsh}, \binits{B.A.}},
\bauthor{\bsnm{Popescu}, \binits{L.}},
\bauthor{\bsnm{Ramos}, \binits{J.P.}},
\bauthor{\bsnm{Rothe}, \binits{S.}},
\bauthor{\bsnm{Seliverstov}, \binits{M.D.}},
\bauthor{\bsnm{Sels}, \binits{S.}},
\bauthor{\bsnm{Stegemann}, \binits{S.}},
\bauthor{\bsnm{Stryjczyk}, \binits{M.}},
\bauthor{\bsnm{Verelst}, \binits{V.}}:
\batitle{Production study of {{Fr}}, {{Ra}} and {{Ac}} radioactive ion beams at
  {{ISOLDE}}, {{CERN}}}.
\bjtitle{Sci Rep}
\bvolume{14}(\bissue{1}),
\bfpage{11033}
(\byear{2024})
\doiurl{10.1038/s41598-024-60331-z}
\end{barticle}
\endbibitem

\bibitem[\protect\citeauthoryear{Johnson et~al.}{2023}]{Johnson2023}
\begin{botherref}
\oauthor{\bsnm{Johnson}, \binits{J.D.}},
\oauthor{\bsnm{Heines}, \binits{M.}},
\oauthor{\bsnm{Bruchertseifer}, \binits{F.}},
\oauthor{\bsnm{Chevallay}, \binits{E.}},
\oauthor{\bsnm{Cocolios}, \binits{T.E.}},
\oauthor{\bsnm{Dockx}, \binits{K.}},
\oauthor{\bsnm{Duchemin}, \binits{C.}},
\oauthor{\bsnm{Heinitz}, \binits{S.}},
\oauthor{\bsnm{Heinke}, \binits{R.}},
\oauthor{\bsnm{Hurier}, \binits{S.}},
\oauthor{\bsnm{Lambert}, \binits{L.}},
\oauthor{\bsnm{Leenders}, \binits{B.}},
\oauthor{\bsnm{Skliarova}, \binits{H.}},
\oauthor{\bsnm{Stora}, \binits{T.}},
\oauthor{\bsnm{Wojtaczka}, \binits{W.}}:
{Resonant laser ionization and mass separation of {{$^{225}$Ac}}}.
Scientific Reports
\textbf{13}(1)
(2023)
\doiurl{10.1038/s41598-023-28299-4}
\end{botherref}
\endbibitem

\bibitem[\protect\citeauthoryear{Lasner and DeMille}{2018}]{Lasner2018}
\begin{barticle}
\bauthor{\bsnm{Lasner}, \binits{Z.}},
\bauthor{\bsnm{DeMille}, \binits{D.}}:
\batitle{{Statistical sensitivity of phase measurements via laser-induced
  fluorescence with optical cycling detection}}.
\bjtitle{Physical Review A}
\bvolume{98}(\bissue{5}),
\bfpage{53823}
(\byear{2018})
\doiurl{10.1103/PhysRevA.98.053823}
\end{barticle}
\endbibitem

\bibitem[\protect\citeauthoryear{Ho et~al.}{2020}]{Ho2020a}
\begin{botherref}
\oauthor{\bsnm{Ho}, \binits{C.J.}},
\oauthor{\bsnm{Devlin}, \binits{J.A.}},
\oauthor{\bsnm{Rabey}, \binits{I.M.}},
\oauthor{\bsnm{Yzombard}, \binits{P.}},
\oauthor{\bsnm{Lim}, \binits{J.}},
\oauthor{\bsnm{Wright}, \binits{S.C.}},
\oauthor{\bsnm{Fitch}, \binits{N.J.}},
\oauthor{\bsnm{Hinds}, \binits{E.A.}},
\oauthor{\bsnm{Tarbutt}, \binits{M.R.}},
\oauthor{\bsnm{Sauer}, \binits{B.E.}}:
{New techniques for a measurement of the electron's electric dipole moment}.
New Journal of Physics
\textbf{22}(5)
(2020)
\doiurl{10.1088/1367-2630/ab83d2}
\end{botherref}
\endbibitem

\bibitem[\protect\citeauthoryear{Arrowsmith-Kron
  et~al.}{2024}]{Opportunities2024Progressb}
\begin{barticle}
\bauthor{\bsnm{Arrowsmith-Kron}, \binits{G.}},
\bauthor{\bsnm{Athanasakis-Kaklamanakis}, \binits{M.}},
\bauthor{\bsnm{Au}, \binits{M.}},
\bauthor{\bsnm{Ballof}, \binits{J.}},
\bauthor{\bsnm{Berger}, \binits{R.}},
\bauthor{\bsnm{Borschevsky}, \binits{A.}},
\bauthor{\bsnm{Breier}, \binits{A.A.}},
\bauthor{\bsnm{Buchinger}, \binits{F.}},
\bauthor{\bsnm{Budker}, \binits{D.}},
\bauthor{\bsnm{Caldwell}, \binits{L.}},
\bauthor{\bsnm{Charles}, \binits{C.}},
\bauthor{\bsnm{Dattani}, \binits{N.}},
\bauthor{\bsnm{Groote}, \binits{R.P.}},
\bauthor{\bsnm{DeMille}, \binits{D.}},
\bauthor{\bsnm{Dickel}, \binits{T.}},
\bauthor{\bsnm{Dobaczewski}, \binits{J.}},
\bauthor{\bsnm{D{\"{u}}llmann}, \binits{C.E.}},
\bauthor{\bsnm{Eliav}, \binits{E.}},
\bauthor{\bsnm{Engel}, \binits{J.}},
\bauthor{\bsnm{Fan}, \binits{M.}},
\bauthor{\bsnm{Flambaum}, \binits{V.}},
\bauthor{\bsnm{Flanagan}, \binits{K.T.}},
\bauthor{\bsnm{Gaiser}, \binits{A.N.}},
\bauthor{\bsnm{Garcia~Ruiz}, \binits{R.F.}},
\bauthor{\bsnm{Gaul}, \binits{K.}},
\bauthor{\bsnm{Giesen}, \binits{T.F.}},
\bauthor{\bsnm{Ginges}, \binits{J.S.M.}},
\bauthor{\bsnm{Gottberg}, \binits{A.}},
\bauthor{\bsnm{Gwinner}, \binits{G.}},
\bauthor{\bsnm{Heinke}, \binits{R.}},
\bauthor{\bsnm{Hoekstra}, \binits{S.}},
\bauthor{\bsnm{Holt}, \binits{J.D.}},
\bauthor{\bsnm{Hutzler}, \binits{N.R.}},
\bauthor{\bsnm{Jayich}, \binits{A.}},
\bauthor{\bsnm{Karthein}, \binits{J.}},
\bauthor{\bsnm{Leach}, \binits{K.G.}},
\bauthor{\bsnm{Madison}, \binits{K.W.}},
\bauthor{\bsnm{Malbrunot-Ettenauer}, \binits{S.}},
\bauthor{\bsnm{Miyagi}, \binits{T.}},
\bauthor{\bsnm{Moore}, \binits{I.D.}},
\bauthor{\bsnm{Moroch}, \binits{S.}},
\bauthor{\bsnm{Navratil}, \binits{P.}},
\bauthor{\bsnm{Nazarewicz}, \binits{W.}},
\bauthor{\bsnm{Neyens}, \binits{G.}},
\bauthor{\bsnm{Norrgard}, \binits{E.B.}},
\bauthor{\bsnm{Nusgart}, \binits{N.}},
\bauthor{\bsnm{Pa{\v{s}}teka}, \binits{L.F.}},
\bauthor{\bsnm{N~Petrov}, \binits{A.}},
\bauthor{\bsnm{Pla{\ss}}, \binits{W.R.}},
\bauthor{\bsnm{Ready}, \binits{R.A.}},
\bauthor{\bsnm{Pascal~Reiter}, \binits{M.}},
\bauthor{\bsnm{Reponen}, \binits{M.}},
\bauthor{\bsnm{Rothe}, \binits{S.}},
\bauthor{\bsnm{Safronova}, \binits{M.S.}},
\bauthor{\bsnm{Scheidenerger}, \binits{C.}},
\bauthor{\bsnm{Shindler}, \binits{A.}},
\bauthor{\bsnm{Singh}, \binits{J.T.}},
\bauthor{\bsnm{Skripnikov}, \binits{L.V.}},
\bauthor{\bsnm{Titov}, \binits{A.V.}},
\bauthor{\bsnm{Udrescu}, \binits{S.-M.}},
\bauthor{\bsnm{Wilkins}, \binits{S.G.}},
\bauthor{\bsnm{Yang}, \binits{X.}}:
\batitle{{Opportunities for fundamental physics research with radioactive
  molecules}}.
\bjtitle{Reports on Progress in Physics}
\bvolume{87}(\bissue{8}),
\bfpage{084301}
(\byear{2024})
\doiurl{10.1088/1361-6633/ad1e39}
\end{barticle}
\endbibitem

\bibitem[\protect\citeauthoryear{Athanasakis-Kaklamanakis
  et~al.}{2025}]{AthKak2025ElectronCorrelation}
\begin{barticle}
\bauthor{\bsnm{Athanasakis-Kaklamanakis}, \binits{M.}},
\bauthor{\bsnm{Wilkins}, \binits{S.G.}},
\bauthor{\bsnm{Skripnikov}, \binits{L.V.}},
\bauthor{\bsnm{Koszor{\'{u}}s}, \binits{A.}},
\bauthor{\bsnm{Breier}, \binits{A.A.}},
\bauthor{\bsnm{Ahmad}, \binits{O.}},
\bauthor{\bsnm{Au}, \binits{M.}},
\bauthor{\bsnm{Bai}, \binits{S.W.}},
\bauthor{\bsnm{Belo{\v{s}}evi{\'{c}}}, \binits{I.}},
\bauthor{\bsnm{Berbalk}, \binits{J.}},
\bauthor{\bsnm{Berger}, \binits{R.}},
\bauthor{\bsnm{Bernerd}, \binits{C.}},
\bauthor{\bsnm{Bissell}, \binits{M.L.}},
\bauthor{\bsnm{Borschevsky}, \binits{A.}},
\bauthor{\bsnm{Brinson}, \binits{A.}},
\bauthor{\bsnm{Chrysalidis}, \binits{K.}},
\bauthor{\bsnm{Cocolios}, \binits{T.E.}},
\bauthor{\bsnm{Groote}, \binits{R.P.}},
\bauthor{\bsnm{Dorne}, \binits{A.}},
\bauthor{\bsnm{Fajardo-Zambrano}, \binits{C.M.}},
\bauthor{\bsnm{Field}, \binits{R.W.}},
\bauthor{\bsnm{Flanagan}, \binits{K.T.}},
\bauthor{\bsnm{Franchoo}, \binits{S.}},
\bauthor{\bsnm{Garcia~Ruiz}, \binits{R.F.}},
\bauthor{\bsnm{Gaul}, \binits{K.}},
\bauthor{\bsnm{Geldhof}, \binits{S.}},
\bauthor{\bsnm{Giesen}, \binits{T.F.}},
\bauthor{\bsnm{Hanstorp}, \binits{D.}},
\bauthor{\bsnm{Heinke}, \binits{R.}},
\bauthor{\bsnm{Imgram}, \binits{P.}},
\bauthor{\bsnm{Isaev}, \binits{T.A.}},
\bauthor{\bsnm{Kyuberis}, \binits{A.A.}},
\bauthor{\bsnm{Kujanp{\"{a}}{\"{a}}}, \binits{S.}},
\bauthor{\bsnm{Lalanne}, \binits{L.}},
\bauthor{\bsnm{Lass{\`{e}}gues}, \binits{P.}},
\bauthor{\bsnm{Lim}, \binits{J.}},
\bauthor{\bsnm{Liu}, \binits{Y.C.}},
\bauthor{\bsnm{Lynch}, \binits{K.M.}},
\bauthor{\bsnm{McGlone}, \binits{A.}},
\bauthor{\bsnm{Mei}, \binits{W.C.}},
\bauthor{\bsnm{Neyens}, \binits{G.}},
\bauthor{\bsnm{Nichols}, \binits{M.}},
\bauthor{\bsnm{Nies}, \binits{L.}},
\bauthor{\bsnm{Pa{\v{s}}teka}, \binits{L.F.}},
\bauthor{\bsnm{Perrett}, \binits{H.A.}},
\bauthor{\bsnm{Raggio}, \binits{A.}},
\bauthor{\bsnm{Reilly}, \binits{J.R.}},
\bauthor{\bsnm{Rothe}, \binits{S.}},
\bauthor{\bsnm{Smets}, \binits{E.}},
\bauthor{\bsnm{Udrescu}, \binits{S.-M.}},
\bauthor{\bsnm{Borne}, \binits{B.}},
\bauthor{\bsnm{Wang}, \binits{Q.}},
\bauthor{\bsnm{Warbinek}, \binits{J.}},
\bauthor{\bsnm{Wessolek}, \binits{J.}},
\bauthor{\bsnm{Yang}, \binits{X.F.}},
\bauthor{\bsnm{Z{\"{u}}lch}, \binits{C.}}:
\batitle{{Electron correlation and relativistic effects in the excited states
  of radium monofluoride}}.
\bjtitle{Nature Communications}
\bvolume{16}(\bissue{1}),
\bfpage{2139}
(\byear{2025})
\doiurl{10.1038/s41467-025-55977-w}
\end{barticle}
\endbibitem

\bibitem[\protect\citeauthoryear{Skripnikov et~al.}{2023}]{Skripnikov2023AcF}
\begin{barticle}
\bauthor{\bsnm{Skripnikov}, \binits{L.V.}},
\bauthor{\bsnm{Oleynichenko}, \binits{A.V.}},
\bauthor{\bsnm{Zaitsevskii}, \binits{A.}},
\bauthor{\bsnm{Mosyagin}, \binits{N.S.}},
\bauthor{\bsnm{Athanasakis-Kaklamanakis}, \binits{M.}},
\bauthor{\bsnm{Au}, \binits{M.}},
\bauthor{\bsnm{Neyens}, \binits{G.}}:
\batitle{{Ab initio study of electronic states and radiative properties of the
  AcF molecule}}.
\bjtitle{The Journal of Chemical Physics}
\bvolume{159}(\bissue{12}),
\bfpage{124301}
(\byear{2023})
\doiurl{10.1063/5.0159888}
\end{barticle}
\endbibitem

\bibitem[\protect\citeauthoryear{Ramos}{2020}]{Ramos2020}
\begin{barticle}
\bauthor{\bsnm{Ramos}, \binits{J.P.}}:
\batitle{{Thick solid targets for the production and online release of
  radioisotopes: The importance of the material characteristics – A review}}.
\bjtitle{Nuclear Instruments and Methods in Physics Research Section B: Beam
  Interactions with Materials and Atoms}
\bvolume{463},
\bfpage{201}--\blpage{210}
(\byear{2020})
\doiurl{10.1016/j.nimb.2019.05.045}
\end{barticle}
\endbibitem

\bibitem[\protect\citeauthoryear{Penescu et~al.}{2010}]{Penescu2010}
\begin{barticle}
\bauthor{\bsnm{Penescu}, \binits{L.}},
\bauthor{\bsnm{Catherall}, \binits{R.}},
\bauthor{\bsnm{Lettry}, \binits{J.}},
\bauthor{\bsnm{Stora}, \binits{T.}}:
\batitle{{Development of high efficiency Versatile Arc Discharge Ion Source at
  CERN ISOLDE}}.
\bjtitle{Review of Scientific Instruments}
\bvolume{81}(\bissue{2}),
\bfpage{02}--\blpage{906}
(\byear{2010})
\doiurl{10.1063/1.3271245}
\end{barticle}
\endbibitem

\bibitem[\protect\citeauthoryear{Cocolios et~al.}{2013}]{Cocolios2013CRIS}
\begin{barticle}
\bauthor{\bsnm{Cocolios}, \binits{T.E.}},
\bauthor{\bsnm{Al~Suradi}, \binits{H.H.}},
\bauthor{\bsnm{Billowes}, \binits{J.}},
\bauthor{\bsnm{Budin{\v{c}}evi{\'{c}}}, \binits{I.}},
\bauthor{\bsnm{Groote}, \binits{R.P.}},
\bauthor{\bsnm{De~Schepper}, \binits{S.}},
\bauthor{\bsnm{Fedosseev}, \binits{V.N.}},
\bauthor{\bsnm{Flanagan}, \binits{K.T.}},
\bauthor{\bsnm{Franchoo}, \binits{S.}},
\bauthor{\bsnm{Garcia~Ruiz}, \binits{R.F.}},
\bauthor{\bsnm{Heylen}, \binits{H.}},
\bauthor{\bsnm{Le~Blanc}, \binits{F.}},
\bauthor{\bsnm{Lynch}, \binits{K.M.}},
\bauthor{\bsnm{Marsh}, \binits{B.A.}},
\bauthor{\bsnm{Mason}, \binits{P.J.R.}},
\bauthor{\bsnm{Neyens}, \binits{G.}},
\bauthor{\bsnm{Papuga}, \binits{J.}},
\bauthor{\bsnm{Procter}, \binits{T.J.}},
\bauthor{\bsnm{Rajabali}, \binits{M.M.}},
\bauthor{\bsnm{Rossel}, \binits{R.E.}},
\bauthor{\bsnm{Rothe}, \binits{S.}},
\bauthor{\bsnm{Simpson}, \binits{G.S.}},
\bauthor{\bsnm{Smith}, \binits{A.J.}},
\bauthor{\bsnm{Strashnov}, \binits{I.}},
\bauthor{\bsnm{Stroke}, \binits{H.H.}},
\bauthor{\bsnm{Verney}, \binits{D.}},
\bauthor{\bsnm{Walker}, \binits{P.M.}},
\bauthor{\bsnm{Wendt}, \binits{K.D.A.}},
\bauthor{\bsnm{Wood}, \binits{R.T.}}:
\batitle{{The Collinear Resonance Ionization Spectroscopy (CRIS) experimental
  setup at CERN-ISOLDE}}.
\bjtitle{Nuclear Instruments and Methods in Physics Research Section B: Beam
  Interactions with Materials and Atoms}
\bvolume{317},
\bfpage{565}--\blpage{569}
(\byear{2013})
\doiurl{10.1016/j.nimb.2013.05.088}
\end{barticle}
\endbibitem

\bibitem[\protect\citeauthoryear{Athanasakis-Kaklamanakis
  et~al.}{2024}]{AthKak2024Lifetime}
\begin{barticle}
\bauthor{\bsnm{Athanasakis-Kaklamanakis}, \binits{M.}},
\bauthor{\bsnm{Wilkins}, \binits{S.G.}},
\bauthor{\bsnm{Lass{\`{e}}gues}, \binits{P.}},
\bauthor{\bsnm{Lalanne}, \binits{L.}},
\bauthor{\bsnm{Reilly}, \binits{J.R.}},
\bauthor{\bsnm{Ahmad}, \binits{O.}},
\bauthor{\bsnm{Au}, \binits{M.}},
\bauthor{\bsnm{Bai}, \binits{S.W.}},
\bauthor{\bsnm{Berbalk}, \binits{J.}},
\bauthor{\bsnm{Bernerd}, \binits{C.}},
\bauthor{\bsnm{Borschevsky}, \binits{A.}},
\bauthor{\bsnm{Breier}, \binits{A.A.}},
\bauthor{\bsnm{Chrysalidis}, \binits{K.}},
\bauthor{\bsnm{Cocolios}, \binits{T.E.}},
\bauthor{\bsnm{Groote}, \binits{R.P.}},
\bauthor{\bsnm{Fajardo-Zambrano}, \binits{C.M.}},
\bauthor{\bsnm{Flanagan}, \binits{K.T.}},
\bauthor{\bsnm{Franchoo}, \binits{S.}},
\bauthor{\bsnm{Ruiz}, \binits{R.F.G.}},
\bauthor{\bsnm{Hanstorp}, \binits{D.}},
\bauthor{\bsnm{Heinke}, \binits{R.}},
\bauthor{\bsnm{Imgram}, \binits{P.}},
\bauthor{\bsnm{Koszor{\'{u}}s}, \binits{A.}},
\bauthor{\bsnm{Kyuberis}, \binits{A.A.}},
\bauthor{\bsnm{Lim}, \binits{J.}},
\bauthor{\bsnm{Liu}, \binits{Y.C.}},
\bauthor{\bsnm{Lynch}, \binits{K.M.}},
\bauthor{\bsnm{McGlone}, \binits{A.}},
\bauthor{\bsnm{Mei}, \binits{W.C.}},
\bauthor{\bsnm{Neyens}, \binits{G.}},
\bauthor{\bsnm{Nies}, \binits{L.}},
\bauthor{\bsnm{Oleynichenko}, \binits{A.V.}},
\bauthor{\bsnm{Raggio}, \binits{A.}},
\bauthor{\bsnm{Rothe}, \binits{S.}},
\bauthor{\bsnm{Skripnikov}, \binits{L.V.}},
\bauthor{\bsnm{Smets}, \binits{E.}},
\bauthor{\bsnm{Borne}, \binits{B.}},
\bauthor{\bsnm{Warbinek}, \binits{J.}},
\bauthor{\bsnm{Wessolek}, \binits{J.}},
\bauthor{\bsnm{Yang}, \binits{X.F.}}:
\batitle{{Radiative lifetime of the {{$A$ $^2\Pi_{1/2}$}} state in RaF with relevance to
  laser cooling}}.
\bjtitle{Physical Review A}
\bvolume{110}(\bissue{1}),
\bfpage{010802}
(\byear{2024})
\doiurl{10.1103/PhysRevA.110.L010802}
\end{barticle}
\endbibitem

\bibitem[\protect\citeauthoryear{Skripnikov et~al.}{2020}]{Skripnikov2020}
\begin{barticle}
\bauthor{\bsnm{Skripnikov}, \binits{L.V.}},
\bauthor{\bsnm{Mosyagin}, \binits{N.S.}},
\bauthor{\bsnm{Titov}, \binits{A.V.}},
\bauthor{\bsnm{Flambaum}, \binits{V.V.}}:
\batitle{{Actinide and lanthanide molecules to search for strong
  CP-violation}}.
\bjtitle{Physical Chemistry Chemical Physics}
\bvolume{22}(\bissue{33}),
\bfpage{18374}--\blpage{18380}
(\byear{2020})
\doiurl{10.1039/d0cp01989e}
\end{barticle}
\endbibitem

\bibitem[\protect\citeauthoryear{Chen et~al.}{2024}]{Chen2024SchiffMoments}
\begin{barticle}
\bauthor{\bsnm{Chen}, \binits{T.}},
\bauthor{\bsnm{Zhang}, \binits{C.}},
\bauthor{\bsnm{Cheng}, \binits{L.}},
\bauthor{\bsnm{Ng}, \binits{K.B.}},
\bauthor{\bsnm{Malbrunot-Ettenauer}, \binits{S.}},
\bauthor{\bsnm{Flambaum}, \binits{V.V.}},
\bauthor{\bsnm{Lasner}, \binits{Z.}},
\bauthor{\bsnm{Doyle}, \binits{J.M.}},
\bauthor{\bsnm{Yu}, \binits{P.}},
\bauthor{\bsnm{Conn}, \binits{C.J.}},
\bauthor{\bsnm{Zhang}, \binits{C.}},
\bauthor{\bsnm{Hutzler}, \binits{N.R.}},
\bauthor{\bsnm{Jayich}, \binits{A.M.}},
\bauthor{\bsnm{Augenbraun}, \binits{B.}},
\bauthor{\bsnm{DeMille}, \binits{D.}}:
\batitle{{Relativistic Exact Two-Component Coupled-Cluster Study of Molecular
  Sensitivity Factors for Nuclear Schiff Moments}}.
\bjtitle{The Journal of Physical Chemistry A}
\bvolume{128}(\bissue{31}),
\bfpage{6540}--\blpage{6554}
(\byear{2024})
\doiurl{10.1021/acs.jpca.4c02640}
\end{barticle}
\endbibitem

\bibitem[\protect\citeauthoryear{Gaul and Berger}{2020}]{gaul:2020}
\begin{barticle}
\bauthor{\bsnm{Gaul}, \binits{K.}},
\bauthor{\bsnm{Berger}, \binits{R.}}:
\batitle{Toolbox approach for quasi-relativistic calculation of molecular
  properties for precision tests of fundamental physics}.
\bjtitle{J. Chem. Phys.}
\bvolume{152}(\bissue{4}),
\bfpage{044101}
(\byear{2020})
\doiurl{10.1063/1.5121483}
{\href{https://arxiv.org/abs/1907.10432}{{arXiv:1907.10432}}}
{[physics.chem-ph]}
\end{barticle}
\endbibitem

\bibitem[\protect\citeauthoryear{Chupp and Ramsey-Musolf}{2015}]{chupp:2015}
\begin{barticle}
\bauthor{\bsnm{Chupp}, \binits{T.}},
\bauthor{\bsnm{Ramsey-Musolf}, \binits{M.}}:
\batitle{Electric dipole moments: A global analysis}.
\bjtitle{Phys. Rev. C}
\bvolume{91},
\bfpage{035502}
(\byear{2015})
\doiurl{10.1103/PhysRevC.91.035502}
\end{barticle}
\endbibitem

\bibitem[\protect\citeauthoryear{Gaul and Berger}{2024}]{Gaul2024GlobalCP_new}
\begin{barticle}
\bauthor{\bsnm{Gaul}, \binits{K.}},
\bauthor{\bsnm{Berger}, \binits{R.}}:
\batitle{Global analysis of $\mathcal{C}\mathcal{P}$-violation in atoms,
  molecules and role of medium-heavy systems}.
\bjtitle{Journal of High Energy Physics}
\bvolume{2024},
\bfpage{100}
(\byear{2024})
\doiurl{10.1007/JHEP08(2024)100}
\end{barticle}
\endbibitem

\bibitem[\protect\citeauthoryear{Degenkolb et~al.}{2024}]{degenkolb:2024}
\begin{barticle}
\bauthor{\bsnm{Degenkolb}, \binits{S.}},
\bauthor{\bsnm{Elmer}, \binits{N.}},
\bauthor{\bsnm{Modak}, \binits{T.}},
\bauthor{\bsnm{M{\"u}hlleitner}, \binits{M.}},
\bauthor{\bsnm{Plehn}, \binits{T.}}:
\batitle{A global view of the edm landscape}.
\bjtitle{arXiv}
\bvolume{hep-ph},
\bfpage{2403}--\blpage{02052}
(\byear{2024})
\doiurl{10.48550/arXiv.2403.02052}
\end{barticle}
\endbibitem

\bibitem[\protect\citeauthoryear{Dobaczewski and Engel}{2005}]{Dobaczewski2005}
\begin{barticle}
\bauthor{\bsnm{Dobaczewski}, \binits{J.}},
\bauthor{\bsnm{Engel}, \binits{J.}}:
\batitle{{Nuclear Time-Reversal Violation and the Schiff Moment of {{$^{225}$Ra}}}}.
\bjtitle{Physical Review Letters}
\bvolume{94}(\bissue{23}),
\bfpage{232502}
(\byear{2005})
\doiurl{10.1103/PhysRevLett.94.232502}
\end{barticle}
\endbibitem

\bibitem[\protect\citeauthoryear{Maples}{1977}]{Maples1977}
\begin{botherref}
\oauthor{\bsnm{Maples}, \binits{C.}}:
{Nuclear data sheets for A = 227}.
Nucl. Data Sheets
\textbf{22}(2)
(1977)
\doiurl{10.1016/S0090-3752(77)80008-2}
\end{botherref}
\endbibitem

\bibitem[\protect\citeauthoryear{de~Vries et~al.}{2015}]{deVries2015Theta}
\begin{barticle}
\bauthor{\bsnm{Vries}, \binits{J.}},
\bauthor{\bsnm{Mereghetti}, \binits{E.}},
\bauthor{\bsnm{Walker-Loud}, \binits{A.}}:
\batitle{{Baryon mass splittings and strong CP violation in SU(3) chiral
  perturbation theory}}.
\bjtitle{Physical Review C}
\bvolume{92}(\bissue{4}),
\bfpage{45201}
(\byear{2015})
\doiurl{10.1103/PhysRevC.92.045201}
\end{barticle}
\endbibitem

\bibitem[\protect\citeauthoryear{Bsaisou et~al.}{2015}]{Bsaisou2015Theta}
\begin{barticle}
\bauthor{\bsnm{Bsaisou}, \binits{J.}},
\bauthor{\bsnm{Vries}, \binits{J.}},
\bauthor{\bsnm{Hanhart}, \binits{C.}},
\bauthor{\bsnm{Liebig}, \binits{S.}},
\bauthor{\bsnm{Mei{\ss}ner}, \binits{U.-G.}},
\bauthor{\bsnm{Minossi}, \binits{D.}},
\bauthor{\bsnm{Nogga}, \binits{A.}},
\bauthor{\bsnm{Wirzba}, \binits{A.}}:
\batitle{{Nuclear electric dipole moments in chiral effective field theory}}.
\bjtitle{Journal of High Energy Physics}
\bvolume{2015}(\bissue{3}),
\bfpage{104}
(\byear{2015})
\doiurl{10.1007/JHEP03(2015)104}
\end{barticle}
\endbibitem

\bibitem[\protect\citeauthoryear{Yamanaka et~al.}{2017}]{Yamanaka2017Theta}
\begin{barticle}
\bauthor{\bsnm{Yamanaka}, \binits{N.}},
\bauthor{\bsnm{Sahoo}, \binits{B.K.}},
\bauthor{\bsnm{Yoshinaga}, \binits{N.}},
\bauthor{\bsnm{Sato}, \binits{T.}},
\bauthor{\bsnm{Asahi}, \binits{K.}},
\bauthor{\bsnm{Das}, \binits{B.P.}}:
\batitle{{Probing exotic phenomena at the interface of nuclear and particle
  physics with the electric dipole moments of diamagnetic atoms: A unique
  window to hadronic and semi-leptonic CP violation}}.
\bjtitle{The European Physical Journal A}
\bvolume{53}(\bissue{3}),
\bfpage{54}
(\byear{2017})
\doiurl{10.1140/epja/i2017-12237-2}
\end{barticle}
\endbibitem

\bibitem[\protect\citeauthoryear{Cho et~al.}{1989}]{Cho1989TlF}
\begin{barticle}
\bauthor{\bsnm{Cho}, \binits{D.}},
\bauthor{\bsnm{Sangster}, \binits{K.}},
\bauthor{\bsnm{Hinds}, \binits{E.A.}}:
\batitle{{Tenfold improvement of limits on T violation in thallium fluoride}}.
\bjtitle{Physical Review Letters}
\bvolume{63}(\bissue{23}),
\bfpage{2559}--\blpage{2562}
(\byear{1989})
\doiurl{10.1103/PhysRevLett.63.2559}
\end{barticle}
\endbibitem

\bibitem[\protect\citeauthoryear{Allmendinger et~al.}{2019}]{allmendinger:2019}
\begin{barticle}
\bauthor{\bsnm{Allmendinger}, \binits{F.}},
\bauthor{\bsnm{Engin}, \binits{I.}},
\bauthor{\bsnm{Heil}, \binits{W.}},
\bauthor{\bsnm{Karpuk}, \binits{S.}},
\bauthor{\bsnm{Krause}, \binits{H.-J.}},
\bauthor{\bsnm{Niederl{\"a}nder}, \binits{B.}},
\bauthor{\bsnm{Offenh{\"a}usser}, \binits{A.}},
\bauthor{\bsnm{Repetto}, \binits{M.}},
\bauthor{\bsnm{Schmidt}, \binits{U.}},
\bauthor{\bsnm{Zimmer}, \binits{S.}}:
\batitle{Measurement of the permanent electric dipole moment of the
  {$^{129}\mathrm{Xe}$} atom}.
\bjtitle{Phys. Rev. A}
\bvolume{100},
\bfpage{022505}
(\byear{2019})
\doiurl{10.1103/PhysRevA.100.022505}
\end{barticle}
\endbibitem

\bibitem[\protect\citeauthoryear{Sachdeva et~al.}{2019}]{sachdeva:2019}
\begin{barticle}
\bauthor{\bsnm{Sachdeva}, \binits{N.}},
\bauthor{\bsnm{Fan}, \binits{I.}},
\bauthor{\bsnm{Babcock}, \binits{E.}},
\bauthor{\bsnm{Burghoff}, \binits{M.}},
\bauthor{\bsnm{Chupp}, \binits{T.E.}},
\bauthor{\bsnm{Degenkolb}, \binits{S.}},
\bauthor{\bsnm{Fierlinger}, \binits{P.}},
\bauthor{\bsnm{Haude}, \binits{S.}},
\bauthor{\bsnm{Kraegeloh}, \binits{E.}},
\bauthor{\bsnm{Kilian}, \binits{W.}},
\bauthor{\bsnm{Knappe-Gr{\"u}neberg}, \binits{S.}},
\bauthor{\bsnm{Kuchler}, \binits{F.}},
\bauthor{\bsnm{Liu}, \binits{T.}},
\bauthor{\bsnm{Marino}, \binits{M.}},
\bauthor{\bsnm{Meinel}, \binits{J.}},
\bauthor{\bsnm{Rolfs}, \binits{K.}},
\bauthor{\bsnm{Salhi}, \binits{Z.}},
\bauthor{\bsnm{Schnabel}, \binits{A.}},
\bauthor{\bsnm{Singh}, \binits{J.T.}},
\bauthor{\bsnm{Stuiber}, \binits{S.}},
\bauthor{\bsnm{Terrano}, \binits{W.A.}},
\bauthor{\bsnm{Trahms}, \binits{L.}},
\bauthor{\bsnm{Voigt}, \binits{J.}}:
\batitle{New limit on the permanent electric dipole moment of
  {$^{129}\mathrm{Xe}$} using {$^{3}\mathrm{He}$} comagnetometry and {SQUID}
  detection}.
\bjtitle{Phys. Rev. Lett.}
\bvolume{123},
\bfpage{143003}
(\byear{2019})
\doiurl{10.1103/PhysRevLett.123.143003}
\end{barticle}
\endbibitem

\bibitem[\protect\citeauthoryear{Zheng et~al.}{2022}]{zheng:2022}
\begin{barticle}
\bauthor{\bsnm{Zheng}, \binits{T.A.}},
\bauthor{\bsnm{Yang}, \binits{Y.A.}},
\bauthor{\bsnm{Wang}, \binits{S.-Z.}},
\bauthor{\bsnm{Singh}, \binits{J.T.}},
\bauthor{\bsnm{Xiong}, \binits{Z.-X.}},
\bauthor{\bsnm{Xia}, \binits{T.}},
\bauthor{\bsnm{Lu}, \binits{Z.-T.}}:
\batitle{Measurement of the electric dipole moment of {$^{171}\mathrm{Yb}$}
  atoms in an optical dipole trap}.
\bjtitle{Phys. Rev. Lett.}
\bvolume{129},
\bfpage{083001}
(\byear{2022})
\doiurl{10.1103/PhysRevLett.129.083001}
\end{barticle}
\endbibitem

\bibitem[\protect\citeauthoryear{Murthy et~al.}{1989}]{murthy:1989}
\begin{barticle}
\bauthor{\bsnm{Murthy}, \binits{S.A.}},
\bauthor{\bsnm{Krause}, \binits{D.}},
\bauthor{\bsnm{Li}, \binits{Z.L.}},
\bauthor{\bsnm{Hunter}, \binits{L.R.}}:
\batitle{New limits on the electron electric dipole moment from cesium}.
\bjtitle{Phys. Rev. Lett.}
\bvolume{63},
\bfpage{965}--\blpage{968}
(\byear{1989})
\doiurl{10.1103/PhysRevLett.63.965}
\end{barticle}
\endbibitem

\bibitem[\protect\citeauthoryear{Regan et~al.}{2002}]{regan:2002}
\begin{barticle}
\bauthor{\bsnm{Regan}, \binits{B.C.}},
\bauthor{\bsnm{Commins}, \binits{E.D.}},
\bauthor{\bsnm{Schmidt}, \binits{C.J.}},
\bauthor{\bsnm{DeMille}, \binits{D.}}:
\batitle{{New Limit on the Electron Electric Dipole Moment}}.
\bjtitle{Phys. Rev. Lett.}
\bvolume{88}(\bissue{7}),
\bfpage{71805}
(\byear{2002})
\doiurl{10.1103/PhysRevLett.88.071805}
\end{barticle}
\endbibitem

\bibitem[\protect\citeauthoryear{Parker et~al.}{2015}]{parker:2015}
\begin{barticle}
\bauthor{\bsnm{Parker}, \binits{R.H.}},
\bauthor{\bsnm{Dietrich}, \binits{M.R.}},
\bauthor{\bsnm{Kalita}, \binits{M.R.}},
\bauthor{\bsnm{Lemke}, \binits{N.D.}},
\bauthor{\bsnm{Bailey}, \binits{K.G.}},
\bauthor{\bsnm{Bishof}, \binits{M.}},
\bauthor{\bsnm{Greene}, \binits{J.P.}},
\bauthor{\bsnm{Holt}, \binits{R.J.}},
\bauthor{\bsnm{Korsch}, \binits{W.}},
\bauthor{\bsnm{Lu}, \binits{Z.-T.}},
\bauthor{\bsnm{Mueller}, \binits{P.}},
\bauthor{\bsnm{O'Connor}, \binits{T.P.}},
\bauthor{\bsnm{Singh}, \binits{J.T.}}:
\batitle{First measurement of the atomic electric dipole moment of
  {$^{225}\mathrm{Ra}$}}.
\bjtitle{Phys. Rev. Lett.}
\bvolume{114},
\bfpage{233002}
(\byear{2015})
\doiurl{10.1103/PhysRevLett.114.233002}
\end{barticle}
\endbibitem

\bibitem[\protect\citeauthoryear{Hudson et~al.}{2002}]{hudson:2002}
\begin{barticle}
\bauthor{\bsnm{Hudson}, \binits{J.J.}},
\bauthor{\bsnm{Sauer}, \binits{B.E.}},
\bauthor{\bsnm{Tarbutt}, \binits{M.R.}},
\bauthor{\bsnm{Hinds}, \binits{E.A.}}:
\batitle{{Measurement of the Electron Electric Dipole Moment Using YbF
  Molecules}}.
\bjtitle{Phys. Rev. Lett.}
\bvolume{89}(\bissue{2}),
\bfpage{23003}
(\byear{2002})
\doiurl{10.1103/PhysRevLett.89.023003}
\end{barticle}
\endbibitem

\bibitem[\protect\citeauthoryear{Hudson et~al.}{2011}]{hudson:2011}
\begin{barticle}
\bauthor{\bsnm{Hudson}, \binits{J.J.}},
\bauthor{\bsnm{Kara}, \binits{D.M.}},
\bauthor{\bsnm{Smallman}, \binits{I.J.}},
\bauthor{\bsnm{Sauer}, \binits{B.E.}},
\bauthor{\bsnm{Tarbutt}, \binits{M.R.}},
\bauthor{\bsnm{Hinds}, \binits{E.A.}}:
\batitle{{Improved measurement of the shape of the electron}}.
\bjtitle{Nature}
\bvolume{473}(\bissue{7348}),
\bfpage{493}
(\byear{2011})
\doiurl{10.1038/nature10104}
\end{barticle}
\endbibitem

\bibitem[\protect\citeauthoryear{Roussy et~al.}{2023}]{roussy:2023}
\begin{barticle}
\bauthor{\bsnm{Roussy}, \binits{T.S.}},
\bauthor{\bsnm{Caldwell}, \binits{L.}},
\bauthor{\bsnm{Wright}, \binits{T.}},
\bauthor{\bsnm{Cairncross}, \binits{W.B.}},
\bauthor{\bsnm{Shagam}, \binits{Y.}},
\bauthor{\bsnm{Ng}, \binits{K.B.}},
\bauthor{\bsnm{Schlossberger}, \binits{N.}},
\bauthor{\bsnm{Park}, \binits{S.Y.}},
\bauthor{\bsnm{Wang}, \binits{A.}},
\bauthor{\bsnm{Ye}, \binits{J.}},
\bauthor{\bsnm{Cornell}, \binits{E.A.}}:
\batitle{An improved bound on the electron's electric dipole moment}.
\bjtitle{Science}
\bvolume{381}(\bissue{6653}),
\bfpage{46}--\blpage{50}
(\byear{2023})
\doiurl{10.1126/science.adg4084}
{\href{https://arxiv.org/abs/https://www.science.org/doi/pdf/10.1126/science.adg4084}{{https://www.science.org/doi/pdf/10.1126/science.adg4084}}}
\end{barticle}
\endbibitem

\bibitem[\protect\citeauthoryear{Cho et~al.}{1991}]{cho:1991}
\begin{barticle}
\bauthor{\bsnm{Cho}, \binits{D.}},
\bauthor{\bsnm{Sangster}, \binits{K.}},
\bauthor{\bsnm{Hinds}, \binits{E.A.}}:
\batitle{Search for time-reversal-symmetry violation in thallium fluoride using
  a jet source}.
\bjtitle{Phys. Rev. A}
\bvolume{44},
\bfpage{2783}--\blpage{2799}
(\byear{1991})
\doiurl{10.1103/PhysRevA.44.2783}
\end{barticle}
\endbibitem

\bibitem[\protect\citeauthoryear{Eckel et~al.}{2013}]{eckel:2013}
\begin{barticle}
\bauthor{\bsnm{Eckel}, \binits{S.}},
\bauthor{\bsnm{Hamilton}, \binits{P.}},
\bauthor{\bsnm{Kirilov}, \binits{E.}},
\bauthor{\bsnm{Smith}, \binits{H.W.}},
\bauthor{\bsnm{DeMille}, \binits{D.}}:
\batitle{{Search for the electron electric dipole moment using
  $\ensuremath{\Omega}$-doublet levels in PbO}}.
\bjtitle{Phys. Rev. A}
\bvolume{87},
\bfpage{052130}
(\byear{2013})
\doiurl{10.1103/PhysRevA.87.052130}
\end{barticle}
\endbibitem

\bibitem[\protect\citeauthoryear{Andreev et~al.}{2018}]{andreev:2018}
\begin{barticle}
\bauthor{\bsnm{Andreev}, \binits{V.}},
\bauthor{\bsnm{Ang}, \binits{D.G.}},
\bauthor{\bsnm{DeMille}, \binits{D.}},
\bauthor{\bsnm{Doyle}, \binits{J.M.}},
\bauthor{\bsnm{Gabrielse}, \binits{G.}},
\bauthor{\bsnm{Haefner}, \binits{J.}},
\bauthor{\bsnm{Hutzler}, \binits{N.R.}},
\bauthor{\bsnm{Lasner}, \binits{Z.}},
\bauthor{\bsnm{Meisenhelder}, \binits{C.}},
\bauthor{\bsnm{O'Leary}, \binits{B.R.}},
\bauthor{\bsnm{Panda}, \binits{C.D.}},
\bauthor{\bsnm{West}, \binits{A.D.}},
\bauthor{\bsnm{West}, \binits{E.P.}},
\bauthor{\bsnm{Wu}, \binits{X.}},
\bauthor{\bsnm{{Collaboration, ACME}}}:
\batitle{Improved limit on the electric dipole moment of the electron}.
\bjtitle{Nature}
\bvolume{562}(\bissue{7727}),
\bfpage{355}--\blpage{360}
(\byear{2018})
\doiurl{10.1038/s41586-018-0599-8}
\end{barticle}
\endbibitem

\bibitem[\protect\citeauthoryear{Chupp et~al.}{2019}]{Chupp2019}
\begin{barticle}
\bauthor{\bsnm{Chupp}, \binits{T.E.}},
\bauthor{\bsnm{Fierlinger}, \binits{P.}},
\bauthor{\bsnm{Ramsey-Musolf}, \binits{M.J.}},
\bauthor{\bsnm{Singh}, \binits{J.T.}}:
\batitle{{Electric dipole moments of atoms, molecules, nuclei, and particles}}.
\bjtitle{Reviews of Modern Physics}
\bvolume{91}(\bissue{1}),
\bfpage{15001}
(\byear{2019})
\doiurl{10.1103/RevModPhys.91.015001}
\end{barticle}
\endbibitem

\bibitem[\protect\citeauthoryear{Ginges and Flambaum}{2004}]{Ginges2004}
\begin{barticle}
\bauthor{\bsnm{Ginges}, \binits{J.S.M.}},
\bauthor{\bsnm{Flambaum}, \binits{V.V.}}:
\batitle{{Violations of fundamental symmetries in atoms and tests of
  unification theories of elementary particles}}.
\bjtitle{Physics Reports}
\bvolume{397}(\bissue{2}),
\bfpage{63}--\blpage{154}
(\byear{2004})
\doiurl{10.1016/j.physrep.2004.03.005}
\end{barticle}
\endbibitem

\bibitem[\protect\citeauthoryear{Ferrer et~al.}{2017}]{Ferrer2017}
\begin{botherref}
\oauthor{\bsnm{Ferrer}, \binits{R.}},
\oauthor{\bsnm{Barzakh}, \binits{A.}},
\oauthor{\bsnm{Bastin}, \binits{B.}},
\oauthor{\bsnm{Beerwerth}, \binits{R.}},
\oauthor{\bsnm{Block}, \binits{M.}},
\oauthor{\bsnm{Creemers}, \binits{P.}},
\oauthor{\bsnm{Grawe}, \binits{H.}},
\oauthor{\bsnm{De~Groote}, \binits{R.}},
\oauthor{\bsnm{Delahaye}, \binits{P.}},
\oauthor{\bsnm{Fl{\'{e}}chard}, \binits{X.}},
\oauthor{\bsnm{Franchoo}, \binits{S.}},
\oauthor{\bsnm{Fritzsche}, \binits{S.}},
\oauthor{\bsnm{Gaffney}, \binits{L.P.}},
\oauthor{\bsnm{Ghys}, \binits{L.}},
\oauthor{\bsnm{Gins}, \binits{W.}},
\oauthor{\bsnm{Granados}, \binits{C.}},
\oauthor{\bsnm{Heinke}, \binits{R.}},
\oauthor{\bsnm{Hijazi}, \binits{L.}},
\oauthor{\bsnm{Huyse}, \binits{M.}},
\oauthor{\bsnm{Kron}, \binits{T.}},
\oauthor{\bsnm{Kudryavtsev}, \binits{Y.}},
\oauthor{\bsnm{Laatiaoui}, \binits{M.}},
\oauthor{\bsnm{Lecesne}, \binits{N.}},
\oauthor{\bsnm{Loiselet}, \binits{M.}},
\oauthor{\bsnm{Lutton}, \binits{F.}},
\oauthor{\bsnm{Moore}, \binits{I.D.}},
\oauthor{\bsnm{Mart{\'{i}}nez}, \binits{Y.}},
\oauthor{\bsnm{Mogilevskiy}, \binits{E.}},
\oauthor{\bsnm{Naubereit}, \binits{P.}},
\oauthor{\bsnm{Piot}, \binits{J.}},
\oauthor{\bsnm{Raeder}, \binits{S.}},
\oauthor{\bsnm{Rothe}, \binits{S.}},
\oauthor{\bsnm{Savajols}, \binits{H.}},
\oauthor{\bsnm{Sels}, \binits{S.}},
\oauthor{\bsnm{Sonnenschein}, \binits{V.}},
\oauthor{\bsnm{Thomas}, \binits{J.C.}},
\oauthor{\bsnm{Traykov}, \binits{E.}},
\oauthor{\bsnm{Van~Beveren}, \binits{C.}},
\oauthor{\bsnm{Van Den~Bergh}, \binits{P.}},
\oauthor{\bsnm{Van~Duppen}, \binits{P.}},
\oauthor{\bsnm{Wendt}, \binits{K.}},
\oauthor{\bsnm{Zadvornaya}, \binits{A.}}:
{Towards high-resolution laser ionization spectroscopy of the heaviest elements
  in supersonic gas jet expansion}.
Nature Communications
\textbf{8}
(2017)
\doiurl{10.1038/ncomms14520}
\end{botherref}
\endbibitem

\bibitem[\protect\citeauthoryear{Ferrer et~al.}{2021}]{Ferrer2021IGLIS}
\begin{barticle}
\bauthor{\bsnm{Ferrer}, \binits{R.}},
\bauthor{\bsnm{Verlinde}, \binits{M.}},
\bauthor{\bsnm{Verstraelen}, \binits{E.}},
\bauthor{\bsnm{Claessens}, \binits{A.}},
\bauthor{\bsnm{Huyse}, \binits{M.}},
\bauthor{\bsnm{Kraemer}, \binits{S.}},
\bauthor{\bsnm{Kudryavtsev}, \binits{Y.}},
\bauthor{\bsnm{Romans}, \binits{J.}},
\bauthor{\bsnm{Bergh}, \binits{P.}},
\bauthor{\bsnm{Van~Duppen}, \binits{P.}},
\bauthor{\bsnm{Zadvornaya}, \binits{A.}},
\bauthor{\bsnm{Chazot}, \binits{O.}},
\bauthor{\bsnm{Grossir}, \binits{G.}},
\bauthor{\bsnm{Kalikmanov}, \binits{V.I.}},
\bauthor{\bsnm{Nabuurs}, \binits{M.}},
\bauthor{\bsnm{Reynaerts}, \binits{D.}}:
\batitle{{Hypersonic nozzle for laser-spectroscopy studies at 17 K
  characterized by resonance-ionization-spectroscopy-based flow mapping}}.
\bjtitle{Physical Review Research}
\bvolume{3}(\bissue{4}),
\bfpage{43041}
(\byear{2021})
\doiurl{10.1103/PhysRevResearch.3.043041}
\end{barticle}
\endbibitem

\bibitem[\protect\citeauthoryear{Lantis et~al.}{2024}]{Lantis2024Jetris}
\begin{barticle}
\bauthor{\bsnm{Lantis}, \binits{J.}},
\bauthor{\bsnm{Claessens}, \binits{A.}},
\bauthor{\bsnm{M{\"{u}}nzberg}, \binits{D.}},
\bauthor{\bsnm{Auler}, \binits{J.}},
\bauthor{\bsnm{Block}, \binits{M.}},
\bauthor{\bsnm{Chhetri}, \binits{P.}},
\bauthor{\bsnm{D{\"{u}}llmann}, \binits{C.E.}},
\bauthor{\bsnm{Ferrer}, \binits{R.}},
\bauthor{\bsnm{Giacoppo}, \binits{F.}},
\bauthor{\bsnm{Guti{\'{e}}rrez}, \binits{M.J.}},
\bauthor{\bsnm{Ivandikov}, \binits{F.}},
\bauthor{\bsnm{Kaleja}, \binits{O.}},
\bauthor{\bsnm{Kieck}, \binits{T.}},
\bauthor{\bsnm{Kim}, \binits{E.}},
\bauthor{\bsnm{Laatiaoui}, \binits{M.}},
\bauthor{\bsnm{Lecesne}, \binits{N.}},
\bauthor{\bsnm{Manea}, \binits{V.}},
\bauthor{\bsnm{Nothhelfer}, \binits{S.}},
\bauthor{\bsnm{Raeder}, \binits{S.}},
\bauthor{\bsnm{Romans}, \binits{J.}},
\bauthor{\bsnm{Romero-Romero}, \binits{E.}},
\bauthor{\bsnm{Roubin}, \binits{A.}},
\bauthor{\bsnm{Savajols}, \binits{H.}},
\bauthor{\bsnm{Sels}, \binits{S.}},
\bauthor{\bsnm{Stemmler}, \binits{M.}},
\bauthor{\bsnm{Van~Duppen}, \binits{P.}},
\bauthor{\bsnm{Walther}, \binits{T.}},
\bauthor{\bsnm{Warbinek}, \binits{J.}},
\bauthor{\bsnm{Wendt}, \binits{K.}},
\bauthor{\bsnm{Yakushev}, \binits{A.}},
\bauthor{\bsnm{Zadvornaya}, \binits{A.}}:
\batitle{{In-gas-jet laser spectroscopy of {{$^{254}$No}} with JetRIS}}.
\bjtitle{Physical Review Research}
\bvolume{6}(\bissue{2}),
\bfpage{23318}
(\byear{2024})
\doiurl{10.1103/PhysRevResearch.6.023318}
\end{barticle}
\endbibitem

\bibitem[\protect\citeauthoryear{Udrescu et~al.}{2024}]{Udrescu2023}
\begin{barticle}
\bauthor{\bsnm{Udrescu}, \binits{S.M.}},
\bauthor{\bsnm{Wilkins}, \binits{S.G.}},
\bauthor{\bsnm{Breier}, \binits{A.A.}},
\bauthor{\bsnm{Athanasakis-Kaklamanakis}, \binits{M.}},
\bauthor{\bsnm{Garcia~Ruiz}, \binits{R.F.}},
\bauthor{\bsnm{Au}, \binits{M.}},
\bauthor{\bsnm{Belo{\v{s}}evi{\'{c}}}, \binits{I.}},
\bauthor{\bsnm{Berger}, \binits{R.}},
\bauthor{\bsnm{Bissell}, \binits{M.L.}},
\bauthor{\bsnm{Binnersley}, \binits{C.L.}},
\bauthor{\bsnm{Brinson}, \binits{A.J.}},
\bauthor{\bsnm{Chrysalidis}, \binits{K.}},
\bauthor{\bsnm{Cocolios}, \binits{T.E.}},
\bauthor{\bsnm{Groote}, \binits{R.P.}},
\bauthor{\bsnm{Dorne}, \binits{A.}},
\bauthor{\bsnm{Flanagan}, \binits{K.T.}},
\bauthor{\bsnm{Franchoo}, \binits{S.}},
\bauthor{\bsnm{Gaul}, \binits{K.}},
\bauthor{\bsnm{Geldhof}, \binits{S.}},
\bauthor{\bsnm{Giesen}, \binits{T.F.}},
\bauthor{\bsnm{Hanstorp}, \binits{D.}},
\bauthor{\bsnm{Heinke}, \binits{R.}},
\bauthor{\bsnm{Koszor{\'{u}}s}, \binits{A.}},
\bauthor{\bsnm{Kujanp{\"{a}}{\"{a}}}, \binits{S.}},
\bauthor{\bsnm{Lalanne}, \binits{L.}},
\bauthor{\bsnm{Neyens}, \binits{G.}},
\bauthor{\bsnm{Nichols}, \binits{M.}},
\bauthor{\bsnm{Perrett}, \binits{H.A.}},
\bauthor{\bsnm{Reilly}, \binits{J.R.}},
\bauthor{\bsnm{Rothe}, \binits{S.}},
\bauthor{\bsnm{Borne}, \binits{B.}},
\bauthor{\bsnm{Vernon}, \binits{A.R.}},
\bauthor{\bsnm{Wang}, \binits{Q.}},
\bauthor{\bsnm{Wessolek}, \binits{J.}},
\bauthor{\bsnm{Yang}, \binits{X.F.}},
\bauthor{\bsnm{Z{\"{u}}lch}, \binits{C.}}:
\batitle{{Precision spectroscopy and laser-cooling scheme of a
  radium-containing molecule}}.
\bjtitle{Nature Physics}
\bvolume{20}(\bissue{2}),
\bfpage{202}--\blpage{207}
(\byear{2024})
\doiurl{10.1038/s41567-023-02296-w}
\end{barticle}
\endbibitem

\bibitem[\protect\citeauthoryear{}{2023}]{DIRAC23}
\begin{botherref}
{DIRAC}, a relativistic ab initio electronic structure program, Release
  {DIRAC23} (2023), written by R.~Bast, A.~S.~P.~Gomes, T.~Saue and L.~Visscher
  and H.~J.~{\relax Aa}.~Jensen, with contributions from I.~A.~Aucar,
  V.~Bakken, C.~Chibueze, J.~Creutzberg, K.~G.~Dyall, S.~Dubillard,
  U.~Ekstr{\"o}m, E.~Eliav, T.~Enevoldsen, E.~Fa{\ss}hauer, T.~Fleig,
  O.~Fossgaard, L.~Halbert, E.~D.~Hedeg{\aa}rd, T.~Helgaker, B.~Helmich--Paris,
  J.~Henriksson, M.~van~Horn, M.~Ilia{\v{s}}, Ch.~R.~Jacob, S.~Knecht,
  S.~Komorovsk{\'y}, O.~Kullie, J.~K.~L{\ae}rdahl, C.~V.~Larsen, Y.~S.~Lee,
  N.~H.~List, H.~S.~Nataraj, M.~K.~Nayak, P.~Norman, A.~Nyvang, G.~Olejniczak,
  J.~Olsen, J.~M.~H.~Olsen, A.~Papadopoulos, Y.~C.~Park, J.~K.~Pedersen,
  M.~Pernpointner, J.~V.~Pototschnig, R.~di~Remigio, M.~Repisky, K.~Ruud,
  P.~Sa{\l}ek, B.~Schimmelpfennig, B.~Senjean, A.~Shee, J.~Sikkema, A.~Sunaga,
  A.~J.~Thorvaldsen, J.~Thyssen, J.~van~Stralen, M.~L.~Vidal, S.~Villaume,
  O.~Visser, T.~Winther, S.~Yamamoto and X.~Yuan (available at
  \url{https://doi.org/10.5281/zenodo.7670749}, see also
  \url{https://www.diracprogram.org})
(2023)
\end{botherref}
\endbibitem

\bibitem[\protect\citeauthoryear{Gaul et~al.}{2024}]{gaul:2024b}
\begin{barticle}
\bauthor{\bsnm{Gaul}, \binits{K.}},
\bauthor{\bsnm{Hutzler}, \binits{N.R.}},
\bauthor{\bsnm{Yu}, \binits{P.}},
\bauthor{\bsnm{Jayich}, \binits{A.M.}},
\bauthor{\bsnm{Ilias}, \binits{M.}},
\bauthor{\bsnm{Borschevsky}, \binits{A.}}:
\batitle{$\mathcal{CP}$-violation sensitivity of closed-shell radium-containing
  polyatomic molecular ions}.
\bjtitle{Phys. Rev. A}
\bvolume{109},
\bfpage{042819}
(\byear{2024})
\doiurl{10.1103/PhysRevA.109.042819}
\end{barticle}
\endbibitem

\bibitem[\protect\citeauthoryear{Dyall}{2016}]{dyall3}
\begin{botherref}
\oauthor{\bsnm{Dyall}, \binits{K.G.}}:
{Relativistic double-zeta, triple-zeta, and quadruple-zeta basis sets for the
  light elements {H}–{Ar}}.
Theor Chem Acc
\textbf{135}
(2016)
\end{botherref}
\endbibitem

\bibitem[\protect\citeauthoryear{Dyall}{2009}]{dyall2}
\begin{barticle}
\bauthor{\bsnm{Dyall}, \binits{K.G.}}:
\batitle{{Relativistic Double-Zeta, Triple-Zeta, and Quadruple-Zeta Basis Sets
  for the {4s}, {5s}, {6s}, and {7s} Elements}}.
\bjtitle{The Journal of Physical Chemistry A}
\bvolume{113},
\bfpage{12638}--\blpage{12644}
(\byear{2009})
\end{barticle}
\endbibitem

\bibitem[\protect\citeauthoryear{Helgaker et~al.}{1997}]{cbs}
\begin{barticle}
\bauthor{\bsnm{Helgaker}, \binits{T.}},
\bauthor{\bsnm{Klopper}, \binits{W.}},
\bauthor{\bsnm{Koch}, \binits{H.}},
\bauthor{\bsnm{Noga}, \binits{J.}}:
\batitle{Basis-set convergence of correlated calculations on water}.
\bjtitle{The Journal of Chemical Physics}
\bvolume{106},
\bfpage{9639}--\blpage{9646}
(\byear{1997})
\end{barticle}
\endbibitem

\bibitem[\protect\citeauthoryear{Hao et~al.}{2018}]{prop1}
\begin{barticle}
\bauthor{\bsnm{Hao}, \binits{Y.}},
\bauthor{\bsnm{Ilia\v{s}}, \binits{M.}},
\bauthor{\bsnm{Eliav}, \binits{E.}},
\bauthor{\bsnm{Schwerdtfeger}, \binits{P.}},
\bauthor{\bsnm{Flambaum}, \binits{V.V.}},
\bauthor{\bsnm{Borschevsky}, \binits{A.}}:
\batitle{Nuclear anapole moment interaction in {BaF} from relativistic
  coupled-cluster theory}.
\bjtitle{Phys. Rev. A}
\bvolume{98},
\bfpage{032510}
(\byear{2018})
\end{barticle}
\endbibitem

\bibitem[\protect\citeauthoryear{Guo et~al.}{2021}]{prop2}
\begin{barticle}
\bauthor{\bsnm{Guo}, \binits{Y.}},
\bauthor{\bsnm{Pa\v{s}teka}, \binits{L.F.}},
\bauthor{\bsnm{Eliav}, \binits{E.}},
\bauthor{\bsnm{Borschevsky}, \binits{A.}}:
\batitle{Ionization potentials and electron affinity of oganesson with
  relativistic coupled cluster method}.
\bjtitle{Advances in Quantum Chemistry}
\bvolume{83},
\bfpage{107}--\blpage{123}
(\byear{2021})
\end{barticle}
\endbibitem

\bibitem[\protect\citeauthoryear{Haase et~al.}{2020}]{prop3}
\begin{barticle}
\bauthor{\bsnm{Haase}, \binits{P.A.B.}},
\bauthor{\bsnm{Eliav}, \binits{E.}},
\bauthor{\bsnm{Ilia\v{s}}, \binits{M.}},
\bauthor{\bsnm{Borschevsky}, \binits{A.}}:
\batitle{Hyperfine structure constants on the relativistic coupled cluster
  level with associated uncertainties}.
\bjtitle{J. Phys. Chem. A}
\bvolume{124},
\bfpage{3157}--\blpage{3169}
(\byear{2020})
\end{barticle}
\endbibitem

\bibitem[\protect\citeauthoryear{Leimbach et~al.}{2020}]{LeiKarGuo20}
\begin{barticle}
\bauthor{\bsnm{Leimbach}, \binits{D.}},
\bauthor{\bsnm{Karls}, \binits{J.}},
\bauthor{\bsnm{Guo}, \binits{Y.}},
\bauthor{\bsnm{Ahmed}, \binits{R.}},
\bauthor{\bsnm{Ballof}, \binits{J.}},
\bauthor{\bsnm{Bengtsson}, \binits{L.}},
\bauthor{\bsnm{Boix~Pamies}, \binits{F.}},
\bauthor{\bsnm{Borschevsky}, \binits{A.}},
\bauthor{\bsnm{Chrysalidis}, \binits{K.}},
\bauthor{\bsnm{Eliav}, \binits{E.}}, \betal:
\batitle{The electron affinity of astatine}.
\bjtitle{Nature communications}
\bvolume{11}(\bissue{1}),
\bfpage{3824}
(\byear{2020})
\end{barticle}
\endbibitem

\bibitem[\protect\citeauthoryear{Kyuberis et~al.}{2024}]{kyuPas23}
\begin{barticle}
\bauthor{\bsnm{Kyuberis}, \binits{A.A.}},
\bauthor{\bsnm{Pasteka}, \binits{L.F.}},
\bauthor{\bsnm{Eliav}, \binits{E.}},
\bauthor{\bsnm{Perrett}, \binits{H.}},
\bauthor{\bsnm{Sunaga}, \binits{A.}},
\bauthor{\bsnm{Udrescu}, \binits{S.M.}},
\bauthor{\bsnm{Wilkins}, \binits{S.G.}},
\bauthor{\bsnm{Ruiz}, \binits{R.F.G.}},
\bauthor{\bsnm{Borschevsky}, \binits{A.}}:
\batitle{Theoretical determination of the ionization potentials of {CaF},
  {SrF}, and {BaF}}.
\bjtitle{Physical Review A}
\bvolume{109},
\bfpage{022813}
(\byear{2024})
\end{barticle}
\endbibitem

\bibitem[\protect\citeauthoryear{Becke}{1993}]{becke:1993}
\begin{barticle}
\bauthor{\bsnm{Becke}, \binits{A.D.}}:
\batitle{{A new mixing of Hartree--Fock and local density-functional
  theories}}.
\bjtitle{The Journal of chemical physics}
\bvolume{98}(\bissue{2}),
\bfpage{1372}--\blpage{1377}
(\byear{1993})
\end{barticle}
\endbibitem

\bibitem[\protect\citeauthoryear{Chang et~al.}{1986}]{chang:1986}
\begin{barticle}
\bauthor{\bsnm{Chang}, \binits{C.}},
\bauthor{\bsnm{Pelissier}, \binits{M.}},
\bauthor{\bsnm{Durand}, \binits{P.}}:
\batitle{Regular two-component {P}auli-like effective {H}amiltonians in {D}irac
  theory}.
\bjtitle{Phys. Scr.}
\bvolume{34},
\bfpage{394}--\blpage{404}
(\byear{1986})
\end{barticle}
\endbibitem

\bibitem[\protect\citeauthoryear{van Lenthe et~al.}{1993}]{lenthe:1993}
\begin{barticle}
\bauthor{\bsnm{Lenthe}, \binits{E.}},
\bauthor{\bsnm{Baerends}, \binits{E.-J.}},
\bauthor{\bsnm{Snijders}, \binits{J.G.}}:
\batitle{Relativistic regular two-component {H}amiltonians}.
\bjtitle{J. Chem. Phys.}
\bvolume{99},
\bfpage{4597}
(\byear{1993})
\doiurl{10.1063/1.466059}
\end{barticle}
\endbibitem

\bibitem[\protect\citeauthoryear{Visscher and Dyall}{1997}]{visscher:1997}
\begin{barticle}
\bauthor{\bsnm{Visscher}, \binits{L.}},
\bauthor{\bsnm{Dyall}, \binits{K.G.}}:
\batitle{{Dirac-Fock atomic electronic structure calculations using different
  nuclear charge distributions}}.
\bjtitle{At. Data Nucl. Data Tables}
\bvolume{67},
\bfpage{207}--\blpage{224}
(\byear{1997})
\end{barticle}
\endbibitem

\bibitem[\protect\citeauthoryear{van W{\"u}llen}{1998}]{wullen:1998}
\begin{barticle}
\bauthor{\bsnm{W{\"u}llen}, \binits{C.}}:
\batitle{Molecular density functional calculations in the regular relativistic
  approximation: {M}ethod, application to coinage metal diatomics, hydrides,
  fluorides and chlorides, and comparison with first-order relativistic
  calculations}.
\bjtitle{J. Chem. Phys.}
\bvolume{109},
\bfpage{392}--\blpage{399}
(\byear{1998})
\doiurl{10.1063/1.476576}
\end{barticle}
\endbibitem

\bibitem[\protect\citeauthoryear{Liu et~al.}{2002}]{liu:2002}
\begin{barticle}
\bauthor{\bsnm{Liu}, \binits{W.}},
\bauthor{\bsnm{W{\"u}llen}, \binits{C.}},
\bauthor{\bsnm{Wang}, \binits{F.}},
\bauthor{\bsnm{Li}, \binits{L.}}:
\batitle{Spectroscopic constants of {MH} and {M}$_2$ ({M} = {Tl}, {E113}, {Bi},
  {E115}): direct comparisons of four- and two-component approaches in the
  framework of relativistic density functional theory}.
\bjtitle{J. Chem. Phys.}
\bvolume{116},
\bfpage{3626}--\blpage{3634}
(\byear{2002})
\end{barticle}
\endbibitem

\bibitem[\protect\citeauthoryear{Gilbert et~al.}{2008}]{gilbert:2008}
\begin{barticle}
\bauthor{\bsnm{Gilbert}, \binits{A.T.B.}},
\bauthor{\bsnm{Besley}, \binits{N.A.}},
\bauthor{\bsnm{Gill}, \binits{P.M.W.}}:
\batitle{Self-consistent field calculations of excited states using the maximum
  overlap method {(MOM)}}.
\bjtitle{J. Phys. Chem. A}
\bvolume{112}(\bissue{50}),
\bfpage{13164}--\blpage{13171}
(\byear{2008})
\doiurl{10.1021/jp801738f}
{\href{https://arxiv.org/abs/https://doi.org/10.1021/jp801738f}{{https://doi.org/10.1021/jp801738f}}}.
\bcomment{PMID: 18729344}
\end{barticle}
\endbibitem

\bibitem[\protect\citeauthoryear{Barca et~al.}{2018}]{barca:2018}
\begin{barticle}
\bauthor{\bsnm{Barca}, \binits{G.M.J.}},
\bauthor{\bsnm{Gilbert}, \binits{A.T.B.}},
\bauthor{\bsnm{Gill}, \binits{P.M.W.}}:
\batitle{Simple models for difficult electronic excitations}.
\bjtitle{J. Chem. Theo. Comp.}
\bvolume{14}(\bissue{3}),
\bfpage{1501}--\blpage{1509}
(\byear{2018})
\doiurl{10.1021/acs.jctc.7b00994}
{\href{https://arxiv.org/abs/https://doi.org/10.1021/acs.jctc.7b00994}{{https://doi.org/10.1021/acs.jctc.7b00994}}}.
\bcomment{PMID: 29444408}
\end{barticle}
\endbibitem

\bibitem[\protect\citeauthoryear{Brück et~al.}{2023}]{bruck:2023}
\begin{barticle}
\bauthor{\bsnm{Brück}, \binits{S.A.}},
\bauthor{\bsnm{Sahu}, \binits{N.}},
\bauthor{\bsnm{Gaul}, \binits{K.}},
\bauthor{\bsnm{Berger}, \binits{R.}}:
\batitle{Quasi-relativistic approach to analytical gradients of parity
  violating potentials}.
\bjtitle{The Journal of Chemical Physics}
\bvolume{158}(\bissue{19}),
\bfpage{194109}
(\byear{2023})
\doiurl{10.1063/5.0141271}
{\href{https://arxiv.org/abs/https://pubs.aip.org/aip/jcp/article-pdf/doi/10.1063/5.0141271/18050295/194109\_1\_5.0141271.pdf}{{https://pubs.aip.org/aip/jcp/article-pdf/doi/10.1063/5.0141271/18050295/194109\_1\_5.0141271.pdf}}}
\end{barticle}
\endbibitem

\bibitem[\protect\citeauthoryear{Colombo~Jofr{\'e}
  et~al.}{2022}]{colombojofre:2022}
\begin{barticle}
\bauthor{\bsnm{Colombo~Jofr{\'e}}, \binits{M.T.}},
\bauthor{\bsnm{Kozio{\l}}, \binits{K.}},
\bauthor{\bsnm{Aucar}, \binits{I.A.}},
\bauthor{\bsnm{Gaul}, \binits{K.}},
\bauthor{\bsnm{Berger}, \binits{R.}},
\bauthor{\bsnm{Aucar}, \binits{G.A.}}:
\batitle{Relativistic and qed corrections to one-bond indirect nuclear
  spin–spin couplings in {X$_2^{2+}$ and X$_3^{2+}$ ions (X = Zn, Cd, Hg)}}.
\bjtitle{J. Chem. Phys.}
\bvolume{157}(\bissue{6}),
\bfpage{064103}
(\byear{2022})
\doiurl{10.1063/5.0095586}
\end{barticle}
\endbibitem

\bibitem[\protect\citeauthoryear{Simpson et~al.}{}]{simpson:2025}
\begin{botherref}
\oauthor{\bsnm{Simpson}, \binits{R.}},
\oauthor{\bsnm{Z\"ulch}, \binits{C.}},
\oauthor{\bsnm{Ng}, \binits{K.B.}},
\oauthor{\bsnm{Belosevic}, \binits{I.}},
\oauthor{\bsnm{Charles}, \binits{C.}},
\oauthor{\bsnm{Justus}, \binits{P.}},
\oauthor{\bsnm{Berger}, \binits{R.}},
\oauthor{\bsnm{Malbrunot-Ettenauer}, \binits{S.}},
\oauthor{\bsnm{Kwiatkowski}, \binits{A.A.}},
\oauthor{\bsnm{Reiter}, \binits{M.P.}},
\oauthor{\bsnm{Ash}, \binits{J.}},
\oauthor{\bsnm{Babcock}, \binits{C.}},
\oauthor{\bsnm{Bergmann}, \binits{J.}},
\oauthor{\bsnm{Brisley}, \binits{E.}},
\oauthor{\bsnm{Cardona}, \binits{J.}},
\oauthor{\bsnm{Chambers}, \binits{C.}},
\oauthor{\bsnm{Czihaly}, \binits{A.}},
\oauthor{\bsnm{Gottberg}, \binits{A.}},
\oauthor{\bsnm{Kakkar}, \binits{S.}},
\oauthor{\bsnm{Lassen}, \binits{J.}},
\oauthor{\bsnm{Milán}, \binits{F.M.}},
\oauthor{\bsnm{Mollaebrahimi}, \binits{A.}},
\oauthor{\bsnm{Radchenko}, \binits{V.}},
\oauthor{\bsnm{Taylor}, \binits{E.}},
\oauthor{\bsnm{Teigelhöfer}, \binits{A.}},
\oauthor{\bsnm{Walls}, \binits{C.}},
\oauthor{\bsnm{Weaver}, \binits{A.}},
\oauthor{\bsnm{Weligampola}, \binits{P.}}:
Formation of gaseous, doubly charged cerium monofluoride CeF$^{2+}$ and its
  sensitivity to new physics.
unpublished
\end{botherref}
\endbibitem

\bibitem[\protect\citeauthoryear{Meyer}{1970}]{meyer:1970}
\begin{barticle}
\bauthor{\bsnm{Meyer}, \binits{R.}}:
\batitle{Trigonometric interpolation method for one-dimensional
  quantum-mechanical problems}.
\bjtitle{J. Chem. Phys.}
\bvolume{52},
\bfpage{2053}--\blpage{2059}
(\byear{1970})
\doiurl{10.1063/1.1673259}
\end{barticle}
\endbibitem

\bibitem[\protect\citeauthoryear{Udrescu et~al.}{2021}]{Udrescu2021}
\begin{barticle}
\bauthor{\bsnm{Udrescu}, \binits{S.M.}},
\bauthor{\bsnm{Brinson}, \binits{A.J.}},
\bauthor{\bsnm{Garcia~Ruiz}, \binits{R.F.}},
\bauthor{\bsnm{Gaul}, \binits{K.}},
\bauthor{\bsnm{Berger}, \binits{R.}},
\bauthor{\bsnm{Billowes}, \binits{J.}},
\bauthor{\bsnm{Binnersley}, \binits{C.L.}},
\bauthor{\bsnm{Bissell}, \binits{M.L.}},
\bauthor{\bsnm{Breier}, \binits{A.A.}},
\bauthor{\bsnm{Chrysalidis}, \binits{K.}},
\bauthor{\bsnm{Cocolios}, \binits{T.E.}},
\bauthor{\bsnm{Cooper}, \binits{B.S.}},
\bauthor{\bsnm{Flanagan}, \binits{K.T.}},
\bauthor{\bsnm{Giesen}, \binits{T.F.}},
\bauthor{\bsnm{Groote}, \binits{R.P.}},
\bauthor{\bsnm{Franchoo}, \binits{S.}},
\bauthor{\bsnm{Gustafsson}, \binits{F.P.}},
\bauthor{\bsnm{Isaev}, \binits{T.A.}},
\bauthor{\bsnm{Koszorus}, \binits{A.}},
\bauthor{\bsnm{Neyens}, \binits{G.}},
\bauthor{\bsnm{Perrett}, \binits{H.A.}},
\bauthor{\bsnm{Ricketts}, \binits{C.M.}},
\bauthor{\bsnm{Rothe}, \binits{S.}},
\bauthor{\bsnm{Vernon}, \binits{A.R.}},
\bauthor{\bsnm{Wendt}, \binits{K.D.A.}},
\bauthor{\bsnm{Wienholtz}, \binits{F.}},
\bauthor{\bsnm{Wilkins}, \binits{S.G.}},
\bauthor{\bsnm{Yang}, \binits{X.F.}}:
\batitle{{Isotope Shifts of Radium Monofluoride Molecules}}.
\bjtitle{Physical Review Letters}
\bvolume{127}(\bissue{3}),
\bfpage{033001}
(\byear{2021})
\doiurl{10.1103/PhysRevLett.127.033001}
\end{barticle}
\endbibitem

\bibitem[\protect\citeauthoryear{Sushkov et~al.}{1984}]{Sushkov1984}
\begin{botherref}
\oauthor{\bsnm{Sushkov}, \binits{O.P.}},
\oauthor{\bsnm{Flambaum}, \binits{V.V.}},
\oauthor{\bsnm{Khriplovich}, \binits{I.B.}}:
{Possibility of investigating P-and T-odd nuclear forces in atomic and
  molecular experiments}.
Zh. Eksp. Teor. Fiz
\textbf{60}(873)
(1984)
\doiurl{[Sov. JETP 60, 873 (1984)]}
\end{botherref}
\endbibitem

\bibitem[\protect\citeauthoryear{Z{\"{u}}lch et~al.}{2022}]{Zulch2022}
\begin{botherref}
\oauthor{\bsnm{Z{\"{u}}lch}, \binits{C.}},
\oauthor{\bsnm{Gaul}, \binits{K.}},
\oauthor{\bsnm{Giesen}, \binits{S.M.}},
\oauthor{\bsnm{Ruiz}, \binits{R.F.G.}},
\oauthor{\bsnm{Berger}, \binits{R.}}:
{Cool molecular highly charged ions for precision tests of fundamental
  physics}.
arXiv:2203.10333
(2022)
\end{botherref}
\endbibitem

\bibitem[\protect\citeauthoryear{Dobaczewski
  et~al.}{2021}]{Dobaczewski2021HFODD}
\begin{barticle}
\bauthor{\bsnm{Dobaczewski}, \binits{J.}},
\bauthor{\bsnm{B{\c{a}}czyk}, \binits{P.}},
\bauthor{\bsnm{Becker}, \binits{P.}},
\bauthor{\bsnm{Bender}, \binits{M.}},
\bauthor{\bsnm{Bennaceur}, \binits{K.}},
\bauthor{\bsnm{Bonnard}, \binits{J.}},
\bauthor{\bsnm{Gao}, \binits{Y.}},
\bauthor{\bsnm{Idini}, \binits{A.}},
\bauthor{\bsnm{Konieczka}, \binits{M.}},
\bauthor{\bsnm{Kortelainen}, \binits{M.}},
\bauthor{\bsnm{Pr{\'{o}}chniak}, \binits{L.}},
\bauthor{\bsnm{Romero}, \binits{A.M.}},
\bauthor{\bsnm{Satu{\l}a}, \binits{W.}},
\bauthor{\bsnm{Shi}, \binits{Y.}},
\bauthor{\bsnm{Werner}, \binits{T.R.}},
\bauthor{\bsnm{Yu}, \binits{L.F.}}:
\batitle{{Solution of universal nonrelativistic nuclear DFT equations in the
  Cartesian deformed harmonic-oscillator basis. (IX) HFODD (v3.06h): a new
  version of the program}}.
\bjtitle{Journal of Physics G: Nuclear and Particle Physics}
\bvolume{48}(\bissue{10}),
\bfpage{102001}
(\byear{2021})
\doiurl{10.1088/1361-6471/ac0a82}
\end{barticle}
\endbibitem

\bibitem[\protect\citeauthoryear{Herczeg}{1988}]{Herczeg1988}
\begin{barticle}
\bauthor{\bsnm{Herczeg}, \binits{P.}}:
\batitle{{T-violation in nuclear interactions — An overview}}.
\bjtitle{Hyperfine Interact.}
\bvolume{43}(\bissue{1-4}),
\bfpage{75}--\blpage{93}
(\byear{1988})
\doiurl{10.1007/BF02398288}
\end{barticle}
\endbibitem

\bibitem[\protect\citeauthoryear{Ring and Schuck}{1980}]{Ring1980}
\begin{bbook}
\bauthor{\bsnm{Ring}, \binits{P.}},
\bauthor{\bsnm{Schuck}, \binits{P.}}:
\bbtitle{The Nuclear Many-Body Problem},
\bedition{1}st edn.,
p. \bfpage{718}.
\bpublisher{Springer},
\blocation{Heidelberg}
(\byear{1980})
\end{bbook}
\endbibitem

\end{thebibliography}


\section*{Acknowledgments}
We would like to thank Andrei Zaitsevskii and
Alexander Oleynichenko for help with electronic-structure calculations. We would like to thank Victor V. Flambaum (University of New South Wales), Elise Wursten (Imperial College London), and Christopher J. Ho (Imperial College London) for useful discussions. Financial support from FWO, as well as from the Excellence of Science (EOS) programme (No. 40007501) and the KU Leuven project C14/22/104, is acknowledged. The STFC consolidated grants ST/V001116/1 and ST/P004423/1 and the FNPMLS ERC grant agreement 648381 are acknowledged. Swedish Research Council (Grants No. 2020-03505) and funding from the European Union’s Horizon 2020 research and innovation program under the Marie Sk\l{}odowska-Curie Grant Agreement No. 861198 are acknowledged. This work was partially supported by the STFC Grant
Nos.~ST/P003885/1 and~ST/V001035/1, and
by a Leverhulme Trust Research Project Grant. K.G. thanks the Fonds der Chemischen Industrie (FCI) for generous funding through a Liebig fellowship. 
This project was partly undertaken on the Viking Cluster,
which is a high performance compute facility provided by the
University of York. We are grateful for computational support from the University of York High Performance Computing
service, Viking and the Research Computing team.
C.Z. and R.B. gratefully acknowledge financial support by the Deutsche Forschungsgemeinschaft (DFG, German Research Foundation) with the project number 445296313. 
C.Z., K.G., and R.B. acknowledge computing time provided at the NHR Center NHR@SW at Goethe-University Frankfurt. This was funded by the Federal Ministry of Education and Research and the state governments participating on the basis of the resolutions of the GWK for national high performance computing at universities (\url{http://www.nhr-verein.de/unsere-partner}). 
A.B. and A.K. thank the Center for Information Technology at the University of Groningen for their support and for providing access to the Peregrine and Hábrók high-performance computing clusters. A.B. is grateful for the Dutch Research Council (NWO) project number Vi.Vidi.192.088.\\ 

\section*{Author contribution}
M.A.-K., M.Au, A.K., C.Z., K.G., H.W., and G.N. led the manuscript preparation. M.A.-K., M.Au, A.K., C.Z., K.G., L.L., J.R.R., \'A.Kosz., S.B., J.B., R.B., A.A.B., T.E.C., R.P.d.G., A.D., K.T.F., S.F., J.D.J., R.F.G.R., D.H., S.K., Y.C.L., K.M.L., A.McG., G.N., M.N., L.N., F.P., S.R., B.v.d.B., J.W., S.G.W., and X.F.Y. performed the experiment. A.K. and A.Bor. performed the relativistic coupled cluster calculations and reviewed the results. C.Z., K.G., R.B. performed the 2c-ZORA-cGHF calculations and the global analysis and reviewed the results. L.V.S. performed the electronic-structure calculations of molecular terms and Franck-Condon factors and reviewed the results., and N.S.M. constructed the pseudopotentials for these electronic-structure calculations and reviewed the results. H.W. and J.Dob. performed the nuclear DFT calculations, and H.W., J.Dob, and W.R. reviewed the results. M.A.-K. and A.A.B. performed the data analysis. All coauthors participated in editing and revising the manuscript.

\section*{Competing interests statement}
The authors declare no competing interest.

\end{document}


\title[Supplementary Information: Laser spectroscopy and CP-violation sensitivity of actinium monofluoride]{Supplementary Information: Laser spectroscopy and CP-violation sensitivity of actinium monofluoride}

\author*[1,2,3]{\fnm{M.} \sur{Athanasakis-Kaklamanakis\orcidlink{0000-0003-0336-5980}}}\email{m.athkak@cern.ch}
\author*[4,5]{\fnm{M.} \sur{Au}\orcidlink{0000-0002-8358-7235}}\email{mia.au@cern.ch}
\author[6]{\fnm{A.} \sur{Kyuberis}\orcidlink{0000-0001-7544-3576}}
\author[7]{\fnm{C.} \sur{Z\"ulch}\orcidlink{0009-0007-2563-1342}}
\author[7,8,9,10]{\fnm{K.} \sur{Gaul}\orcidlink{0000-0002-6990-6949}}
\author[11]{\fnm{H.} \sur{Wibowo}\orcidlink{0000-0003-4093-0600}}
\author[12]{\fnm{L.} \sur{Skripnikov}\orcidlink{0000-0002-2062-684X}}

\author[2,1]{\fnm{L.} \sur{Lalanne}\orcidlink{0000-0003-1207-9038}}
\author[13,4]{\fnm{J.~R.} \sur{Reilly}}
\author[1,2]{\fnm{\'A.} \sur{Koszor\'us}\orcidlink{0000-0001-7959-8786}}
\author[2]{\fnm{S.} \sur{Bara}\orcidlink{0000-0001-7129-3486}}
\author[9,5]{\fnm{J.} \sur{Ballof}\orcidlink{0000-0003-0874-500X}}
\author[7,1]{\fnm{R.} \sur{Berger}}
\author[4]{\fnm{C.} \sur{Bernerd}\orcidlink{0000-0002-2183-9695}}
\author[6]{\fnm{A.} \sur{Borschevsky}\orcidlink{0000-0002-6558-1921}}
\author[14,15]{\fnm{A.~A.} \sur{Breier}\orcidlink{0000-0003-1086-9095}}
\author[4]{\fnm{K.} \sur{Chrysalidis}\orcidlink{0000-0003-2908-8424}}
\author[2]{\fnm{T.~E.} \sur{Cocolios}\orcidlink{0000-0002-0456-7878}}
\author[2]{\fnm{R.~P.} \sur{de Groote}\orcidlink{0000-0003-4942-1220}}
\author[2]{\fnm{A.} \sur{Dorne}}
\author[11]{\fnm{J.} \sur{Dobaczewski}\orcidlink{0000-0002-4158-3770}}
\author[2]{\fnm{C.~M.} \sur{Fajardo Zambrano}\orcidlink{0000-0002-6088-6726
}}
\author[13]{\fnm{K.~T.} \sur{Flanagan}\orcidlink{0000-0003-0847-2662}}
\author[16,17]{\fnm{S.} \sur{Franchoo}}
\author[2]{\fnm{J.~D.} \sur{Johnson}\orcidlink{0000-0003-4397-5732}}
\author[18,19]{\fnm{R.~F.} \sur{Garcia Ruiz}\orcidlink{0000-0002-2926-5569}}
\author[20]{\fnm{D.} \sur{Hanstorp}}
\author[21]{\fnm{S.} \sur{Kujanp\"a\"a}\orcidlink{0000-0002-5709-3442}}
\author[22]{\fnm{Y.~C.} \sur{Liu}}
\author[13]{\fnm{K.~M.} \sur{Lynch}\orcidlink{0000-0001-8591-2700}}
\author[13]{\fnm{A.} \sur{McGlone}\orcidlink{0000-0003-4424-865X}}
\author[12]{\fnm{N.~S.} \sur{Mosyagin}\orcidlink{0000-0002-9158-494X}}
\author*[2]{\fnm{G.} \sur{Neyens}\orcidlink{0000-0001-8613-1455}}\email{gerda.neyens@kuleuven.be}
\author[20]{\fnm{M.} \sur{Nichols}\orcidlink{0000-0003-3693-7295}}
\author[1]{\fnm{L.} \sur{Nies}\orcidlink{0000-0003-2448-3775}}
\author[18]{\fnm{F.} \sur{Pastrana}\orcidlink{0009-0008-7469-7513}}
\author[4]{\fnm{S.} \sur{Rothe}}
\author[23,24]{\fnm{W.} \sur{Ryssens}\orcidlink{0000-0002-4775-4403}}
\author[2]{\fnm{B.} \sur{van den Borne}\orcidlink{0000-0003-3348-7276}}
\author[4,13]{\fnm{J.} \sur{Wessolek}}
\author[18,19]{\fnm{S.~G.} \sur{Wilkins}\orcidlink{0000-0001-8897-7227}}
\author[22]{\fnm{X.~F.} \sur{Yang}\orcidlink{0000-0002-1633-4000}}

\affil[1]{\orgdiv{Experimental Physics Department}, \orgname{CERN}, \orgaddress{{Geneva}, \postcode{CH-1211}, \country{Switzerland}}}

\affil[2]{\orgdiv{Instituut voor Kern- en Stralingsfysica}, \orgname{KU Leuven}, \orgaddress{{Leuven}, \postcode{B-3001}, \country{Belgium}}}

\affil[3]{\orgdiv{Centre for Cold Matter}, \orgname{Imperial College London}, \orgaddress{{London}, \postcode{SW7 2AZ}, \country{United Kingdom}}}

\affil[4]{\orgdiv{Systems Department}, \orgname{CERN}, \orgaddress{{Geneva}, \postcode{CH-1211}, \country{Switzerland}}}

\affil[5]{\orgdiv{Department of Chemistry}, \orgname{Johannes Gutenberg-Universit\"{a}t Mainz}, \orgaddress{{Mainz}, \postcode{55099}, \country{Germany}}}

\affil[6]{\orgdiv{Van Swinderen Institute of Particle Physics and Gravity}, \orgname{University of Groningen}, \orgaddress{{Groningen}, \postcode{9712 CP}, \country{Netherlands}}}

\affil[7]{\orgdiv{Fachbereich Chemie}, \orgname{Philipps-Universit\"{a}t Marburg}, \postcode{35032}, \orgaddress{{Marburg},  \country{Germany}}}

\affil[8]{\orgname{Helmholtz Institute Mainz}, \postcode{55099}, \orgaddress{Mainz}, \country{Germany}}

\affil[9]{\orgdiv{GSI Helmholtzzentrum für Schwerionenforschung GmbH}, \orgname{GSI}, \postcode{64291}, \orgaddress{{Darmstadt},  \country{Germany}}}

\affil[10]{\orgdiv{Institut f\"ur Physik}, \orgname{Johannes Gutenberg-Universit\"{a}t Mainz}, \orgaddress{{Mainz}, \postcode{55099}, \country{Germany}}}

\affil[11]{\orgdiv{School of Physics, Engineering and Technology}, \orgname{University of York}, \orgaddress{{York}, \postcode{YO10 5DD}, \country{United Kingdom}}}

\affil[12]{\orgdiv{Affiliation covered by a cooperation agreement with CERN at the time of the experiment}}

\affil[13]{\orgdiv{Department of Physics and Astronomy}, \orgname{The University of Manchester}, \orgaddress{{Manchester}, \postcode{M13 9PL}, \country{United Kingdom}}}

\affil[14]{\orgdiv{Institut f\"ur Physik und Astronomie},  \orgname{Technische Universit\"at Berlin}, \orgaddress{\postcode{10623}, \country{Germany}}}

\affil[15]{\orgdiv{Laboratory for Astrophysics, Institute of Physics}, \orgname{University of Kassel}, \orgaddress{{Kassel}, \postcode{34132}, \country{Germany}}}

\affil[16]{\orgname{Laboratoire Ir\`{e}ne Joliot-Curie}, \orgaddress{{Orsay}, \postcode{F-91405}, \country{France}}}

\affil[17]{\orgname{University Paris-Saclay}, \orgaddress{{Orsay}, \postcode{F-91405}, \country{France}}}

\affil[18]{\orgdiv{Department of Physics}, \orgname{Massachusetts Institute of Technology}, \orgaddress{{Cambridge}, \postcode{MA 02139}, \country{USA}}}

\affil[19]{\orgdiv{Laboratory for Nuclear Science}, \orgname{Massachusetts Institute of Technology}, \orgaddress{{Cambridge}, \postcode{MA 02139}, \country{USA}}}

\affil[20]{\orgdiv{Department of Physics}, \orgname{University of Gothenburg}, \orgaddress{{Gothenburg}, \postcode{SE-41296}, \country{Sweden}}}

\affil[21]{\orgdiv{Department of Physics}, \orgname{University of Jyväskylä}, \orgaddress{{Jyväskylä}, \postcode{40351}, \country{Finland}}}

\affil[22]{\orgdiv{School of Physics and State Key Laboratory of Nuclear Physics and Technology}, \orgname{Peking University}, \orgaddress{{Beijing}, \postcode{100971}, \country{China}}}

\affil[23]{\orgdiv{Institut d’Astronomie et d’Astrophysique}, \orgname{Universit\'{e} libre de Bruxelles}, \orgaddress{Brussels} \postcode{1050}, \country{Belgium}}

\affil[24]{\orgdiv{Brussels Laboratory of the Universe - BLU-ULB}, \orgname{Universit\'{e} libre de Bruxelles}, \orgaddress{Brussels} \postcode{1050}, \country{Belgium}}

\maketitle

\section{Suppl. Note: Data analysis and spectral assignment}
\begin{figure}[h]
    \centering
    \includegraphics[width=0.49\textwidth]{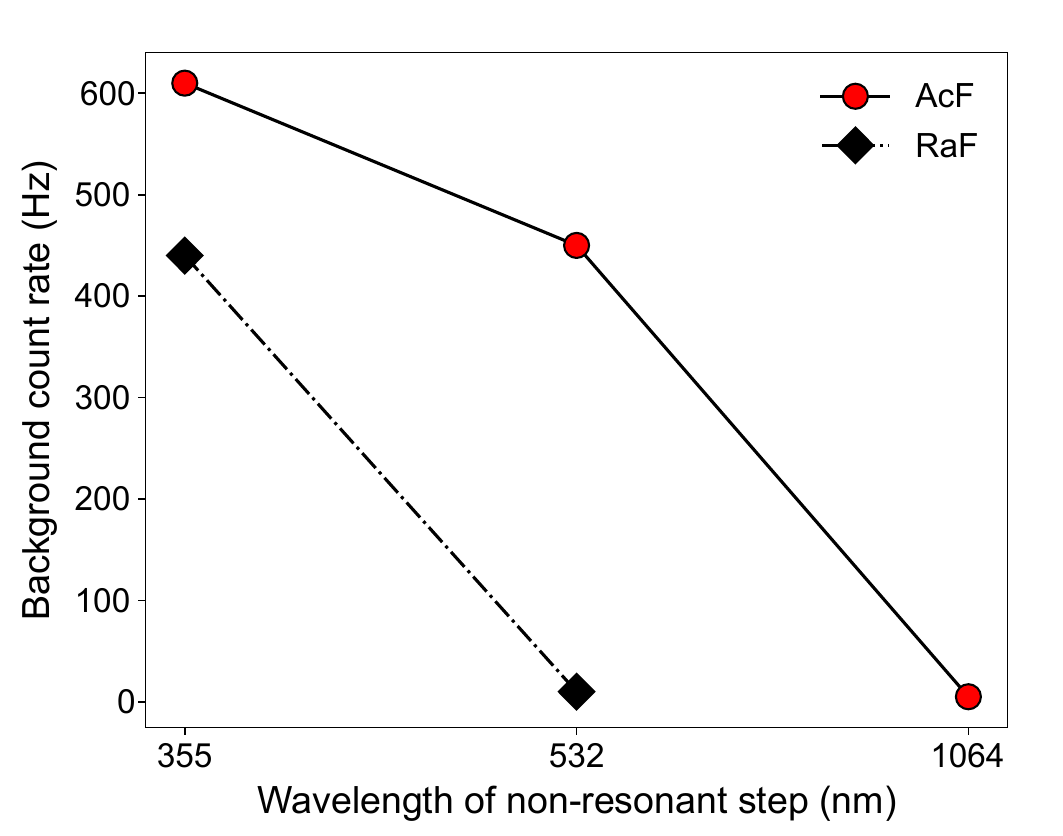}
    \caption{Ion background induced by the non-resonant ionization step without a resonant transition in AcF and RaF, as a function of the wavelength of the non-resonant step.}
    \label{fig:figBackground}
\end{figure}

The spectroscopy of the $(8)^1\Pi \leftarrow X$~$^1\Sigma^+$ transition was complicated by the high background ion rate induced by the 355-nm non-resonant ionization step. When illuminating the $^{227}$AcF beam with 532-nm light, the reduction in the non-resonantly ionized background was significantly smaller than previously seen in $^{226}$RaF (see Suppl. Fig.~\ref{fig:figBackground}). Only when 1064-nm light was used for non-resonant ionization was the background level significantly reduced. This implies the possible existence of a metastable state in AcF that lies within a range of 9,398 to 28,169\,cm$^{-1}$ below the ionization potential, and which is likely populated in the charge-exchange process.

The ions incident on the MagneToF single-ion counter were digitized with a cronologic TimeTagger4 time-to-digital converter (TDC) with a resolution of 400\,ps. Data preparation was performed with Python. The raw data obtained from the asynchronously recording devices (wavemeter, digital multimeter, TDC) were firstly sorted into time-correlated data frames. The wavemeter reading for the excitation laser was Doppler-shifted to the molecular rest frame using the acceleration voltage recorded by the digital multimeter. Taking advantage of the time resolution of the TDC, a time-of-flight filter was applied to the data frame of each scan to discard all data points that did not correspond to a molecule interacting with the lasers in the interaction region.

\begin{figure}[h]
    \centering
    \includegraphics[width=0.69\textwidth]{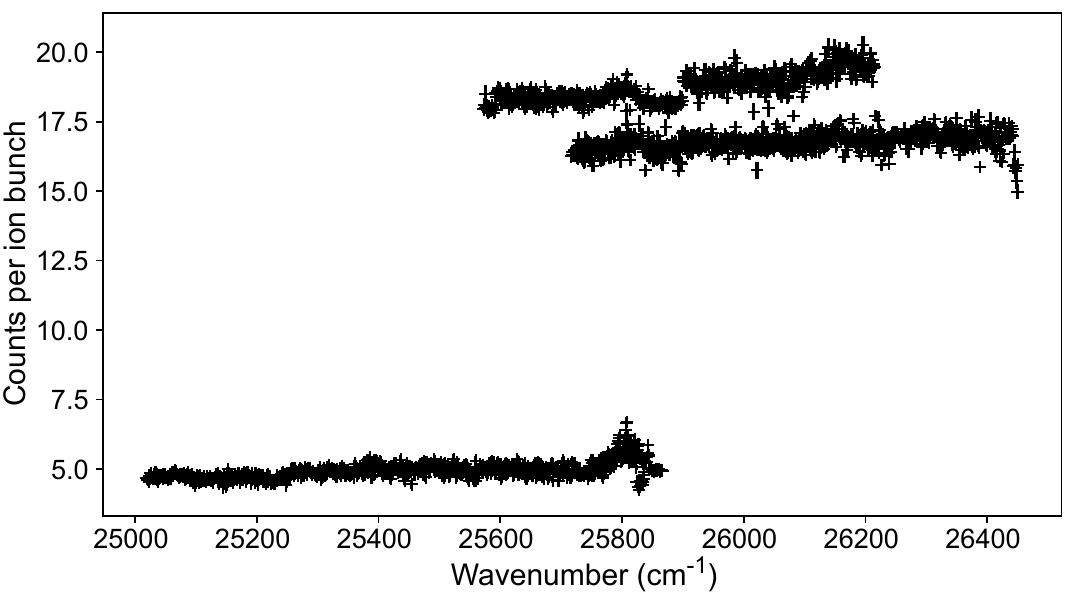}
    \caption{Scans using the two-step laser scheme (Fig.~2a of main text) that comprise the spectrum in Fig.~2c of the main text, without applying vertical offsets.}
    \label{fig:figLongScans}
\end{figure}

\begin{figure}[h]
    \centering
    \includegraphics[width=0.59\textwidth]{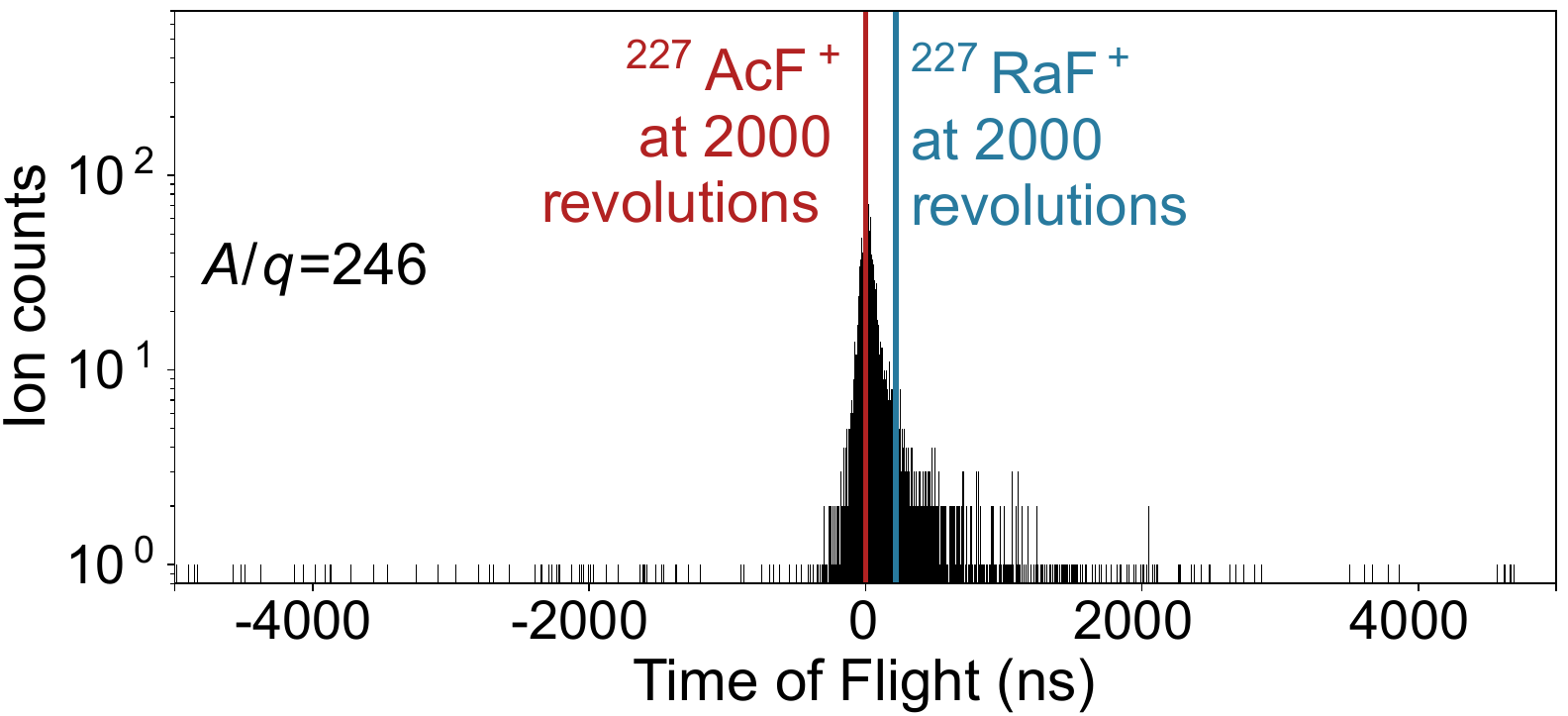}
    \caption{Broader view of the spectrum obtained with multi-reflection time-of-flight mass spectrometry (Fig.~1f of the main text) to confirm the chemical and isotopic purity of the ion beam.}
    \label{fig:figLongMRTOFscan}
\end{figure}
The ion counts and the excitation wavenumber were then binned into a histogram to produce the laser spectra. For each wavenumber bin, the corresponding ion counts were summed and divided by the number of ion bunches ejected from the Paul trap that were included in the bin. The $y$-error corresponding to each bin was determined by taking the square root of the total ion counts in the bin divided by the number of bunches.

The binned spectrum of the $(8)^1\Pi \leftarrow X$~$^1\Sigma^+$ transition (Fig.~2b of main text) was exported from the Python code and imported into the {\sc pgopher} software~\cite{Western2017} to be analyzed with the contour-fitting routine. The temperature of the beam was set to $T_{\rm{rot}}=1,200(80)$\,K, determined using $^{226}$RaF as explained in \textit{Methods}. Considering that no prior experimental information is known about these states in AcF, the molecular constants of both the upper and lower states could not be fitted simultaneously without extensive assumptions. Therefore, the molecular constants of the vibrational states of $X$~$^1\Sigma^+$ were derived from the Fock-space relativistic coupled cluster (FS-RCC) calculations of the harmonic vibrational frequency and the equilibrium internuclear distance from Ref.~\cite{Skripnikov2023AcF} and kept fixed in the fit. The FS-RCC calculations were also used to extract predicted constants for the vibrational states of $(8)^1\Pi$, which were used as starting points for the contour fitting. 

To assign a vibrational transition to each peak, the temperature of 1,200(80)\,K was taken as an approximate value of the vibrational temperature as well. Using
\begin{equation}
    P_{\rm{vib}}(v) = \big( 1 - e^{-\hbar \tilde{\omega} /kT} \big) e^{-\hbar \tilde{\omega} v/kT}
\end{equation}
where $\tilde{\omega}$ is the harmonic vibrational wavenumber of the AcF$^+$ ground state determined with relativistic coupled cluster (RCC) theory~\cite{Skripnikov2023AcF} and $v$ is the vibrational quantum number, the fraction of the AcF$^+$ beam in each vibrational state prior to neutralization was estimated. At 1,200\,K, only $\sim$48$\%$ of the population was expected in the vibrational ground state $v=0$. To estimate the vibrational distribution in AcF after neutralization from AcF$^+$, the Franck-Condon factors (FCFs) between the ground states of AcF$^+$ and AcF were calculated with RCC (see Supplementary Table~\ref{tab:FCF_neutral_ion}). The population fraction in each vibrational state of $X$~$^1\Sigma^+$ in AcF was estimated by considering all diagonal and off-diagonal transitions and the initial populations of the vibrational states of AcF$^+$.

Finally, to estimate the expected relative intensities of the different vibrational bands observed in the spectrum, the FCFs for the $(8)^1\Pi \leftarrow X$~$^1\Sigma^+$ transition were calculated as well (see Supplementary Table~\ref{tab:FCF_gs_excited}). As the observed bands are at least partially overlapping, a preliminary fit was performed to disentangle the band intensities. The observed peaks were then assigned according to the estimated intensities and fitted again. The final results were obtained from a fit with reduced chi-squared of $\chi_{\rm{r}}^2 = 3.4$ and are listed in Supplementary Table~\ref{tab:fitted_constants}.

\begin{table}[h]
\caption{Franck-Condon factors between the ground state of the neutral AcF molecule and the ground state of the AcF$^+$ ion. Calculated with the CCSD(T) method.} \label{tab:FCF_neutral_ion}
\begin{tabular}{ccccc}
\hline\hline
AcF / AcF$^+$ & $v_{\rm{AcF^+}} = 0$     & 1     & 2     & 3     \\
\hline
$v_{\rm{AcF}} = 0$          & 0.637 & 0.292 & 0.062 & 0.008 \\
1          & 0.280 & 0.188 & 0.358 & 0.142 \\
2          & 0.068 & 0.326 & 0.020 & 0.305 \\
3          & 0.012 & 0.146 & 0.259 & 0.007\\
\hline\hline
\end{tabular}
\end{table}

\begin{table}[h]
\caption{Franck-Condon factors between the $X$~$^1\Sigma^+$ ground state of the neutral AcF molecule and the $(8)^1\Pi$ excited electronic state. Calculated with the CCSD(T) method.} \label{tab:FCF_gs_excited}
\begin{tabular}{cccccc}
\hline\hline
$X$~$^1\Sigma^+$ / $(8)^1\Pi$ & $v_{\Pi} = 0$     & 1     & 2     & 3    & 4     \\
\hline
$v_{\Sigma} = 0$    & 0.95930 & 0.03804 & 0.00245 & 0.00019 & 0.00002\\
1                   & 0.04066 & 0.88193 & 0.06969 & 0.00695 & 0.00068\\
2                   & 0.00004 & 0.07988 & 0.80889 & 0.09409 & 0.01527\\
3                   & 0.00000 & 0.00009 & 0.11848 & 0.75370 & 0.09779\\
4                   & 0.00000 & 0.00005 & 0.00002 & 0.14104 & 0.75263\\
\hline\hline
\end{tabular}
\end{table}

\begin{table}[h]
\caption{Molecular constants used in the analysis of the $(8)^1\Pi \leftarrow X$~$^1\Sigma^+$ spectrum. The constants for the $X$~$^1\Sigma^+$ state (denoted with the subscript $l$) are determined from the ab initio CCSD(T) calculations and are kept fixed in the contour fitting. The constants for the $(8)^1 \Pi$ state (denoted with the subscript $u$) are fitted and the error in parentheses represents 1$\sigma$ statistical uncertainty scaled with the reduced-$\chi$ of the fit. The error in square brackets is a systematic uncertainty due to the linewidth set in the fitting process.} \label{tab:fitted_constants}
\begin{tabular}{ccccccc}
\hline\hline
$v$ & $B_l$  & $E_l$ & $D_l$   & $B_u$   & $E_u$    & $D_u$     \\\hline
0 & 0.21572 & 0    & $6.4 \times 10^{-7}$ & 0.21100(17) & 25,834.10(31)[60] & $6(2) \times 10^{-7}$ \\
1 & 0.21520 & 541  & $6.4 \times 10^{-7}$ & 0.21085(15) & 26,357.72(31)[60] & $6(2) \times 10^{-7}$ \\
2 & 0.21469 & 1082 & $6.4 \times 10^{-7}$ & 0.21100(11) & 26,891.45(20)[60] & $6(2) \times 10^{-7}$ \\
3 & 0.21417 & 1623 & $6.4 \times 10^{-7}$ & 0.21128(12) & 27,420.08(21)[60] & $6(2) \times 10^{-7}$ \\
4 & 0.21365 & 2164 & $6.4 \times 10^{-7}$ & 0.21235(20) & 27,942.55(33)[60] & $6(2) \times 10^{-7}$ \\
\hline\hline
\end{tabular}
\end{table}

The spectrum in Fig.~2c of the main text that shows the measured ion count rate across the full wavenumber range scanned during the experiment was constructed by combining several smaller spectra, shown in Suppl. Fig.~\ref{fig:figLongScans}. The individual spectra were obtained at different times during the experimental campaign, with different molecular beam intensities and output power of the 355-nm laser. As a result, the background, which depends on the molecular beam intensity since it is dominated by non-resonant ionization due to the 355-nm laser, differs between spectra. To improve visibility of local and global trends across the full scan range, an offset was applied to each smaller spectrum to arrive at the version shown in Fig.~2c of the main text. The spectrum of the $(8)$~$^1\Pi \leftarrow X$~$^1\Sigma^+$ transition (Fig.~2b in the main text) appears in all spectra of Suppl. Fig~\ref{fig:figLongScans} that cover this wavenumber range, evidenced by the prevalence of a structure above background.

\section{Suppl. Note: Relativistic coupled cluster calculations}

We performed an extensive investigation of the effects of different computational parameters on the calculated $W_\mathcal{S}$, focusing on the employed basis sets, the treatment of electron correlation, and relativity. All the calculations were performed at the calculated equilibrium bond length of $R_e=3.974\,a_0$. This equilibrium geometry was obtained via structure optimization carried out at the DC-CCSD(T) level, correlating 50 electrons with active space from $-20 \, E_\mathrm{h}$ to $50 \, E_\mathrm{h}$ (corresponding to $4f$, $5s$, $5p$, $5d$, $6s$, $6p$, $7s$ for Ac and $2s$, $2p$ for F). The $R_e$ optimization was performed using the s-aug-dyall.cv$n$z ($n$=2,3,4) basis sets and the results were extrapolated to the complete basis set limit (CBSL). For basis set extrapolation, we used the scheme of Helgaker~\textit{et al.}~\cite{cbs} (H-CBSL).

\subsection{Basis sets}
From previous works~\cite{zulch2023,gaul:2024b}, it is known that the calculated Schiff moment enhancement factors are extremely sensitive to the quality of the wave function close to the nucleus. This region is poorly described in the original Dyall basis sets~\cite{dyall2,dyall3}. The exponential factors $\zeta_i$ of the Gaussians corresponding to the steepest $s$ and $p$ functions are related by relatively large ratios $\zeta_i / \zeta_j $ of order 4, but the largest exponential factors are not steep enough to penetrate the nucleus. It is thus usually necessary to include additional basis functions with exponents that are in the range of the exponent of the Gaussian nuclear charge density distribution. The additional exponents were determined using a modified version of the procedure described in Ref.~\cite{gaul:2024b}.

To improve the description of the nuclear region, we modified the original Dyall.cv3z basis set of Ac in two steps. Firstly, we densified the basis near the nucleus by adding one additional $s$ and $p$ exponents interpolated between the two steepest existing functions, thereby reducing the ratio between their exponents and improving the numerical stability of the wavefunction in the steep region. Secondly, we added three more $s$ and two $p$ exponents obtained by extending the pattern of the steepest exponents in the original basis, thus providing a more flexible description of the rapidly varying electronic density near the nucleus. In total, this resulted in a 4$s$3$p$ extension to the original $s$/$p$ space. This approach ensures a more precise evaluation of the Schiff moment sensitivity by capturing the fine details of the wavefunction in the nuclear region. The calculated values of $W_\mathcal{S}$ are presented in Supplementary Table~\ref{Ws_comp_exp}, while Supplementary Table~\ref{tab:exponents:2} lists the manually added exponents. These calculations were carried out correlating 28 electrons with a virtual-space cut-off at 50\,$E_\mathrm{h}$.

 \begin{table}[h]
    \setlength{\tabcolsep}{5pt}
        \centering
        \caption{Influence of the basis set quality on the sensitivity to the nuclear Schiff moment, $W_\mathcal{S}$ ($e/4\pi\epsilon_0 a_0^4$), in AcF, calculated at the levels of DHF, CCSD, and CCSD(T) with 2c-ZORA-cGHF as comparison. The basis set labeled '4s5p11d' refers to the basis set outlined in Table~\ref{tab:exponents_spd}.}
        \label{Ws_comp_exp}
        \begin{tabular}{lccllcc}
        \hline
        Basis set & Steeper & Densified & $W_\mathcal{S}$ [2c-ZORA-cGHF] &$W_\mathcal{S}$ [DHF] & $W_\mathcal{S}$ [CCSD] & $W_\mathcal{S}$ [CCSD(T)] \\     
        \hline
            cv2z&&&--8554&--5649&--4385&--4362\\
            cv3z&&&--12246&--9610&--7515&--7568\\
            cv4z&&&--12705&--9828&--7696&--7721\\
            cv3z&3s&yes&--10272&--10022&--7831&--7892\\
            cv3z&4s&yes&--10268&--10072&--7881&--7942\\
            cv3z&3s3p&yes&--10834&--10294&--8039&--8104\\
            cv3z&3s4p&yes&--10836&--10291&--8035&--8101\\
            cv3z&4s3p&yes&--10832&--10356&--8101&--8167\\  
            s-aug-cv3z&4s5p11d &-&--10057&--10061\\
            \textbf{s-aug-cv3z}&\textbf{4s3p}&\textbf{yes}&\textbf{--10852}&\textbf{--10361}&\textbf{--8141}&\textbf{--8206}\\
        \hline
        \end{tabular}
    \end{table}

\begin{table}
\setlength{\tabcolsep}{20pt}
\caption{Initial segment of the modified Dyall's core-valence basis set of triple-$\zeta$ quality for actinium in coupled cluster calculations. Functions highlighted in \textbf{bold} correspond to additional exponents.}
\label{tab:exponents:2}
\centering
\renewcommand\cellalign{lc}
\begin{tabular}{c c}
\toprule
    & {Ac} \\
\midrule
\multirow{6}{*}{$s$}
& \textbf{4.22393038 × 10\textsuperscript{8}} \\
& \textbf{2.17931394 × 10\textsuperscript{8}} \\
& \textbf{1.12440519 × 10\textsuperscript{8}} \\
& 5.80130752 × 10\textsuperscript{7} \\
& \textbf{2.99315311 × 10\textsuperscript{7}} \\
& 1.54430109 × 10\textsuperscript{7} \\
\midrule
\multirow{5}{*}{$p$}
& \textbf{1.79205583 × 10\textsuperscript{8}} \\
& \textbf{9.57182568 × 10\textsuperscript{7}} \\
& 5.11255540 × 10\textsuperscript{7} \\
& \textbf{2.73074579 × 10\textsuperscript{7}} \\
& 1.45856074 × 10\textsuperscript{7} \\
\bottomrule
\end{tabular}
\end{table}

The convergence of the calculated $W_\mathcal{S}$ values with respect to the other basis set parameters, such as diffuse and high exponent functions was carefully analyzed. As seen in Supplementary Table~\ref{Ws_comp_exp}, increasing the steepness and density of the basis functions has an effect of about 4~$\%$ on the calculated values. Another, even more important parameter, is inclusion of diffuse functions (designated s-aug-cv3z and d-aug-cv3z), where the first layer of augmentation increases the calculated $W_\mathcal{S}$ by about 8$\%$; however, the effect of adding a second augmentation layer is very small, and we can thus consider the results converged with respect to diffuse functions at the s-aug-cv3z level. 
Based on the observations above, we select s-aug-cv3z+4s3p as the final basis set, as it provides the best compromise between computational cost and accuracy; these results are presented in bold font in Supplementary Table~\ref{Ws_comp_exp}. 

\subsection{Relativistic effects}
To study the influence of relativistic effects, we performed calculations at the full four-component Dirac–Coulomb (DC) and Dirac–Coulomb–Gaunt (DCG) levels, adding the Gaunt interaction to the Hamiltonian~\cite{TroBas2020}; furthermore, we also tested the performance of the exact two-component (X2C) Hamiltonian~\cite{KneRep22,Knecht2012}. The current implementation in {\sc DIRAC} is limited to including the Gaunt interaction in the construction of the Fock matrix, and works for Hartree–Fock and DFT calculations. The Hartree-Fock method was used in this investigation. The results are summarized in Supplementary Table~\ref{Ws_comp_rel}. We find that relativistic effects are covered at a level of more than 99$\%$ by using a two-component Hamiltonian. However, since we used the coupled cluster approach for the calculation of $W_\mathcal{S}$ the choice of Hamiltonian was not the limiting factor in terms of the computational expense and thus the four-component Hamiltonian was employed. Since the Gaunt interaction cannot be included in the correlated calculations using {\sc DIRAC}, the Gaunt contribution calculated within the DHF approach is used as an order-of-magnitude estimate of the effect of the Breit contribution to the calculated $W_\mathcal{S}$ in the uncertainty evaluation.

\begin{table}
    \setlength{\tabcolsep}{18pt}
    \caption{Influence of treatment of relativistic effects on $W_\mathcal{S}$ ($e/4\pi\epsilon_0 a_0^4$) in the $X$~$^1\Sigma^+$ ground state of AcF calculated at the DHF level using the dyall.cv3z basis set.}
    \label{Ws_comp_rel}
    \begin{tabular}{lcc}
    \hline
    Method   & $W_\mathcal{S}$\\
    \hline
    X2C      & --9545\\  
    4C       & --9610\\
    4C+Gaunt & --9644\\
    \hline
    \end{tabular}
    \end{table}

\subsection{Electron correlation effects}
The impact of the perturbative treatment of triple-excitation amplitudes (T) on $W_\mathcal{S}$ is shown to be rather small across all basis sets presented in Supplementary Table~\ref{Ws_comp_exp}, not exceeding a 1$\%$ and justifying the neglect of the higher order excitations. 
To estimate the effect of the limited active space, we calculated the difference between the results obtained when correlating electrons occupying orbitals with energies above $-5$~$E_\mathrm{h}$ (corresponding to correlating 28 electrons) and a virtual space cut-off of 50\,$E_\mathrm{h}$, and those obtained correlating all electrons with a corresponding virtual space cut-off of 4000\,$E_\mathrm{h}$ using dyall.cv3z basis set. 
The final value was corrected to the obtained difference of 165.98~$e/4\pi\epsilon_0 a_0^4$ (Table~\ref{Ws_comp_elcorr}). 

\begin{table}
    \setlength{\tabcolsep}{18pt}
    \caption{Influence of the active and the virtual space on the $W_\mathcal{S}$ ($e/4\pi\epsilon_0 a_0^4$) in the $X$~$^1\Sigma^+$ ground state of AcF calculated at the CCSD(T) level using the dyall.cv3z basis set.}
    \label{Ws_comp_elcorr}
    \begin{tabular}{lcccc}
    \hline
    Basis set  &  Number of correlated electrons&Virtual cut-off& $W_\mathcal{S}$\\
    \hline
     cv3z&28&50&--7568\\
     cv3z&28&50&--7563\\
     cv3z&98&4000&--7402\\
    \hline
    \end{tabular}
    \end{table}

\subsection{Molecular Geometry and Vibrational Effects}
We compared $W_\mathcal{S}$ obtained at different bond lengths, calculated at the coupled cluster level (using the dyall.cv3z without additional exponents and correlating 28 electrons with a virtual-space cut-off at 50\,$E_\mathrm{h}$). The results are presented in Supplementary Table~\ref{Ws-vibr}. The calculated value of $W_\mathcal{S}$ is strongly affected by $R$, meaning that vibrational effects become important. 
We calculated the vibrational correction was obtained using the VibCal routine from {\sc DIRAC}, using the values at different radii obtained within the ZORA-cGHF method (Table~\ref{tab:dvrprops}), as it provided a denser set of $W_\mathcal{S}(R)$ values. This calculation was performed for $R_e$ = 3.974\,$a_0$ and found to be 2.42$\%$ of the total $W_\mathcal{S}$ value. This correction has a positive sign and aligns with the trend of $W_\mathcal{S}$ variation with bond length, as presented in Supplementary Table~\ref{Ws-vibr}. We do not correct the recommended final $W_\mathcal{S}$ value by this contribution, calculated using a lower-level theoretical description, but rather include it in the uncertainty evaluation below.  

 \begin{table}
    \setlength{\tabcolsep}{35pt}
        \centering
        \caption{Influence of the interatomic bond length $R$ ($a_0$) on the calculated value of $W_\mathcal{S}$ ($e/4\pi\epsilon_0 a_0^4$) in the $X$~$^1\Sigma^+$ ground state of AcF at the CCSD(T) level.}
        \label{Ws-vibr}
        \begin{tabular}{cccc}
        \hline
               $R$  & $W_\mathcal{S}$   \\
               \hline
              3.500&$-17286$\\
              3.750&$-13005$\\
              3.964&$-7764$\\
              3.974&$-7568$\\
              3.984&$-7323$\\
              4.000&$-6897$\\
              4.038&$-5923$\\
        \hline
        \end{tabular}
    \end{table}

\subsection{Final value}

The final value of $W_\mathcal{S}$, corrected for the incompleteness of the active space and for the difference between the singly and the doubly augmented results, is shown in Supplementary Table~\ref{Final_Value_Ws}. 
The comparison of the present results (obtained with the final basis set \textbf{s-aug-cv3z}$+$\textbf{4s3p} including all corrections and correlating 28 electrons with a virtual-space cut-off at 50\,$E_\mathrm{h}$) to the earlier values from Refs.~\cite{Skripnikov_2020,Chen2024SchiffMoments} is presented in Supplementary Table~\ref{Ws_comp}. Overall, these three results are in good agreement with each other, as can be expected from the fact that in all cases the coupled cluster method was used. The main methodological differences between the three calculations are in the treatment of relativity (4C framework here vs. generalized relativistic effective core pseudopotentials in Ref.~\cite{Skripnikov_2020} and X2C in Ref.~\cite{Chen2024SchiffMoments}) and the choice of basis set. Another important factor is, however, the equilibrium bond length, taken as 4.0\,$a_0$ in both earlier works. That value was obtained in Ref.~\cite{Skripnikov_2020} via scalar relativistic CCSD calculations. In the present work, we used an equilibrium bond length of $R_e=3.974\,a_0$, optimized at the DC-CCSD(T) level. This difference can partially explain the discrepancy between our value and the results from Ref.~\cite{Chen2024SchiffMoments}, where the computational method is very similar to that employed here.

\begin{table}
\setlength{\tabcolsep}{5pt}
        \centering
        \caption{Final theoretical value of $W_\mathcal{S}$ ($e/4\pi\epsilon_0 a_0^4$) including active space and augmentation corrections. The DC-CCSD and DC-CCSD(T) results are obtained at the s-aug-dyall.cv3z+4s3p basis sets, correlating 28 electrons and including virtual orbitals up to 50\,$E_\mathrm{h}$.}\label{Final_Value_Ws}
        \begin{tabular}{lll}
        \hline
    Method & $W_\mathcal{S}$  \\
  \hline
    DC-CCSD     &   $-8141$\\
    DC-CCSD(T)  &   $-8206$\\
    +$\Delta$active space &\phantom{--}166\\
    +$\Delta$augmentation&\phantom{--}23\\
        \hline
       Final result &$-8017$\\
             \hline
        \end{tabular}
    \end{table} 

 \begin{table}[h]
    \setlength{\tabcolsep}{30pt}
        \centering
        \caption{Comparison of the equilibrium bond length $R_e$\,($a_0$) and the sensitivity factor $W_\mathcal{S}$ ($e/4\pi\epsilon_0 a_0^4$) of the $^1\Sigma^+$ ground state of AcF to the nuclear Schiff moment calculated with the CCSD(T) method in this work (TW) and in literature.}
        \label{Ws_comp}
        \begin{tabular}{cccc}
        \hline
         $R_e$ ($a_0$)    & $W_\mathcal{S}$  & Ref.\\
         \hline
          4.0           &   $-8240$   & ~\cite{Skripnikov_2020}\\
          4.0           &   $-7399$   & ~\cite{Chen2024SchiffMoments}\\
          3.974         &   $-8017$   & TW \\
        \hline
        \end{tabular}
    \end{table}

\subsection{Uncertainty evaluation}
The main sources of uncertainty in the calculations are the incompleteness of the employed basis set, the approximations in treating electron correlation, the missing higher-order relativistic effects, and the uncertainty in the equilibrium geometry. These are assumed to be largely independent of each other and hence are investigated separately. A summary of these sources of uncertainty can be found in Supplementary Table~\ref{Tab:uncertainty-all}.

\subsubsection{Basis set: cardinality}
We approximate the cardinality incompleteness error as half of the difference between the cv4z and cv3z results. 

\subsubsection{Basis set: augmentation}
The uncertainty due to the possibly insufficient number of diffuse functions is evaluated as the difference between the results obtained using the doubly augmented and the singly augmented dyall.cv3z basis sets.

\subsubsection{Basis set: steep functions}
We estimated the uncertainty due to the neglect of the necessary steep functions by computing $W_\mathcal{S}$ on the level of 2c-ZORA-cGHF and DHF with the basis set s-aug-cv3z+4s5p11d that has very dense exponents up to approximately $10^{9}$ (Supplementary Table~\ref{tab:exponents_spd}). This basis set is considered converged for $W_\mathcal{S}$ as there is no large difference between the values computed with 2c-ZORA-cGHF and DHF and additional functions up to $10^{11}$ do not change the values. The uncertainty contribution is taken as the relative difference between the value of $W_\mathcal{S}$ computed  with the s-aug-cv3z+4s5p11d basis set and the final s-aug-cv3z+4s3p basis set (Supplementary Table~\ref{Ws_comp_exp}), which is $\approx3\%$. These calculations were performed within the DHF approach, due to the prohibitive expense of coupled cluster calculations with such a large basis.  We want to highlight here the importance of steep functions for the angular momentum d, which were not found to be of significant importance for other molecular species in previous studies, including Refs.~\cite{zulch2023,gaul:2024b}. 

\subsubsection{Correlation: virtual space}
We corrected our calculated values for the effect of correlating all the electrons; we thus do not consider an uncertainty from this source. To evaluate the uncertainty stemming from using a virtual space of limited size, we use the difference between $W_\mathcal{S}$ results calculated with a virtual space cut-off or 50\,$E_\mathrm{h}$ and 4000\,$E_\mathrm{h}$. 
These calculations were performed with the dyall.cv3z basis set and correlating 28 electrons. 

\subsubsection{Relativity: QED effects} 

We use the Gaunt contribution, calculated at the DHF level (Supplementary Table~\ref{Ws_comp_rel}), as an order-of-magnitude estimate of the missing higher order relativistic effects. 

\subsection{Geometry Uncertainty}
The calculated value of $W_\mathcal{S}$ is highly sensitive to the chosen equilibrium bond length. Since the equilibrium bond length has not yet been experimentally determined, it is necessary to account for this uncertainty in our analysis. In our earlier work Ref.~\cite{KyuPasEli25} we used the same procedure as was employed here to optimize the geometries of the lighter homologs of AcF, ScF, YF, and LaF. For ScF and LaF, accurate experimental geometries are known, and the relativistic coupled cluster equilibrium bond lengths were found to agree within a few \AA with the measured value. 

Thus, to estimate the effect of geometry variation on $W_\mathcal{S}$ calculated here, we took the difference between its values computed at $R_e\pm$0.01\,$a_0$, a range that exceeds the theoretical uncertainty in equilibrium bond length found in Ref.~\cite{KyuPasEli25}. This difference is then used as an estimate of the uncertainty associated with the choice of equilibrium bond length.

The vibrational contribution was obtained using the Vibcal module of DIRAC, based on ZORA calculations of $W_\mathcal{S}$ at different bond lengths. Given the harmonic approximation and the limited accuracy of the underlying ZORA-cGHF calculations, we decided to continue with a conservative approach: the resulting contributions were not added directly to the final value, but were incorporated into the overall uncertainty estimate.

The final uncertainty is obtained using the root sum of squares  expression, combining the above terms and assuming that they are independent. Our conservative uncertainty is about 7 percent, dominated by the basis set incompleteness and the uncertainty in the equilibrium geometry.  

   \begin{table}[h]
    \setlength{\tabcolsep}{5pt}
        \centering
        \caption{Estimated uncertainties of the sensitivity $W_\mathcal{S}$ ($e/(4\pi\epsilon_0 a_0^4)$) of AcF to the nuclear Schiff moment of Ac.}
        \label{Tab:uncertainty-all}
        \begin{tabular}{llccc}
        \hline
        Category    & Error source   &  Absolute change in $W_\mathcal{S}$ & Uncertainty contribution ($\%$) \\
        \hline
        Basis set   & Cardinality   & \phantom{--}76.14    & 0.95 \\
                    & Augmentation  & \phantom{--}22.59     & 0.28\\
                    & steep functions & \phantom{--}300.95  & 3.75\\
        Correlation & Virtual space   & \phantom{--}5.16   & 0.06\\ 
        Relativity  & Gaunt interaction  & \phantom{--}47.54 & 0.59\\
        Vibrational correction &         & \phantom{--}192.41 & 2.42\\
        Geometry variance& &\phantom{--}440& 5.49\\
        \hline
        Total uncertainty&&\phantom{--}574 & 7.16\\
        \hline
         \end{tabular}
    \end{table}

\section{Suppl. Note: 2c-ZORA-cGHF}

\begin{table}
\caption{Additional exponents used in Dyall's core valence basis set of triple-$\zeta$ quality dyall.cv3z for Ac in the 2c-ZORA computations.}
\label{tab:exponents}
\begin{tabular}{
c
l
}
\toprule
    & {Ac} \\
\midrule
\multirow{8}{*}{s}
& $ 5.8013075200\times 10^{9}$ \\
& $ 1.9337691733\times 10^{9}$ \\
& $ 6.4458972444\times 10^{8}$ \\
& $ 2.1486324148\times 10^{8}$ \\
& $ 7.1621080494\times 10^{7}$ \\
& $ 6.4817077847\times 10^{7}$ \\
& $ 3.6728043050\times 10^{7}$ \\
& $ 8.8769410766\times 10^{-3}$ \\
\\
\multirow{8}{*}{p}
& $ 5.1125554000\times 10^{9}$ \\
& $ 1.7041851333\times 10^{9}$ \\
& $ 5.6806171111\times 10^{8}$ \\
& $ 1.8935390370\times 10^{8}$ \\
& $ 6.3117967901\times 10^{7}$ \\
& $ 5.7121760951\times 10^{7}$ \\
& $ 3.2855580700\times 10^{7}$ \\
& $ 6.3855296969\times 10^{-3}$ \\
\\
\multirow{1}{*}{d}
& $ 1.4655539211\times 10^{-2}$ \\
\\
\multirow{1}{*}{f}
& $ 4.2024059298\times 10^{-2}$ \\
\\
\multirow{1}{*}{g}
& $ 6.8022308289\times 10^{-2}$ \\%
\bottomrule
\end{tabular}
\end{table}

\begin{table}
\caption{Additionally added and removed exponents to and from Dyall's core valence basis set of triple-$\zeta$ quality with single augmentation (s-aug-dyall.cv3z) for Ac. Only the diffuse functions of the basis are taken until the factor between two exponents is less than $2.5$ and functions are added in an even-tempered manner with a factor of $2.5$ until an exponent of magnitude $1\times10^9$ is reached for the angular momenta s, p and d. For the augmentation a factor of $3.0$ was taken. In Tab.~\ref{Ws_comp_exp} the label s-aug-cv3z+4s5p11d was used.}
\label{tab:exponents_spd}
\begin{tabular}{
c l l
}
\toprule
    & {Added exponents} & {Removed exponents} \\
\midrule
\multirow{8}{*}{s}
& $ 1.2280975098\times 10^{9}$  & $ 5.8013075200\times 10^{7}$  \\
& $ 4.9123900391\times 10^{8}$  & $ 1.5443010900\times 10^{7}$  \\
& $ 1.9649560156\times 10^{8}$  & $ 5.2852762300\times 10^{6}$  \\
& $ 7.8598240625\times 10^{7}$  &   \\
& $ 3.1439296250\times 10^{7}$  &   \\
& $ 1.2575718500\times 10^{7}$  &   \\
& $ 5.0302874000\times 10^{6}$  &   \\
& $ 8.8769410766\times 10^{-3}$ &   \\
\\
\multirow{12}{*}{p}
& $ 1.8719484305\times 10^{9}$  & $ 5.1125554000\times 10^{7}$\\
& $ 7.4877937222\times 10^{8}$  & $ 1.4585607400\times 10^{7}$\\
& $ 2.9951174889\times 10^{8}$  & $ 4.5338065000\times 10^{6}$\\
& $ 1.1980469955\times 10^{8}$  & $ 1.5179413500\times 10^{6}$\\
& $ 4.7921879822\times 10^{7}$  & $ 5.3893866500\times 10^{5}$\\
& $ 1.9168751929\times 10^{7}$  & $ 2.0114827600\times 10^{5}$\\
& $ 7.6675007715\times 10^{6}$  \\
& $ 3.0670003086\times 10^{6}$  \\
& $ 1.2268001234\times 10^{6}$  \\
& $ 4.9072004937\times 10^{5}$  \\
& $ 1.9628801975\times 10^{5}$  \\
& $ 6.3855296969\times 10^{-3}$ \\
\\
\multirow{15}{*}{d}
& $ 1.0673084371\times 10^{9}$ & $ 9.5122399400\times 10^{9}$\\
& $ 4.2692337483\times 10^{8}$ & $ 2.2502156300\times 10^{8}$\\
& $ 1.7076934993\times 10^{8}$ & $ 7.3395734800\times 10^{8}$\\
& $ 6.8307739973\times 10^{7}$ \\
& $ 2.7323095989\times 10^{7}$ \\
& $ 1.0929238396\times 10^{7}$ \\
& $ 4.3716953583\times 10^{6}$ \\
& $ 1.7486781433\times 10^{6}$ \\
& $ 6.9947125732\times 10^{5}$ \\
& $ 2.7978850293\times 10^{5}$ \\
& $ 1.1191540117\times 10^{5}$ \\
& $ 4.4766160469\times 10^{4}$ \\
& $ 1.7906464188\times 10^{4}$ \\
& $ 7.1625856750\times 10^{3}$ \\
& $ 1.4655539211\times 10^{-2}$ \\
\\
\multirow{1}{*}{f}
& $ 4.2024059298\times 10^{-2}$ \\
\\
\multirow{1}{*}{g}
& $ 6.8022308289\times 10^{-2}$ \\%
\bottomrule
\end{tabular}
\end{table}

Supplementary~Table~\ref{tab:exponents} lists the additional exponents used in Dyall's core valence basis set used for the actinium atom in the 2c-ZORA calculations. Calculated changes in the sensitivity to $P,T$-odd properties as a function of vibrational state in the $X$~$^1\Sigma^+$ electronic ground state are shown in Supplementary~Table~\ref{tab:ptodd_vib}. The potential wavenumbers and the enhancement factors of the $P,T$-odd properties discussed here as a function of the internuclear distance used in the DVR are listed in Table~\ref{tab:dvrprops}.

\begin{figure}
    \centering
    \includegraphics[width=0.9\textwidth]{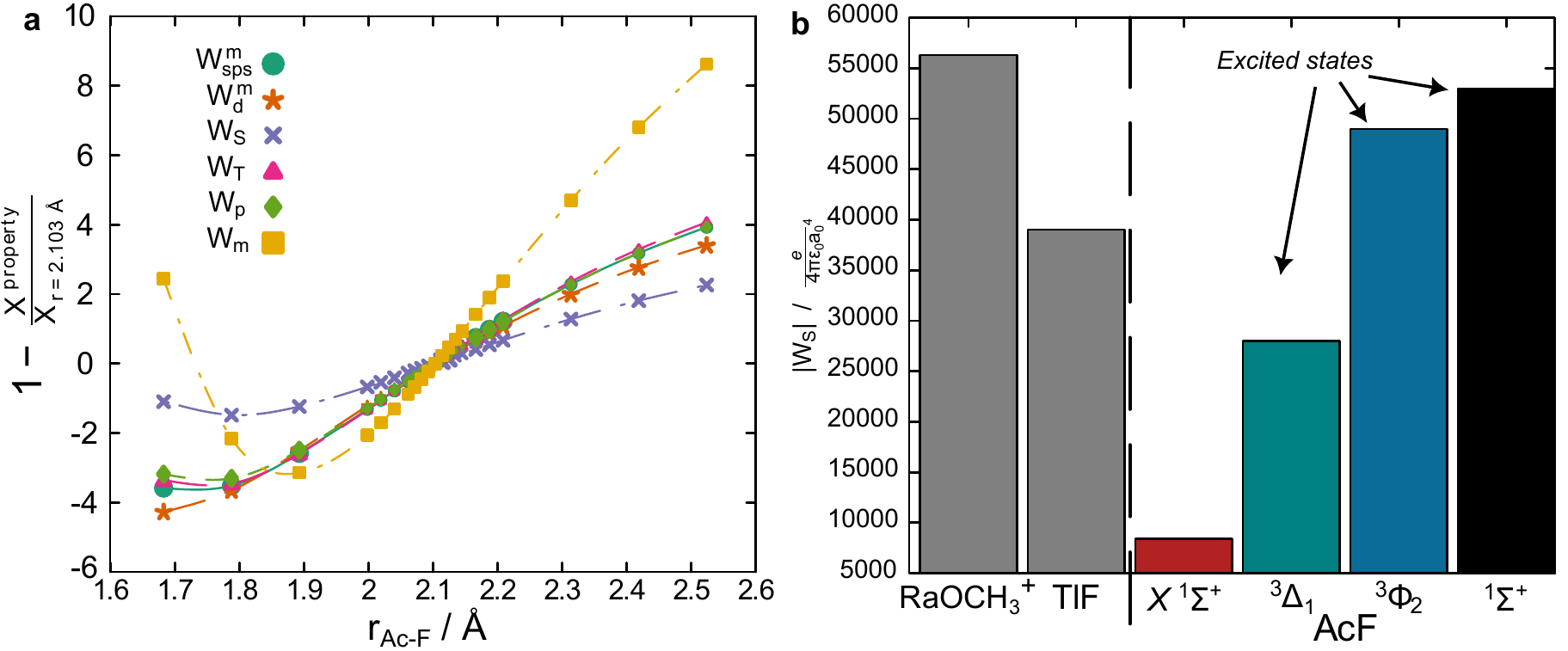}
    \caption{\textbf{(a)} Molecular sensitivity factors $W_i$ for $P,T$-violating properties as a function of bond length in the $X$~$^1\Sigma^+$ ground state of AcF computed on the level of 2c-ZORA-cGKS-BHandH. The equilibrium bond length is $R_e=2.1035\,\AA=3.98\,a_0$. 
    \textbf{(b)} Comparison of the molecular sensitivity to the nuclear Schiff moment ($W_\mathcal{S}$) for different electronic states in AcF computed at the level of 2c-ZORA-cGHF based on Table~4 in \textit{Methods}, with literature values for TlF \cite{Gaul2024GlobalCP_new,Chen2024SchiffMoments,Abe2020TlF,Hubert2022TlF} and RaOCH$_3^+$ \cite{gaul:2024b}. $W_\mathcal{S}$ increases in the excited electronic states of AcF studied herein, but the suitability of excited states for precision experiments is dependent on their lifetime, among other factors.}
    \label{fig:figMethods2cZORAcGHF}
\end{figure}

\begin{table}[!htb]
\centering
\caption{Sensitivity factors to $P,T$-odd properties for the seven lowest vibrational states of the $X$~$^1\Sigma^+$ ground state in AcF, obtained at the level of 2c-ZORA-cGKS employing the BHandH functional. The zero-point vibrational corrections to the electronic properties are given as gs-$\nu$(0) at the equilibrium bond length $R_\mathrm{e}=2.1035\,\AA=3.98\,a_0$, obtained at the level of 2c-ZORA-cGKS-BHandH.}
\label{tab:ptodd_vib}
\begin{tabular}{
c
S[table-format=-5.1,round-mode=figures,round-precision=3]
S[table-format=-1.4e-1,round-mode=figures,round-precision=3]
S[table-format=-1.4e-1,round-mode=figures,round-precision=3]
S[table-format=-6.1,round-mode=figures,round-precision=3]
S[table-format=-5.1,round-mode=figures,round-precision=3]
S[table-format=-2.3,round-mode=figures,round-precision=3]
S[table-format=-2.3,round-mode=figures,round-precision=3]
}
\toprule
{$\nu$} 
& {$E_\mathrm{vib}\ /\ \Delta\tilde{\nu}_\mathrm{T}$}
& {$W_\mathrm{sps}^\mathrm{m}\ /\ \mathrm{peV}$}
& {$W_\mathrm{d}^\mathrm{m}\ /\ \mathrm{GV/cm}$}
& {$W_\mathcal{S}\ /\ \frac{e}{4\pi\epsilon_0 a_0^4}$}
& {$W_\mathrm{T}\ /\ \mathrm{peV}$}
& {$W_\mathrm{p} / \mathrm{feV}$}
& {$W_\mathrm{m} / \frac{\mathrm{kV}\eta_\mathrm{p}}{\mathrm{cm}\mu_\mathrm{N}}$}
\\
\midrule 
0  & 277   &  -0.00280506 & -0.00113873 & -8454.96 & -1.848 & -7.546  &29.18\\
1  & 830   &  -0.00258898 & -0.00106954 & -8049.8 & -1.697 & -6.949   &22.41\\
2  & 1381  &  -0.00238448 & -0.00100422 & -7664.79 & -1.554 & -6.383  &15.93\\
3  & 1929  &  -0.00216133 & -0.00093189 & -7250.377 & -1.398 & -5.769 &9.103\\
4  & 2471  &  -0.00188879 & -0.00084153 & -6756.9985 & -1.209 & -5.023&1.244\\
5  & 3003  &  -0.00159953 & -0.00074498 & -6236.3916 & -1.009 & -4.232&-6.97\\
6  & 3524  &  -0.00134235 & -0.00065976 & -5768.06 &  -.830 & -3.528  &-14.49\\
\midrule 
gs-$\nu$(0)& & -0.00012   & -0.000041   & -235.25   & -.089  & -0.35  &3.82\\
\bottomrule
\end{tabular}
\end{table}

\begin{table}[!htb]
\centering
\caption{Potential wavenumber $\Delta\tilde{\nu}$ and sensitivity factors $W_i$ to $P,T$-odd properties as a function of the internuclear distance $R_\mathrm{e}$ of the $X$~$^1\Sigma^+$ ground state in AcF, obtained at the level of 2c-ZORA-cGKS employing the BHandH functional as used in the computation of the vibrational corrections via a DVR.}
\label{tab:dvrprops}
\begin{tabular}{
c
S[table-format=-5.1,round-mode=figures,round-precision=3]
S[table-format=-1.4e-1,round-mode=figures,round-precision=3]
S[table-format=-1.4e-1,round-mode=figures,round-precision=3]
S[table-format=-6.1,round-mode=figures,round-precision=3]
S[table-format=-5.1,round-mode=figures,round-precision=3]
S[table-format=-2.3,round-mode=figures,round-precision=3]
S[table-format=-2.3,round-mode=figures,round-precision=3]
}
\toprule
 {$R_\mathrm{e}\ /\ a_0$}
& {$\Delta\tilde{\nu}\ /\ \mathrm{cm}^{-1}$} 
& {$W_\mathrm{sps}^\mathrm{m}\ /\ \mathrm{peV}$}
& {$W_\mathrm{d}^\mathrm{m}\ /\ \mathrm{GV/cm}$}
& {$W_\mathcal{S}\ /\ \frac{e}{4\pi\epsilon_0 a_0^4}$}
& {$W_\mathrm{T}\ /\ \mathrm{peV}$}
& {$W_\mathrm{p} / \mathrm{feV}$}
& {$W_\mathrm{m} / \frac{\mathrm{kV}\eta_\mathrm{p}}{\mathrm{cm}\mu_\mathrm{N}}$}
\\
\midrule 
3.1800  &  27443.56 & -0.01344  & -0.00622  & -18236.78 &  -8438.9 &  -33.0620 &  -47.5 \\
3.3788  &  13026.60 & -0.01324  & -0.00549  & -21617.51 &  -8632.8 &  -34.0163 &  104.1 \\
3.5775  &   4901.91 & -0.01051  & -0.00414  & -19445.74 &  -6970.7 &  -27.6009 &  136.7 \\
3.7763  &   1042.20 & -0.00679  & -0.00262  & -14540.74 &  -4529.6 &  -18.0660 &  101.0 \\
3.8161  &    646.33 & -0.00601  & -0.00232  & -13405.29 &  -4009.6 &  -16.0286 &   89.1 \\
3.8558  &    352.47 & -0.00523  & -0.00203  & -12244.26 &  -3487.7 &  -13.9829 &   76.2 \\
3.8956  &    151.86 & -0.00446  & -0.00174  & -11066.34 &  -2966.9 &  -11.9399 &   62.4 \\
3.9154  &     84.07 & -0.00407  & -0.00160  & -10472.75 &  -2706.9 &  -10.9197 &   55.2 \\
3.9353  &     36.76 & -0.00369  & -0.00146  & -9879.164 &  -2449.4 &   -9.9087 &   47.9 \\
3.9552  &      9.04 & -0.00331  & -0.00132  & -9284.486 &  -2192.6 &   -8.9004 &   40.5 \\
3.9751  &      0.00 & -0.00293  & -0.00118  & -8690.353 &  -1937.4 &   -7.8983 &   33.0 \\
3.9949  &      8.75 & -0.00256  & -0.00104  & -8096.220 &  -1684.0 &   -6.9031 &   25.4 \\
4.0148  &     34.45 & -0.00219  & -0.00091  & -7503.721 &  -1432.5 &   -5.9150 &   17.6 \\
4.0347  &     76.29 & -0.00182  & -0.00078  & -6912.855 &  -1183.2 &   -4.9350 &    9.9 \\
4.0546  &    133.52 & -0.00146  & -0.00065  & -6326.891 &   -936.9 &   -3.9671 &    2.1 \\
4.0943  &    291.47 & -0.00074  & -0.00039  & -5159.318 &   -449.5 &   -2.0507 &  -13.7 \\
4.1341  &    503.12 & -0.00005  & -0.00015  & -4008.628 &     26.5 &   -0.1786 &  -29.5 \\
4.1738  &    763.24 &  0.00063  &  0.00009  & -2877.542 &    491.1 &    1.6492 &  -45.3 \\
4.3726  &   2637.09 &  0.00375  &  0.00118  & 2427.7225 &   2627.0 &   10.0590 & -122.0 \\
4.5713  &   5151.64 &  0.00639  &  0.00209  & 7048.9997 &   4438.9 &   17.2015 & -191.4 \\
4.7701  &   7993.87 &  0.00858  &  0.00284  & 10963.416 &   5941.3 &   23.1291 & -251.4 \\
\bottomrule
\end{tabular}
\end{table}

\section{Suppl. Note: Nuclear density functional theory}
In accordance with Refs.~\cite{Dobaczewski2014, Dobaczewski2018}, we summarize the important relations needed to estimate the intrinsic nuclear Schiff moment of $^{227}$Ac and its uncertainty as follows. In the linear regression analysis, the relation between the intrinsic Schiff moments $S_{\rm{int}}$ and the intrinsic electric octupole moment $Q_{30}$ is modeled according to
\begin{equation}
    S_{\rm{int}}=a+bQ_{30},
    \label{Schiff-octupole}
\end{equation}
where the best values of coefficients $a$ and $b$, i.e., $\Bar{a}$ and $\Bar{b}$, are determined by the minimization of the penalty function $\chi^{2}$,
\begin{equation}
    \chi^{2}=\sum_{i}\left[S_{\rm{int}}(i)-a-bQ_{30}(i)\right]^{2},
\end{equation}
where $i=1,\;...,\;N_{d}$. Here $N_{d}=7$ is the number of the functionals employed in this work. For each functional $i$, the calculated intrinsic Schiff and octupole moments are denoted by $S_{\rm{int}}(i)$ and $Q_{30}(i)$, respectively. Supplementary~Table~\ref{tab:Intrinsic Schiff and Octupole moments} summarizes the intrinsic Schiff, $S_{\rm{int}}(i)$, and octupole, $Q_{30}(i)$, moments for each functional $i$. The best values of coefficients $a$ and $b$ are given by
\begin{equation}\label{eq:Regression coeffs a and b}
    \Bar{a}=\langle S_{\rm{int}}\rangle -\Bar{b}\langle Q_{30}\rangle\;\;\;\;\;\text{and}\;\;\;\;\;\Bar{b}=\frac{\langle S_{\rm{int}}Q_{30}\rangle-\langle S_{\rm{int}}\rangle\langle Q_{30}\rangle}{\langle\left[Q_{30}(i)\right]^{2}\rangle-\langle Q_{30}\rangle^{2}},
\end{equation}
where the averaged quantities, $\langle\mathcal{O}\rangle$, are defined as follows:
\begin{align}
    \langle S_{\rm{int}}\rangle&=\frac{1}{N_{d}}\sum_{i}S_{\rm{int}}(i),\;\;\;\langle Q_{30}\rangle=\frac{1}{N_{d}}\sum_{i}Q_{30}(i),\nonumber\\
    \langle S_{\rm{int}}Q_{30}\rangle&=\frac{1}{N_{d}}\sum_{i}S_{\rm{int}}(i)Q_{30}(i),\;\;\;\langle\left[Q_{30}(i)\right]^{2}\rangle=\frac{1}{N_{d}}\sum_{i}\left[Q_{30}(i)\right]^{2}.
\end{align}

\begin{figure}[h]
    \centering
    \includegraphics[width=0.59\textwidth]{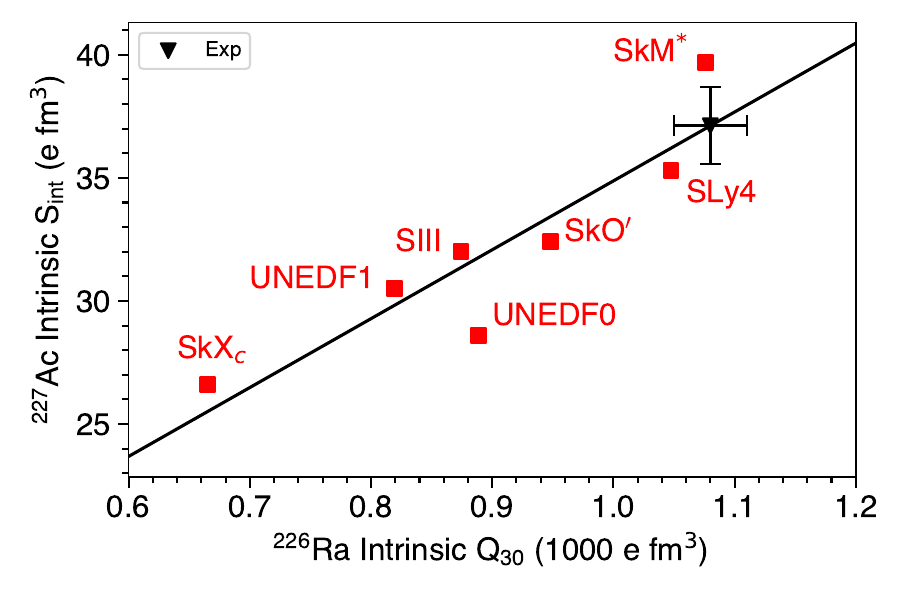}
    \caption{The unconstrained predictions (red squares) of the Skyrme functionals SkX$_{c}$~\cite{AlexBrown1998}, SkM$^{\ast}$~\cite{Bartel1982a}, UNEDF1~\cite{Kortelainen2012}, SIII~\cite{Beiner1975a}, SkO$'$~\cite{Reinhard1999}, SLy4~\cite{Chabanat1998a}, and UNEDF0~\cite{Kortelainen2010} for the intrinsic octupole moments in $^{226}$Ra and the intrinsic Schiff moments in $^{227}$Ac. The inverse black triangle indicates the measured octupole moment of $^{226}$Ra, $Q_{30}(\text{exp})=1080(30)\,e$\,fm$^{3}$~\cite{Dobaczewski2018}. The estimated error of the intrinsic Schiff moment and experimental error of the measured octupole moment are depicted by vertical and horizontal bars, respectively.}
    \label{fig:S0_Ac227}
\end{figure}

\begin{table}[h]
\begingroup

\setlength{\tabcolsep}{10pt} 
\renewcommand{\arraystretch}{1.5} 
\caption{Intrinsic octupole moments, $Q_{30}$ (in 1000\,$e$\,fm$^{3}$), of $^{226}$Ra and intrinsic Schiff moments, $S_{\rm{int}}$ (in $e$\,fm$^{3}$), of $^{227}$Ac determined for Skyrme functionals SkX$_{c}$, SkM$^{\ast}$, UNEDF1, SIII, SkO$'$, SLy4, and UNEDF0.} \label{tab:Intrinsic Schiff and Octupole moments}
\begin{tabular}{ccc}
\hline
Functional ($i$) & $Q_{30} (i)$ & $S_{\rm{int}}(i)$\\
\hline
SkX$_{c}$ & 0.6654 & 26.6\\
SkM$^{\ast}$ & 1.0763 & 39.7\\
UNEDF1 & 0.8194 & 30.5\\
SIII & 0.8742 & 32.0\\
SKO$'$ & 0.9482 & 32.4\\
SLy4 & 1.0475 & 35.3\\
UNEDF0 & 0.8887 & 28.6\\
\hline
\end{tabular}
\endgroup
\end{table}

\noindent Since the error estimates for the calculated moments are not determined, the penalty function $\chi^{2}$ must be normalized by the Birge factor $s$ defined according to 
\begin{equation}
    \chi^{2}_{\text{norm}}=\frac{\chi^{2}}{s}\;\;\;\text{for}\;\;\;s=\frac{\chi^{2}_{0}}{N_{d}-N_{p}},
\end{equation}
where $\chi^{2}_{0}$ stands for the value of the penalty function evaluated using the best values $\Bar{a}$ and $\Bar{b}$, and $N_{p}$ is the number of parameters, which is 2 in this work. Accordingly, the covariance matrix of the fit is defined as
\begin{equation}
    \mathcal{C}=s\mathcal{M}^{-1},
\end{equation}
where $\mathcal{M}$ is the Hessian matrix with components given by
\begin{equation}
    \mathcal{M}_{ij}=\left.\frac{1}{2}\frac{\partial}{\partial p_{i}}\frac{\partial}{\partial p_{j}}\chi^{2}\right|_{\textbf{p}=\Bar{\textbf{p}}},
\end{equation}
where $\textbf{p}\equiv\left\{p_{i}\right\}$, with $i=1,\;...,\;N_{p}$, is a set of parameters. The components of the covariance matrix $\mathcal{C}$ are given by
\begin{align}
    \mathcal{C}_{aa}&=\frac{s}{N_{d}}\frac{\langle \left[Q_{30}\right]^{2}\rangle}{\langle\left[Q_{30}\right]^{2}\rangle-\langle Q_{30}\rangle^{2}},\label{eq:Caa}\\
    \mathcal{C}_{ab}&=\mathcal{C}_{ba}=-\frac{s}{N_{d}}\frac{\langle Q_{30}\rangle}{\langle\left[Q_{30}\right]^{2}\rangle-\langle Q_{30}\rangle^{2}},\label{eq:Cab}\\
    \mathcal{C}_{bb}&=\frac{s}{N_{d}}\frac{1}{\langle\left[Q_{30}\right]^{2}\rangle-\langle Q_{30}\rangle^{2}}\label{eq:Cbb}.
\end{align}
Supplementary~Table~\ref{tab:Regression coeffs a and b} summarizes the regression coefficients $\Bar{a}$ and $\Bar{b}$ together with the components of covariance matrix, i.e., $\mathcal{C}_{aa}$, $\mathcal{C}_{ab}$, $\mathcal{C}_{ba}$, and $\mathcal{C}_{bb}$.

\begin{table}[h]
\begingroup

\setlength{\tabcolsep}{10pt} 
\renewcommand{\arraystretch}{1.5} 
\caption{Regression coefficients $\bar{a}$ and $\bar{b}$, Eq.~\eqref{eq:Regression coeffs a and b}, and components of the covariance matrix, i.e., $\mathcal{C}_{aa}$, $\mathcal{C}_{ab}$, $\mathcal{C}_{ba}$, and $\mathcal{C}_{bb}$, Eqs.~\eqref{eq:Caa}-\eqref{eq:Cbb}.} \label{tab:Regression coeffs a and b}
\begin{tabular}{ccccc}
\hline
$\bar{a}$ & $\bar{b}$ & $\mathcal{C}_{aa}$ & $\mathcal{C}_{ab}=\mathcal{C}_{ba}$ & $\mathcal{C}_{bb}$\\
\hline
6.8899 & 27.9872 & 29.9701 & -32.5268 & 36.0282\\
\hline
\end{tabular}
\endgroup
\end{table}

\noindent The estimated intrinsic Schiff moment, $S_{\rm{int}}(\text{est})$, is obtained by inserting the experimental octupole moment, 
\begin{equation}
    Q_{30}(\text{exp})\equiv(\bar{Q}_{30}(\text{exp})\pm\Delta Q_{30}(\text{exp}))\; [1000\;e\;\text{fm}^{3}]=(1.080\pm 0.030)\; [1000\;e\;\text{fm}^{3}],
\end{equation}
into Eq. \eqref{Schiff-octupole}:
\begin{equation}
    S_{\text{int}}(\text{est})=\bar{a}+\bar{b}\cdot\bar{Q}_{30}(\text{exp}).
    \label{Schiff est - Octupole exp}
\end{equation}
The theoretical uncertainty $\Delta S_{\rm{int}}(\text{the})$ in the Schiff moment $S_{\rm{int}}(\text{est})$ is related to the components of the covariance matrix $\mathcal{C}$ according to
\begin{equation}
    \left[\Delta S_{\rm{int}}(\text{the})\right]^{2}=\mathcal{C}_{aa}+2\mathcal{C}_{ab}\bar{Q}_{30}(\text{exp})+\mathcal{C}_{bb}\left[\bar{Q}_{30}(\text{exp})\right]^{2}.
\end{equation}
The experimental uncertainty $\Delta S_{\rm{int}}(\text{exp})$ in the estimated Schiff moment $S_{\rm{int}}(\text{est})$ is given according to Eq. \eqref{Schiff est - Octupole exp} by
\begin{equation}
    \Delta S_{\rm{int}}(\text{exp})=\bar{b}\cdot\Delta Q_{30}(\text{exp}).
\end{equation}
The total uncertainty $\Delta S_{\rm{int}}(\text{est})$ is calculated as
\begin{equation}
    \Delta S_{\rm{int}}(\text{est})=\sqrt{\left[\Delta S_{\rm{int}}(\text{the})\right]^{2}+\left[\Delta S_{\rm{int}}(\text{exp})\right]^{2}}.
\end{equation}
The estimated intrinsic Schiff moment of $^{227}$Ac is $37.1(16)\;e$\,fm$^{3}$. 

The coefficients $a_{0}$, $a_{1}$, $a_{2}$, $b_{1}$, and $b_{2}$ of the laboratory Schiff moment, $S_{\text{lab}}$, of $^{227}$Ac are also determined using linear regression analysis from the measured octupole moment of $^{226}$Ra. For each functional $i$, the coefficients $a_{0}(i)$, $a_{1}(i)$, $a_{2}(i)$, $b_{1}(i)$, and $b_{2}(i)$ are calculated using Eq.~13 of the main text, and the results are listed in Supplementary~Table~\ref{tab:Intrinsic octupole moments and coefficients of S}. The best-fit values $\Bar{a}$ and $\Bar{b}$ of the regression coefficients together with the elements of the covariance matrix, i.e., $\mathcal{C}_{aa}$, $\mathcal{C}_{ab}$, $\mathcal{C}_{ba}$, and $\mathcal{C}_{bb}$ for each coefficient of the laboratory Schiff moment, $S_{\text{lab}}$, are summarized in Supplementary~Table~\ref{tab:Regression coefficients a and b and covariance matrix of coefficients of S}. Finally, the coefficients of the laboratory Schiff moment and their uncertainties can be obtained as listed in Table~1 of the main text.

\begin{table}[h]
\begingroup

\setlength{\tabcolsep}{10pt} 
\renewcommand{\arraystretch}{1.5} 
\caption{Intrinsic octupole moments , $Q_{30}$ (in 1000\,$e$\,fm$^{3}$), of $^{226}$Ra and the coefficients of laboratory Schiff moment $S$, i.e., $a_{0}$, $a_{1}$, $a_{2}$, $b_{1}$, and $b_{2}$ (all in $e$\,fm$^{3}$), of $^{227}$Ac determined for Skyrme functionals SkX$_{c}$, SkM$^{\ast}$, UNEDF1, SIII, SkO$'$, SLy4, and UNEDF0.} \label{tab:Intrinsic octupole moments and coefficients of S}
\begin{tabular}{ccccccc}
\hline
Functional ($i$) & $Q_{30} (i)$ & $a_{0}(i)$ & $a_{1}(i)$ & $a_{2}(i)$ & $b_{1}(i)$ & $b_{2}(i)$\\
\hline
SkX$_{c}$ & 0.6654 & $6.18689817$ & $-18.4596944$ & $6.83460511$ & $-0.00673417$ & $-0.02998421$\\
SkM$^{\ast}$ & 1.0763 & $4.72201539$ & $-11.36047506$ & $14.53041758$ & $-0.55010364$ & $0.87819081$\\
UNEDF1 & 0.8194 & $6.50793708$ & $-20.95717418$ & $9.18944974$ & $-0.01520532$ & $-0.07080941$\\
SIII & 0.8742 & $6.64312003$ & $-14.49375863$ & $4.00226241$ & $0.06625816$ & $-0.20207772$\\
SKO$'$ & 0.9482 & $3.70025544$ & $-17.11450181$ & $4.29846393$ & $0.0300548$ & $-0.19463846$\\
SLy4 & 1.0475 & $3.49336508$ & $-15.45979809$ & $2.66548083$ & $0.27570809$ & $-0.25444395$\\
UNEDF0 & 0.8887 & $8.01758036$ & $-19.66594059$ & $11.83476339$ & $-0.43935948$ & $0.57810763$\\
\hline
\end{tabular}
\endgroup
\end{table}

\begin{table}[h]
\begingroup

\setlength{\tabcolsep}{10pt} 
\renewcommand{\arraystretch}{1.5} 
\caption{Same as in Supplementary~Table~\ref{tab:Regression coeffs a and b} but for the coefficients of laboratory Schiff moment, $S_{\text{lab}}$, i.e., $a_{0}$, $a_{1}$, $a_{2}$, $b_{1}$, and $b_{2}$ (all in $e$\,fm$^{3}$), of $^{227}$Ac.} \label{tab:Regression coefficients a and b and covariance matrix of coefficients of S}
\begin{tabular}{cccccc}
\hline
 & $a_{0}$ & $a_{1}$ & $a_{2}$ & $b_{1}$ & $b_{2}$\\
\hline
$\Bar{a}$ & $11.9326$ & $-30.9222$ & $4.3099$ & $0.2701$ & $-0.8016$\\
$\Bar{b}$ & $-7.0030$ & $15.6564$ & $3.6688$ & $-0.4003$ & $0.9993$\\
$\mathcal{C}_{aa}$ & $15.8100$ & $51.3788$ & $163.8945$ & $0.7084$ & $1.5112$\\
$\mathcal{C}_{ab}$ & $-17.1587$ & $-55.7619$ & $-177.8762$ & $-0.7688$ & $-1.6401$\\
$\mathcal{C}_{bb}$ & $19.0058$ & $61.7645$ & $197.0242$ & $0.8516$ & $1.8166$\\
\hline
\end{tabular}
\endgroup
\end{table}

For each functional, the standard form factor $f(r)$ of the zero-range density-dependent pairing force, i.e., \cite{Dobaczewski2004}
\begin{equation}\label{eq:pairing formfactor}
    f(r)=V_{0}+V_{1}\rho^{\alpha}(r),
\end{equation}
is used in the HFB calculations. The second term of Eq. \eqref{eq:pairing formfactor} is omitted in this work. The parameters $V_{0}$ for neutrons and protons, i.e., $V_{0,n}$ and $V_{0,p}$, are adjusted to reproduce the experimental pairing gaps of $^{229}$Th and $^{227}$Ac, respectively. Supplementary Table \ref{tab:Neutron and proton pairing strengths} provides the values of parameters $V_{0,n}$ and $V_{0,p}$ for Skyrme functionals SkX$_{c}$, SkM$^{\ast}$, UNEDF1, SIII, SkO$'$, SLy4, and UNEDF0.  

\begin{table}[h]
\begingroup

\setlength{\tabcolsep}{10pt} 
\renewcommand{\arraystretch}{1.5} 
\caption{Neutron and proton pairing strengths, $V_{0,n}$ and $V_{0,p}$, adjusted for Skyrme functionals SkX$_{c}$, SkM$^{\ast}$, UNEDF1, SIII, SkO$'$, SLy4, and UNEDF0.} \label{tab:Neutron and proton pairing strengths}
\begin{tabular}{ccc}
\hline
 Functional & $V_{0,n}$ & $V_{0,p}$\\
\hline
SkX$_{c}$ & $139.02$ & $173.63$ \\
SkM$^{\ast}$ & $181.46$ & $216.25$ \\
UNEDF1 & $145.35$ & $169.80$ \\
SIII & $181.15$ & $220.19$ \\
SKO$'$ & $163.82$ & $184.34$ \\
SLy4 & $207.76$ & $231.89$ \\
UNEDF0 & $130.70$ & $156.45$\\
\hline
\end{tabular}
\endgroup
\end{table}

\begin{landscape}
\begin{figure}[h!]
    \centering
    \includegraphics[width=1.4\textwidth]{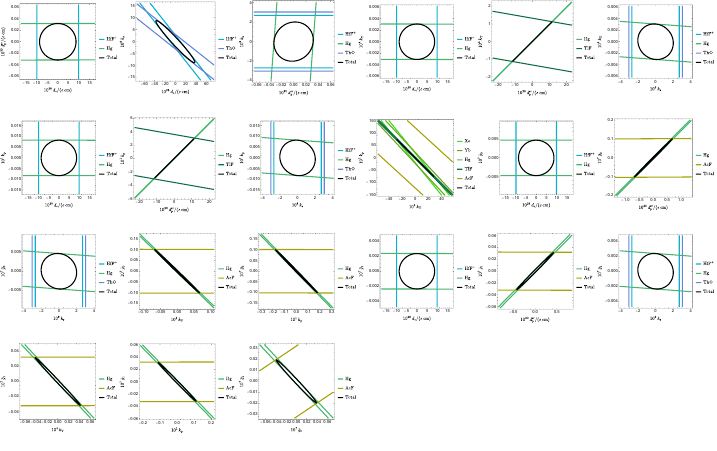}
    \caption{
    Cuts of all two-dimensional subspaces of the full
 seven-dimensional \textit{CP}-odd parameter space for an experimental precision of a proposed $^{227}$AcF experiment of 1\,mHz. The electronic sensitivity factors for $^{227}$AcF used in the analysis, computed at the level of ZORA-cGKS-BHandH, are shown in Table~4 of the main text, the calculated nuclear structure parameters are given in Table~1 of the main text, and for the volume interaction with the short-range contribution to the proton EDM we use $R_\mathrm{vol} \approx 3.22\,\mathrm{fm}^2$, as discussed in the main text. A theoretical uncertainty of 20\%\ is considered, and the molecular sensitivity factors to all \textit{CP}-odd properties in the hyperspace are scaled by a factor of $0.8$ to account for a worst-case scenario. The impact of a proposed $^{227}$AcF experiment is shown in a global analysis including existing experiments using $^{129}$Xe~\cite{allmendinger:2019,sachdeva:2019}, $^{171}$Yb~\cite{zheng:2022}, $^{133}$Cs~\cite{murthy:1989}, $^{199}$Hg~\cite{Graner2016}, $^{205}$Tl~\cite{regan:2002}, $^{255}$Ra~\cite{parker:2015,Bishof2016}, $^{174}$YbF~\cite{hudson:2002,hudson:2011}, $^{180}$HfF$^{+}$~\cite{roussy:2023}, $^{205}$TlF~\cite{cho:1991}, $^{207}$PbO~\cite{eckel:2013}, and $^{232}$ThO~\cite{andreev:2018}, employing the experimental data given in the respective references within the global analysis. All electronic structure parameters for these experiments are taken from Ref.~\cite{Gaul2024GlobalCP_new} and were combined, where available, with nuclear structure data from Ref.~\cite{Chupp2019}. In other cases, rough estimates of nuclear structure parameters from Ref.~\cite{Gaul2024GlobalCP_new} were employed. The seven-dimensional ellipsoid is computed at the 95$\%$ confidence level.
 }
    \label{fig:figGlobalAnalysis_SI1}
\end{figure}

\begin{figure}[h!]
    \centering
    \includegraphics[width=1.4\textwidth]{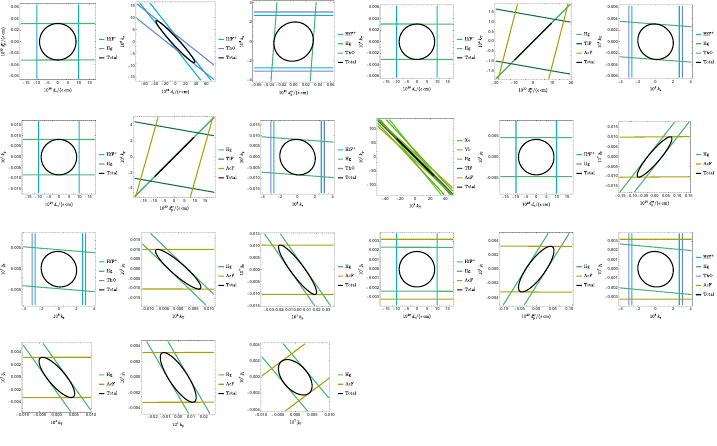}
    \caption{
    Cuts of all two-dimensional subspaces of the full
 seven-dimensional \textit{CP}-odd parameter space for an experimental precision of a proposed $^{227}$AcF experiment of 0.1\,mHz. The electronic sensitivity factors for $^{227}$AcF used in the analysis, computed at the level of ZORA-cGKS-BHandH, are shown in Table~4 of the main text, the calculated nuclear structure parameters are given in Table~1 of the main text, and for the volume interaction with the short-range contribution to the proton EDM we use $R_\mathrm{vol} \approx 3.22\,\mathrm{fm}^2$, as discussed in the main text. A theoretical uncertainty of 20\%\ is considered,  and the molecular sensitivity factors to all \textit{CP}-odd properties in the hyperspace are scaled by a factor of $0.8$ to account for a worst-case scenario. The impact of a proposed $^{227}$AcF experiment is shown in a global analysis including existing experiments using $^{129}$Xe~\cite{allmendinger:2019,sachdeva:2019}, $^{171}$Yb~\cite{zheng:2022}, $^{133}$Cs~\cite{murthy:1989}, $^{199}$Hg~\cite{Graner2016}, $^{205}$Tl~\cite{regan:2002}, $^{255}$Ra~\cite{parker:2015,Bishof2016}, $^{174}$YbF~\cite{hudson:2002,hudson:2011}, $^{180}$HfF$^{+}$~\cite{roussy:2023}, $^{205}$TlF~\cite{cho:1991}, $^{207}$PbO~\cite{eckel:2013}, and $^{232}$ThO~\cite{andreev:2018}, employing the experimental data given in the respective references within the global analysis. All electronic structure parameters for these experiments are taken from Ref.~\cite{Gaul2024GlobalCP_new} and were combined, where available, with nuclear structure data from Ref.~\cite{Chupp2019}. In other cases, rough estimates of nuclear structure parameters from Ref.~\cite{Gaul2024GlobalCP_new} were employed. The seven-dimensional ellipsoid is computed at the 95$\%$ confidence level.
 }
    \label{fig:figGlobalAnalysis_SI2}
\end{figure}
\end{landscape}
